\documentclass[11pt,a4paper]{amsart}

\usepackage{amsfonts}
\usepackage{amsthm}
\usepackage{amsmath}
\usepackage{amssymb}
\usepackage{amscd}
\usepackage{appendix}
\usepackage{bm}
\usepackage{bbm}
\usepackage{braket}
\usepackage{caption}
\usepackage{enumitem}
\usepackage[mathscr]{eucal}
\usepackage{faktor}
\usepackage{float}
\usepackage[hidelinks]{hyperref}
\usepackage{indentfirst}
\usepackage[latin2]{inputenc}
\usepackage{marvosym}
\usepackage{mathtools}
\usepackage{multicol}
\usepackage{t1enc}
\usepackage{tcolorbox}
\usepackage{tikz}
\usepackage{xcolor}

\definecolor{myred}{RGB}{255,204,203}

\newcommand{\overbar}[1]{\mkern +2.5mu\overline{\mkern-0.3mu#1\mkern+2.5mu}\mkern 0.5mu}

\def\C{\mathcal{C}}

\numberwithin{equation}{section}
\usepackage[margin=0.9in]{geometry}

\DeclareMathOperator\U{U}
\DeclareMathOperator\SU{SU}

\DeclareMathOperator\Hom{Hom}
\DeclareMathOperator\End{End}
\DeclareMathOperator\tr{tr}
\DeclareMathOperator\id{id}
\DeclareMathOperator\nul{nul}

\DeclareMathOperator\TY{TY}

\theoremstyle{plain}
\newtheorem{thm}{Theorem}[section]
\newtheorem{lemma}[thm]{Lemma}
\newtheorem{cor}[thm]{Corollary}
\newtheorem{prop}[thm]{Proposition}

\theoremstyle{definition}

\newtheorem{remark}[thm]{Remark}
\newtheorem{question}[thm]{Question}
\newtheorem{ex}[thm]{Example}
\newtheorem{heuristic}[thm]{Heuristic}
\newtheorem{convention}[thm]{Convention}
\newtheorem{approach}[thm]{Approach}
\newtheorem{caveat}[thm]{Caveat}

\title{Topological Quantum Teleportation and Superdense Coding -- Without Braiding}
\author[Sachin J. Valera]{Sachin J. Valera $^{\dagger}$}
\thanks{$^{\dagger}$\textit{Center for Quantum and Topological Systems, New York University Abu Dhabi}}

\begin{document}

\begin{abstract}
We present the teleportation and superdense coding protocols for a family of anyon theories coming from Tambara-Yamagami categories, of which the lowest rank theories describe Ising anyons. In contrast to the usual approach to anyonic computation, we relax the requirement that we should be able to braid anyons. This is motivated by the goal of designing basic protocols that require less control over quasiparticles, and which may therefore be amenable to realisation in near-term systems. Since these implementations are braid-free, they are also compatible with Majorana modes on a 1d quantum wire.
\vspace{-8mm}
\end{abstract}

\maketitle

\setcounter{tocdepth}{2}
\tableofcontents

\vspace{-8mm}

\section{\textbf{Introduction}}

\subsection{Motivation}
\label{motivesec}
Anyons are localised quasiparticles, expected to arise in $(2+1)$-dimensional condensed matter systems. An anyon can be either \textit{abelian} or \textit{nonabelian}. When the state space of two anyons has dimension $d>1$, both anyons are said to be nonabelian; else, if $d=1$ then at least one of the anyons is abelian. A sequence of particle exchanges (henceforth referred to as \textit{braiding})\footnote{Viewed in $(2+1)$-dimensions, exchanging anyons amounts to braiding their worldlines.} will realise the action of a $\U(d)$-operator on their state space. The ability to control systems hosting nonabelian anyons holds particular allure, as it would amount to manipulating (topologically protected) \textit{qudits}.\footnote{A \textit{qudit} is the $d$-ary generalisation of a qubit (i.e. a quantum state in a $d$-dimensional space).}\\
It is predicted that anyons should manifest in a variety of settings. One of the most pursued directions is via the fractional quantum Hall effect, whose low-energy excitations were argued by Halperin and Arovas \textit{et al.} (1984) to be anyons \cite{halperin,arovas}. To date, it appears that the only widely accepted evidence for the existence of anyons comes from the $\nu=1/3$ filling of the fractional quantum Hall effect, where Nakamura \textit{et al.} and Bartolomei \textit{et al.} (2020) observed signatures of \textit{abelian} anyons \cite{nakamura, bartolomei}.\\
In the \textit{nonabelian} case, perhaps the most promising candidate for near-term realisation is the \textit{Ising anyon} (see for instance the experiments of Willett \textit{et al.} (2023) which probe $\nu=5/2,7/2$ \cite{willett}). A closely related quasiparticle is the \textit{Majorana zero mode} (MZM) (see the review in \cite{dSmzm}), which has two distinct incarnations:
\begin{enumerate}[label=(\roman*)]
    \item MZMs arise as Ising anyons in $(2+1)$-d systems. They are a realisation of a non-invertible simple object in a \textit{braided} unitary fusion category $\TY(\mathbb{Z}_{2})$ (see Section \ref{tyanyonsec}).
    \item MZMs arise as quasiparticles protected by an energy gap in $(1+1)$-d systems. The idea of trapping MZMs at the ends of $1$d quantum wires is due to Kitaev (2001) \cite{kitaev01}. Here, MZMs have the same fusion rules as Ising anyons, and are a realisation of a non-invertible simple object in a \textit{non-braided} unitary fusion category $\TY(\mathbb{Z}_{2})$.\footnote{Schemes have been proposed for realising the nonabelian braiding statistics of MZMs which are confined to move in $1$d, by setting up networks of wires, or by simulating their braiding via a series of measurements \cite{dSmzm,bondersonMO}.} 
\end{enumerate}

\noindent At present, experimental evidence for the existence of Ising anyons and MZMs remains inconclusive. This doubt has emphasised the value of the following diagnostic\footnote{This diagnostic is a paraphrasing of quotes from \cite[Chapter 37.4]{simonbook} and \cite[28:45-29:10]{wangvideo}.} -- 

\begin{tcolorbox}
    The realisation and manipulation of topological qubits could demonstrate the existence of MZMs in the underlying system.
\end{tcolorbox}

\noindent The main results of this paper will be applicable to both incarnations of MZMs described above.

\subsection{Problem and approach} With the above in mind, we set about the task of designing simple quantum protocols using nonabelian anyons; namely, \textit{teleportation} and its dual protocol, \textit{superdense coding}. These foundational protocols seem like an obvious choice: they are simple, and their successful realisation is a surefire indicator that we have manipulated quantum resources. For instance, superdense coding immediately places us outside the paradigm of classical information theory: it demonstrates that an entanglement-assisted quantum channel is able to send $2$ bits per channel use (in comparison with Shannon's classical bound of $1$ bit per channel use). Both protocols rely on the presence of a \textit{maximally entangled pair of qudits}, which we call an \textit{e-dit}. 
\vspace{-0.5mm}
\begin{subequations}
\label{benloix}
    \begin{align}
        p& \ \text{e-dits} + 2p \ \text{cdits} \rhd p \ \text{qudits} \label{bloi1} \\
        p& \ \text{e-dits} + p \ \text{qudits} \rhd 2p \ \text{cdits} \label{bloi2}
    \end{align}
\end{subequations}

\noindent The content of the teleportation and superdense coding protocols is captured by (\ref{bloi1})-(\ref{bloi2}) respectively (cf. \textit{Bennett's laws}). A \textit{cdit} (also called a \textit{dit}) is the $d$-ary generalisation of a classical bit, and $X+Y\rhd Z$ may be read as "resources $X$ and $Y$ can be used to send resource $Z$". We briefly summarise the protocols for $p=1$. In the following, let $A$ and $B$ denote disjoint regions of a system, respectively containing qudits $q_{1}$ and $q_{2}$ which are maximally entangled.
\begin{enumerate}[label=(\alph*)]
\item \textit{Teleportation}. A third qudit $q_{0}$ with density operator $\rho_{0}$ is contained in $A$, and can be sent to $B$ as follows: $q_{0},q_{1}$ are measured in an entangled basis, yielding cdits $i,j$ which are sent to $B$. A unitary (conditioned on $i,j$) is locally applied to $q_{2}$, whence $q_{2}$ has state $\rho_{0}$.  
\item \textit{Superdense coding}. A unitary (conditioned on chosen cdits $i,j$) is locally applied to $q_1$. Then, $q_{1}$ is sent to $B$, and $q_{1},q_{2}$ are measured in an entangled basis, yielding cdits $i,j$.
\end{enumerate}
$A$ and $B$ are typically represented by people named \textit{Alice} and \textit{Bob}, who possess (rather than engulf) the qudits. One might also designate a central region or middle-man, \textit{Charlie}, who distributes the maximally entangled qudits between Alice and Bob (either beforehand or on-demand).\footnote{There is also the possibility that Charlie is one of Alice or Bob, or that Alice and Bob share the halves of their e-dits \textit{in situ} before they part.} We adopt this presentation in the sequel. The case with the most obvious practical applications is $d=2$, where we have bits, qubits, ebits. For instance, we can encode a qubit in (the fusion state of) a pair of Ising anyons (or MZMs). 

\subsubsection{Dispensing with braiding} Our task is to design teleportation and superdense coding protocols using nonabelian anyons. Suppose we have at our disposal a system of anyons whose braiding operations are universal for computation, e.g. Fibonacci anyons. Then one might question the value of constructing explicit implementations of these protocols, as the required quantum gates for both circuits can be compiled as braids via the Solovay-Kitaev algorithm.\\
However, should we wish to use these protocols as a proof of concept or as a diagnostic in near-term systems, then they should be designed to be \textit{as simple and economical as possible}. A drastic step in this direction is not only to spurn intricate braids,\footnote{For example, see the $2$-qubit CNOT gate as realised by the braiding of Fibonacci anyons \cite[Figure 3]{Bonesteel}. Both of our protocols require CNOT gates.} but to remove the need to braid quasiparticles altogether. Our main contribution then, is a \textit{braid-free implementation} of both protocols. This should significantly lessen the the required degree of control over our system: in principle, we should only need the ability to
\begin{enumerate}[label=(\arabic*)]
\item \textit{Pair-create} quasiparticles from the vacuum \label{simpop1} 
\item \textit{Fuse} adjacent quasiparticles
\item \textit{Measure} fusion outcomes \label{simpop3}
\end{enumerate}

\subsubsection{Tambara-Yamagami theories} By seeking to dispense with braiding during anyon teleportation, we show in Section \ref{tynatsec} that we are naturally led to consider a family of anyon theories $\TY(\mathbb{Z}_{2}^{n})$ for $n\geq1$, which we call the \textit{Tambara-Yamagami theories} (Section \ref{tyanyonsec}). Each of these theories contains a single nonabelian anyon which we call a \textit{Tambara-Yamagami anyon}. A pair of Tambara-Yamagami anyons encodes a qudit with $d=2^{n}$, and so braid-free $d$-ary implementations of both protocols are presented for $n\geq1$. The simplest case $n=1$ corresponds to the Ising theories, and is the most relevant for physical realisation. Since the above operations \ref{simpop1}-\ref{simpop3} solely rely on the fusion structure of the Tambara-Yamagami theories, our implementation for $n=1$ is also compatible with MZMs on $1$d quantum wires. We thus make contact with our initial motivation from Section \ref{motivesec}.\\
The simplicity of our implementations for both protocols hinges on several convenient properties inherent to the fusion structure of Tambara-Yamagami categories, e.g. pairs of Tambara-Yamagami anyons can be maximally entangled by recoupling alone (Section \ref{maxentangsec}). In the case of Ising anyons, we also see that the braided versions of the protocols can be implemented economically, using braids of negligible length. For instance, in the most costly scenario, the $N$-anyon teleportation protocol for Ising anyons requires Bob to perform an $(N+2)$-braid of length $2N$ (Section \ref{brcorsec}).

\subsection{Main results and outline of paper} We briefly outline the contents of the sections that follow, and state the main theorems therein.\\
\noindent Sections \ref{prelims}-\ref{standsec} constitute the preliminaries. In Section \ref{prelims}, we review the algebraic theory of anyons  and its accompanying graphical calculus. In Section \ref{tyanyonsec}, we introduce the anyon theories that will be of primary interest; namely, the Tambara-Yamagami theories. In Section \ref{standsec}, we introduce notation and terminology for operators and diagrams that appear frequently in the sequel.\\
\noindent Sections \ref{telechap}-\ref{techsec} contain the novel content, with the main results in Sections \ref{telechap}-\ref{sdcchap}. In Section \ref{telechap}, we determine a procedure for teleporting the state of $N$ Tambara-Yamagami anyons (of which Ising anyons are a special case). Similarly, we find a braid-free implementation of the $d$-ary superdense coding protocol in Section \ref{sdcchap}. In particular, both procedures only rely on having capabilities \ref{simpop1}-\ref{simpop3} listed above. This is summarised by the following theorems.\\ 

\noindent\textbf{Theorem \ref{brfreetelethm} (Braid-free qudit teleportation using Tambara-Yamagami anyons)}. \phantom{boo}\\
\textit{Consider a Tambara-Yamagami theory of rank $d+1$, where $d=2^{n}$. The fusion state of $N$ Tambara-Yamagami anyons (that is, an $N/2$-qudit state, where $N$ is even) can be teleported via the procedure shown in Figure \ref{palominoscruff}. In particular, no braiding is required.}\\
\vspace{-2mm}

\noindent\textbf{Theorem \ref{brfreesdcthm} (Braid-free $d$-ary superdense coding using Tambara-Yamagami anyons)}. \textit{Consider a Tambara-Yamagami theory of rank $d+1$ where $d=2^{n}$.  We can realise the $d$-ary superdense coding protocol using Tambara-Yamagami anyons as shown in Figure \ref{sdcbrfreediag}. In particular, no braiding is required.}\\
\vspace{-3mm}

\noindent Animated examples of the braid-free \textit{$1$-qubit} teleportation and superdense coding procedures are found at \cite{sjvyt1,sjvyt2}. These examples illustrate the processes shown in Figures \ref{TQT-1dit} and \ref{sdcbrfreediag}, but with a key variation applied: instead of splitting fermions from Ising anyons, fermions are pair-created from the vacuum, and so only operations \ref{simpop1}-\ref{simpop3} are required. Indeed, there are various freedoms in how we can implement the protocols from Theorems \ref{brfreetelethm} and \ref{brfreesdcthm}, along with their braided counterparts for Ising anyons from Corollaries \ref{brtelecor} and \ref{brsdccor}. We elaborate on these in the main exposition.

\begin{caveat}[\textbf{Braid-free measurements}]
In braid-free implementations of the protocols, we assume that the measurement of topological charge can be performed directly (without braiding). For example, this might be done by bringing two anyons close together and measuring their energy. In contrast, many proposals for charge measurement involve Mach-Zehnder--like interferometry, wherein one or more probe anyons effectively wind around the region whose charge is to be determined \cite{interf1,interf2,interf3,bonderson-thesis}. Ising anyons are particularly amenable to the latter type of measurement, as the charge of a pair can be determined with just one probe anyon (see \cite{sv24} for a worked example).
\end{caveat}

\noindent In Section \ref{techsec}, we establish some properties of the operators introduced in Section \ref{standsec}, which in turn allows us to examine the individual components of the above protocols in more detail. We highlight the following three corollaries for Ising anyons: braid-free Pauli gates (Section \ref{paulisansbrsec}), and braided versions of the teleportation and superdense coding protocols (Figures \ref{palominoscruff-2}-\ref{sdcbrdiag}, Section \ref{brcorsec}). 

\subsection{Relation to previous work} 
\begin{itemize}
    \item Huang \textit{et al.} (2021) performed a quantum simulation (using superconducting qubits) of $1$-qubit teleportation in a system of MZMs \cite{huang}. In the simulated system, each logical qubit is encoded in a pair of MZMs at the ends of a Kitaev chain, and a modified teleportation circuit is implemented by braiding the MZMs.  
    \item Xu \& Zhou (2022) presented a multi-qubit teleportation protocol for Ising anyons that uses braiding operations \cite{xuzhou}. In Corollary \ref{brtelecor}, we recover a variation of their result where no braiding operations are required anywhere other than during Bob's correction step. 
    \item Abramsky \& Coecke (2004) viewed teleportation as an equivalence of string diagrams in the category of Hilbert spaces \cite{abracoke}; see also \cite{kindergarten,vicaryheunen}. In the traditional circuit formulation, teleportation is not necessarily an obvious phenomenon. Whereas in the categorical setting, it is the physical interpretation of an elementary graphical identity; namely, the `snake equations' for rigid monoidal categories. It is apparent that teleportation manifests as a zigzag-like flow of quantum information in spacetime, and in the anyonic setting, we see that this flow is tied to the pivotal structure of the underlying fusion category. Our approach to determining braid-free anyon teleportation protocols does not require knowledge of the skeletal data for the underlying anyon theories (that is, explicit matrix representations of morphisms), but instead leverages the utility of their graphical calculus. In this sense, our exposition formulates teleportation in a way that is similar to the spirit of \cite{abracoke}.
\end{itemize}

\subsection*{Acknowledgements} The author would like to thank Dmitri Nikshych and David Penneys for pointing him towards results on Tambara-Yamagami categories which are key to the exposition of this paper. The author is grateful for the support of Tamkeen under the NYU Abu Dhabi Research Institute grant CG008. This work is dedicated to M. Hilby.

\section{\textbf{Review of Algebraic Theory of Anyons}}
\label{prelims}

\noindent We review the algebraic theory of anyons (at least to the extent needed for our main exposition) and its accompanying graphical calculus: this framework is called a `braided $6j$ fusion system'.\footnote{No category theory is needed to read this section -- but as an aside, a braided 6j fusion system is also known as a skeleton of a braided fusion category, where an explicit choice of basis is made on all triangular Hom-spaces.} A comprehensive introduction is found in \cite[Parts I-II]{simonbook}. Useful summaries of said framework are also given in \cite[Chapter 2]{bonderson-thesis}, \cite[Chapter 4]{wangbook} and \cite[Appendix E]{kitaev06}. \\

\noindent A theory of anyons has an underlying finite set of labels $\mathfrak{L}=\{0,a,b,\ldots\}$ that represent their distinct possible `types' or \textit{(topological) charges}, where trivial label $0$ represents the vacuum. Two anyons of charge $a$ and $b$ can generally have total charge $a\times b=\sum_{c}N^{ab}_{c}c$, where the \textit{fusion coefficients} $N^{ab}_{c}\in\mathbb{Z}_{\geq0}$ are finite constants specified by the given theory. The summation indicates that the total charge of two or more anyons is \textit{possibly} a superposition of charges. It always holds that $\sum_{c}N^{ab}_{c}\geq1$ and $N^{a0}_{b}=N^{0a}_{b}=\delta_{ab}$. Fusion `$\times$' is \textit{commutative} and \textit{associative}. Each $a\in\mathfrak{L}$ has a unique \textit{dual} charge $\bar{a}$ such that $N^{a\bar{a}}_{0}=N^{\bar{a}a}_{0}=1$. Coefficients $N^{ab}_{c}$ satisfy symmetries (\ref{twissta}).\footnote{Symmetry (i) comes from the commutativity of fusion, while (ii) comes from associativity together with the existence of a dual. Physically, (iii) is a manifestation of CPT symmetry (see Section \ref{lbopsec}).}
        \begin{equation}
            \text{(i)} \ N^{ab}_{c}=N^{ba}_{c} \ \ , \ \ \text{(ii)} \ N^{ab}_{c}=N^{b\bar{c}}_{\bar{a}}=N^{\bar{c}a}_{\bar{b}} \ \ , \ \ \text{(iii)} \ N^{ab}_{c}=N^{\bar{b}\bar{a}}_{\bar{c}} 
            \label{twissta}
        \end{equation}
\noindent The set $\mathfrak{L}$ together with the associated fusion coefficients is referred to as the underlying \textit{anyon model} $(\mathfrak{L},\times)$ or \textit{fusion rules} of the theory.\\

\noindent Above, we mentioned that for a given pair of anyons, although the charge quantum number of each constituent particle may be fixed (say, $a$ and $b$), their joint charge may exist in a superposition. Such a superposition is described by a normalised state vector \eqref{fus-state} called the \textit{fusion state} of $a$ and $b$, where $c$ runs over the elements of $\mathfrak{L}$ such that $N^{ab}_{c}\neq0$, and $\mu=1,\ldots,N^{ab}_{c}$. At any given time, the amplitudes $\gamma_{c,\mu}$ will depend on the history of particles $a$ and $b$. The kets in \eqref{fus-state} (referred to as \textit{fusion channels}) define an orthonormal basis for the state space $V^{ab}$ of $a$ and $b$, and the probability that a charge measurement on the pair results in an outcome $(c,\mu)$ is $|\gamma_{c,\mu}|^{2}$. For a given $c$, the variable $\mu$ indexes the number of distinguishable ways in which $a$ and $b$ can fuse to $c$, and the subspace spanned by these fusion channels is denoted by  $V^{ab}_{c}$. Note that there is a $\U(N^{ab}_{c})$-freedom associated to choosing the fusion channels in $V^{ab}_{c}$, which is further discussed in Section \ref{gaugesec}.
\begin{equation} \sum_{c,\mu}\gamma_{c,\mu}\ket{ab\rightarrow c;\mu} 
\label{fus-state}
\end{equation}
\noindent Fusion channels $\{\ket{ab\rightarrow c;\mu}\}_{\mu}$ can be thought of as projection maps from $a\times b$ to summands $c$. Conversely, for any charge $c$, we may consider its inclusion maps $\{\bra{ab\rightarrow c;\mu}\}_{\mu}$ into any pair $a \times b$ when $N^{ab}_{c}\neq0$. These bras are referred to as \textit{splitting channels}, and physically correspond to splitting a particle $c$ into a pair of particles $a$ and $b$ in the initial fusion state $\ket{ab\rightarrow c;\mu}$. Any process requires the initialisation of at least one particle-antiparticle pair from the vacuum. \\

\noindent The physical operations that can be performed on a system of anyons are summarised in \ref{anyac1}-\ref{anyac4} below, each of which are further discussed in the remainder of this review. Anyons can
\vspace{1mm}
    \begin{enumerate}[label=(\arabic*)]
        \item \textit{Fuse}, i.e. a projective measurement of the fusion channel of two adjacent anyons.  \label{anyac1}
        \item \textit{Split}, i.e. one anyon can be split into a pair of adjacent anyons whose initial fusion channel is determined by the splitting channel. \label{anyac2}
        \item \textit{Braid}, i.e. a sequence of particle exchanges represented as worldlines in (2+1)-dimensions. \label{anyac3}
        \item \textit{Twist}, i.e. a $2\pi$ (clockwise or anticlockwise) self-rotation. \label{anyac4}    
    \end{enumerate}
    \vspace{1mm}
\noindent Given that our primary goal is to dispense with braiding when realising the protocols of interest, operations \ref{anyac3}-\ref{anyac4} are not of much relevance to the main results of this paper (with the exception of Section \ref{brcorsec}, where we recover the braided versions of these protocols for Ising anyons).  Nonetheless, they are included below for completeness of our basic review. On the other hand, pivotal structure and leg-bending operators (Section \ref{lbopsec}) may be concepts that are less familiar, but which are central to our main exposition.  

\subsection{Fusion spaces and quantum dimension}
\noindent The fusion Hilbert space of $a$ and $b$ is given by $V^{ab}= \bigoplus_{c} V^{ab}_{c}$ where $\dim(V^{ab}_{c})=N^{ab}_{c}$. Measuring the charge of such a pair projects their state onto a $1$-dimensional subspace of $V^{ab}_{c}$. The dual space of $V^{ab}_{c}$ is denoted $V^{c}_{ab}$, and its orthogonal basis elements can be thought of as the distinguishable ways in which a pair $a$ and $b$ may be initialised from $c$. Spaces of the form $V^{ab}_{c}, V^{c}_{ab}$ are called \textit{triangular}, and trivalent vertices represent their orthonormal basis elements.\footnote{The multiplicity index (typically Greek, given here by $\mu$) next to the vertex is omitted when $N^{ab}_{c}=1$.} 
\vspace{-1mm}
\begin{equation}
\raisebox{-8mm}{
\includegraphics[width=0.5\textwidth]{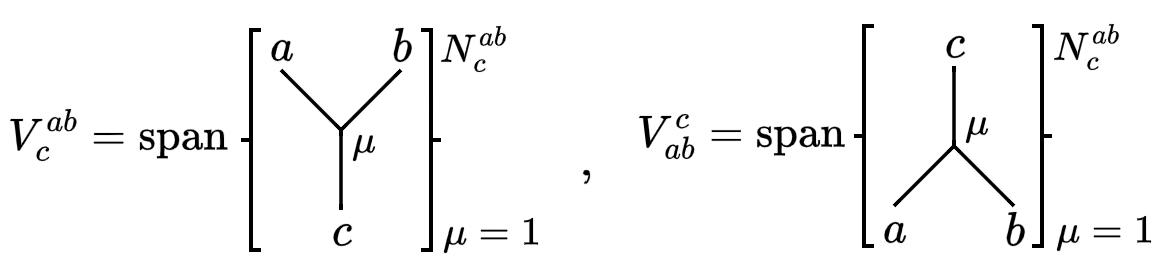}
}
\label{taro}
\end{equation}

\noindent Our convention is that diagrams should be parsed from \textit{top-to-bottom} (i.e.
the time axis runs \textit{downwards}). Any given line will be accompanied by a label $a\in\mathfrak{L}$ (unless it is obvious what the label should be) and should be interpreted as the worldline of an anyon with charge $a$; when $a=0$, the worldline is either dashed or omitted altogether. We may sometimes direct worldlines for clarity, although directing a worldline with a self-dual label is superfluous. In (\ref{fyxl}), we give diagrammatic expressions in $V^{ab}$ for (i) orthogonality, and (ii) the completeness relation (i.e. identity operator); given $a,b$ fixed, note that the set of all elements of form (iii) (which we refer to as \textit{jumping jacks}) defines an orthonormal basis of $\End(V^{ab})$ with respect to the trace inner product.
\vspace{-1mm}
\begin{equation}
\raisebox{-10mm}{
\includegraphics[width=0.7\textwidth]{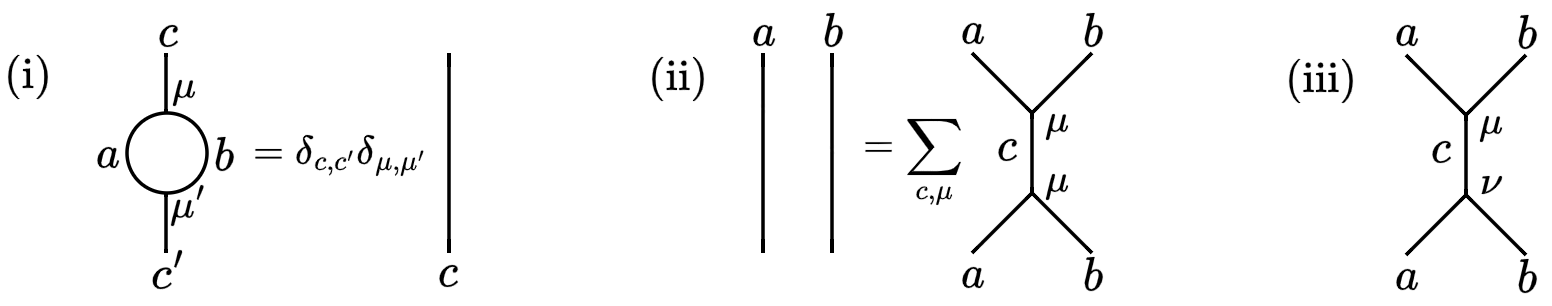}
}
\label{fyxl}
\end{equation}
The amplitude of a loop labelled by $a$ is the \textit{quantum dimension} $d_{a}>0$ of $a$.\footnote{That $d_{a}>0$ follows from the positive-definiteness of the inner product, as $d_{a}$ is a squared norm on $V^{a\bar{a}}_{0}$ or $V^{\bar{a}a}_{0}$. A Hermitian inner product structure is postulated on all vector spaces, as we are working with a quantum system.} We will later see that the quantum dimensions are encoded in the so-called $F$-symbols of a theory (Remark \ref{lavat}\ref{qdimfeq}). The quantum dimensions satisfy (\ref{qdimeq2})-(\ref{qdimeq1}).\footnote{Identity \eqref{qdimeq2} follows from the so-called pivotal structure of a theory: see Remark \ref{lavat}\ref{smeagol}.}$^{,}$\footnote{From (\ref{qdimeq2}) and the Frobenius-Perron theorem for nonnegative matrices, one can deduce that $d_{a}$ is the largest positive eigenvalue of matrix $N^{a}$ where $[N^{a}]_{bc}:=N^{ab}_{c}$.\label{fpdimfoot}}$^{,}$\footnote{Identity \eqref{qdimeq1} follows from combining the observation in footnote \ref{fpdimfoot} with the observation that $N^{\bar{a}}=(N^{a})^{T}$; the latter is seen using symmetries \eqref{twissta}. Alternatively, \eqref{qdimeq1} is a consequence of the `\textit{spherical property}', which says that the left and right trace of any diagram coincide (see e.g. \cite[Chapter 14.8.2]{simonbook}). The spherical property is possessed by all unitary 6j fusion systems (a unitary fusion category has a canonical spherical structure \cite{ENO,bartlett}). } Note that $d_0 =1$ by \eqref{qdimeq2}. 

\vspace{-12mm}

\begin{subequations}\label{fulleqdim}
    \begin{multicols}{2}
    \begin{equation}
            d_{a}d_{b}=\sum_{c}N^{ab}_{c}d_{c}   \label{qdimeq2}
        \end{equation}\break 
        \begin{equation}
            \raisebox{-4mm}{
\includegraphics[width=0.275\textwidth]{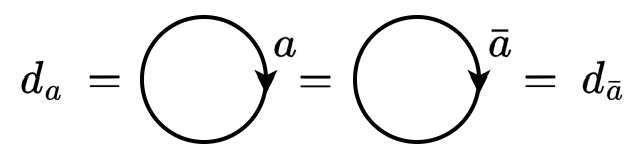}
}   \label{qdimeq1}
        \end{equation}       
    \end{multicols}
\end{subequations}
\vspace{-5mm}

\begin{remark}[\textbf{Abelian vs. nonabelian anyon}]
    A charge $a$ is called \textit{abelian} if $\sum_{c}N^{ab}_{c}=1$ for all $b\in\mathfrak{L}$, or equivalently, if
    \begin{equation}
        a\times \bar{a}=\bar{a}\times a =0
        \label{invobeq}
    \end{equation}
    Else, $a$ is called \textit{nonabelian}.\footnote{The nomenclature reflects the fact that an exchange operator (for an anyon $x\in\mathfrak{L}$ with any another anyon) acts on a $1$-dimensional space when $x$ is abelian: the resulting evolution is a phase factor, and thus commutes with other unitaries acting on the system. If $x$ is nonabelian, an exchange evolution is possibly a higher-dimensional unitary.}$^{,}$\footnote{Conformal field theorists refer to abelian anyons as `simple currents'. Meanwhile, in the language of fusion categories, a simple object $a$ which satisfies fusion rule \eqref{invobeq} is called an `invertible object'.} We see from \eqref{fulleqdim} that $d_{a}=1$ for $a$ abelian; else, $d_{a}\geq\sqrt{2}$.
\label{abvnab}    
\end{remark}

\begin{remark}[\textbf{Normalisation convention}] Unless stated otherwise, all trivalent vertices (\ref{taro}) implicitly carry a normalisation factor of $(d_{c}/d_{a}d_{b})^{1/4}$.\end{remark}

\noindent We write the fusion space of $k\geq2$ anyons $a_{1},\ldots,a_{k}$ as in (\ref{multifusdef}); the possible values for their total charge is independent of the order in which they are fused (by associativity), and are indexed here by $c$. Such a space can be understood by decomposing it into triangular spaces as in (\ref{canef}).
\begin{equation}
V^{a_{1},\ldots,a_{k}}=\bigoplus_{c}V^{a_{1},\ldots,a_{k}}_{c}
\label{multifusdef}
\end{equation}
\noindent A collection of $k$ anyons can be pairwise fused in $C_{k-1}$ different ways, where $C_{k}$ is the $k^{th}$ Catalan number.\footnote{$C_{k-1}$ can also be thought of as counting distinct parenthesisations of a length $k$ string. E.g. $|\{(ab)c,a(bc)\}|=2$.} Each distinct sequence defines a \textit{fusion basis}.\footnote{Graphically, a $k$-particle fusion basis is a full binary tree with $k$ leaves. There are two trees for $k=3$ -- see (\ref{fmovedef}).} Such a basis specifies a decomposition of the $k$-anyon space, and amounts to a choice of measurement basis. By associativity of fusion, all such decompositions are isomorphic, e.g. there are two fusion bases when $k=3$.
\begin{equation}
V^{abc}_{d}\cong\bigoplus_{e}V^{ab}_{e}\otimes V^{ec}_{d}\cong\bigoplus_{f}V^{af}_{d}\otimes V^{bc}_{f}
\label{canef}
\end{equation}

\subsection{F and R-matrices} A change of fusion basis is realised by (some sequence of) so-called \textit{$F$-matrices}, which are unitary here since they transform between orthonormal bases. An $F$-matrix \textit{recouples} a triple of objects, i.e. $F^{abc}_{d}:\bigoplus_{e}V^{ab}_{e}\otimes V^{ec}_{d}\xrightarrow{\sim}\bigoplus_{f}V^{af}_{d}\otimes V^{bc}_{f}$ is given by (\ref{fmovedef}), and its inverse is written $G^{abc}_{d}$. Also note that $(F^{abc}_{d})^{*}:\bigoplus_{e}V^{e}_{ab}\otimes V^{d}_{ec}\xrightarrow{\sim}\bigoplus_{f}V^{d}_{af}\otimes V^{f}_{bc}$.
\begin{equation}
\raisebox{-8.5mm}{
\includegraphics[width=0.73\textwidth]{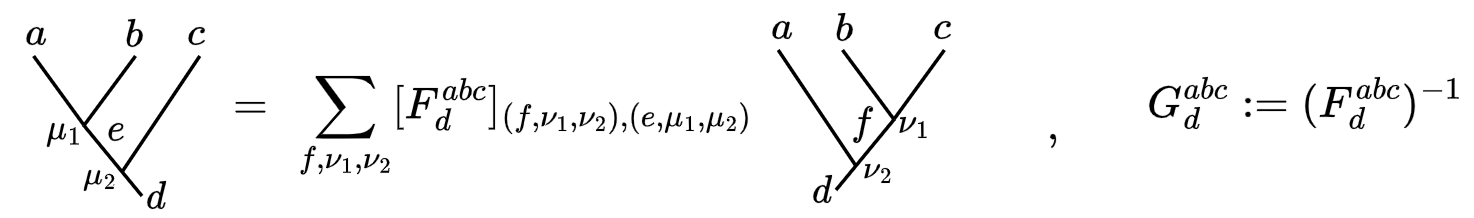}}
\label{fmovedef}
\end{equation}

\noindent A clockwise exchange of two anyons $a$ and $b$ is described by the \textit{$R$-matrix} $R^{ab}:V^{ab}\xrightarrow{\sim} V^{ba}$, where $R^{ab}=\bigoplus_{c}R^{ab}_{c}$. Diagrammatically, $R^{ab}_{c}:V^{ab}_{c}\xrightarrow{\sim} V^{ba}_{c}$ is given by (\ref{strobius})(i). Then $R^{ab}$ is given by (\ref{strobius})(ii), which can be seen by stacking the crossing on top of the identity operator for $V^{ba}$. 
\vspace{-1mm}
\begin{equation}
\raisebox{-8mm}{
\includegraphics[width=0.9\textwidth]{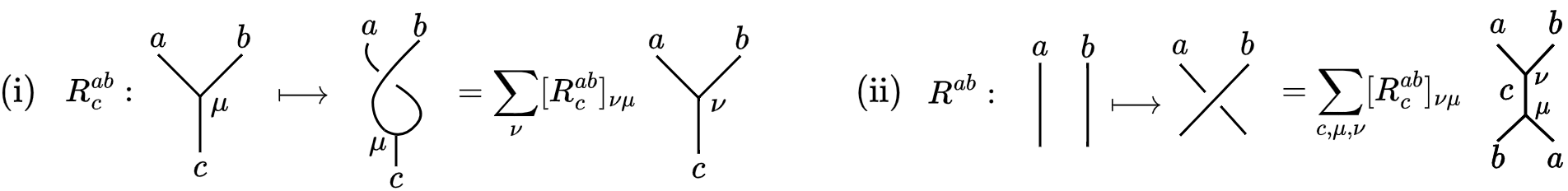}}
\label{strobius}
\end{equation}
\noindent An anticlockwise exchange of $a$ and $b$ is given by $(R^{-1})^{ab}:=(R^{ba})^{-1}$, and we always have that $R^{a0}=R^{0a}=1$.  Since $R$-matrices describe evolutions of quantum states, we require that they are unitary: given the unitarity of $F$-matrices, this turns out to be automatically satisfied \cite{Gal14}.\footnote{Technically, $R^{ab}$ is a linear isometry since it maps between two different spaces for $a\neq b$. However, its matrix representation is unitary and so we do not heed this distinction.}$^{,}$\footnote{In the context of braiding anyons, the unitary evolution postulate of quantum mechanics is obviated by the postulate that state spaces are Hilbert: demanding that all triangular spaces are Hermitian inner product spaces is equivalent to requiring that all $F$-matrices are unitary. Then by \cite{Gal14}, the unitarity of $R$-matrices is immediate.}$^{,}$\footnote{In other words, given a set of unitary $F$-matrices (that solve the pentagon equation for some anyon model $(\mathfrak{L},\times)$) as input for the hexagon equations, each possible solution set of $R$-matrices will be unitary. See Section \ref{penthexsec}. \label{galnote}} 

\begin{remark} The entries of $F$ and $R$-matrices are respectively called $F$ and $R$-\textit{symbols}.\end{remark}

\subsection{Conservation of charge} 
\label{chargeconsec}
Given a spatial region of a system with total charge $c\in\mathfrak{L}$, its total charge cannot be altered by processes that are internal (i.e. without particles entering or leaving) to this region. Conservation of charge is expressed by \eqref{chargeconeq}, where the coupon $\pi$ denotes an arbitrary process consisting of operations \ref{anyac1}-\ref{anyac4}. This follows from orthogonality relation \eqref{fyxl}(i).\footnote{In the tensor categorical description of anyon theories (see Section \ref{anytheosec} for brief remarks and references), anyons are simple objects and charge conservation is Schur's lemma.} 

\vspace{-6mm}
\begin{equation}
\raisebox{-9mm}{\includegraphics[width=0.25\textwidth]{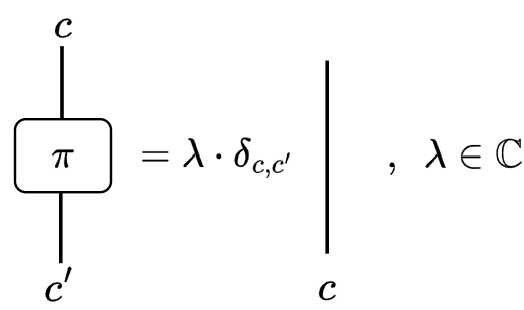}}
\label{chargeconeq}
\end{equation}

\noindent Physically, local conservation of charge comes from the characterisation of topological charge as a \textit{(local) superselection sector}.\footnote{The state space of a given region decomposes into direct summands indexed by the total charge $c\in\mathfrak{L}$. Each of these summands is a `superselection sector', which means that observables localised within the region cannot measure superpositions over distinct summands. As we have seen, $F$ and $R$-matrices respect this decomposition and do not mix between the sectors local to the particles on which they act. For further reading, see e.g. \cite{sv24,val21}.} That these sectors are `local' reflects that interactions with anyons from outside the region can still result in evolutions which mix between the sectors.

\subsection{Pentagon and hexagon equations}
\label{penthexsec}
\begin{figure}[H]
    \hspace{-17.5mm}
    \includegraphics[width=1.1\textwidth]{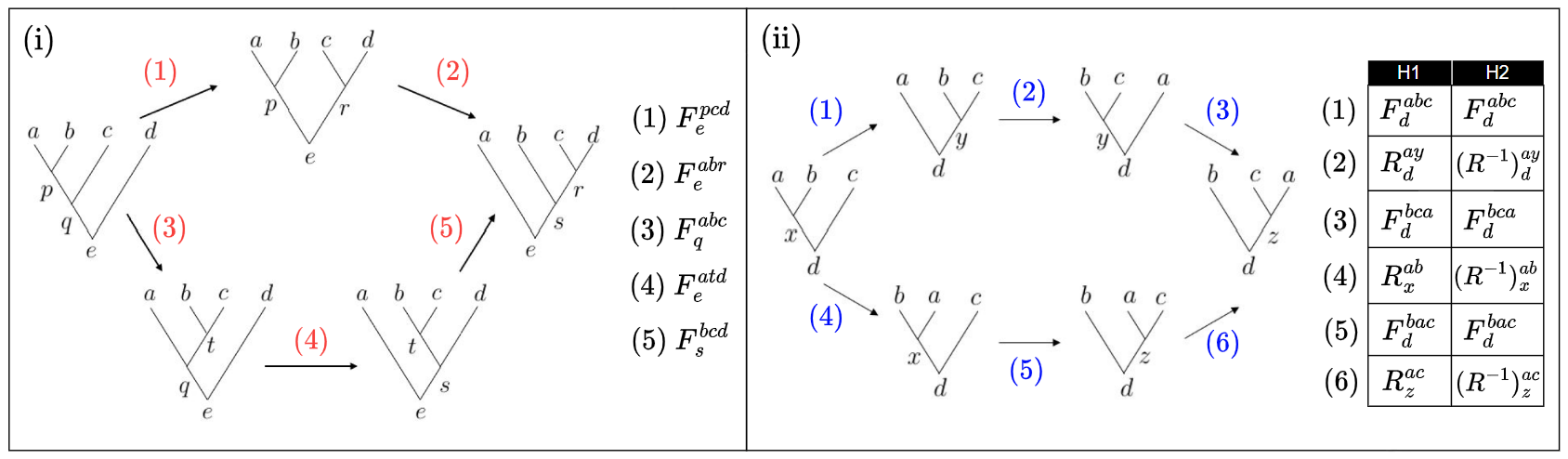}
    \hspace{-17.5mm}
    \vspace{-3mm}
    \caption{Arrows map between fusion spaces, all of which have a specified decomposition. The pentagon and hexagon axioms respectively demand that the diagrams in (i) and (ii) commute for all possible labellings.}
    \label{penthexeqs}
\end{figure}

In order for the $F$ and $R$-matrices to be consistently defined, they must satisfy compatibility conditions known as the \textit{pentagon} and \textit{hexagon} axioms. There are two distinct hexagon axioms, which we respectively denote by H1 and H2 in Figure \ref{penthexeqs}.
\vspace{-3mm}
\vspace{-3mm}
\noindent By fixing a choice of fusion state in the inital and final spaces (given a permissible choice of labels) in (i)-(ii) above, we obtain an entry-wise form of the compatibility conditions. This results in a system of equations called the\textit{ pentagon and hexagon equations}. For instance, if the triangular spaces in (i)-(ii) are all $1$-dimensional for the given choices of labels, these equations take form (\ref{mfpenteq})-(\ref{mfhexeqs}).\footnote{For a \textit{multiplicity-free} theory (i.e. all fusion coefficients take value in $\{0,1\}$), its consistency equations are given by (\ref{mfpenteq})-(\ref{mfhexeqs}) for all possible label choices.} In the presence of triangular spaces with dimension greater than $1$, they become considerably more unwieldy, e.g. see \cite[Appendix D]{val21}. 
\begin{equation}
[F^{abr}_{e}]_{sp}[F^{pcd}_{e}]_{rq}=\sum_{t}[F^{bcd}_{s}]_{rt}[F^{atd}_{e}]_{sq}[F^{abc}_{q}]_{tp}
    \label{mfpenteq}
\end{equation}
\vspace{-3mm}
\begin{subequations}
\label{mfhexeqs}
    \begin{align}
        \sum_{y} [F^{bca}_{d}]_{zy}[R^{ay}_{d}][F^{abc}_{d}]_{yx} &= [R^{ac}_{z}][F^{bac}_{d}]_{zx}[R^{ab}_{x}] \\
        \sum_{y} [F^{bca}_{d}]_{zy}[(R^{-1})^{ay}_{d}][F^{abc}_{d}]_{yx} &= [(R^{-1})^{ac}_{z}][F^{bac}_{d}]_{zx}[(R^{-1})^{ab}_{x}]
    \end{align}
\end{subequations}
\vspace{-2mm}
\begin{remark}\phantom{boo}
\begin{enumerate}[label=(\roman*)]
\item The pentagon and hexagon equations place very stringent constraints on whether a set of fusion rules $(\mathfrak{L},\times)$ will give rise to a theory of anyons. These equations can be viewed as a machine that takes as input $(\mathfrak{L},\times)$, and whose outputs (i.e. solutions, if any exist) define theories of anyons (specifically, when all of the $F$ and $R$-matrices of a given output are unitary). Solving these equations is typically very difficult (especially if $|\mathfrak{L}|$ is large or if there are fusion coefficients greater than $1$). See also the discussion in Section \ref{anytheosec}.
\item The hexagon axioms ensure that (a) diagrams enjoy equivalence under the $2^{nd}$ and $3^{rd}$ Reidemeister moves, and (b) strands may slide over (or under) trivalent vertices.
\item Technically, there is another compatibility condition called the \textit{triangle} axiom which ensures that fusion with the vacuum is trivial.\footnote{A further discussion of the triangle axiom (or `fundamental triangle equation') is found in \cite[Section E.1.2]{kitaev06}.} This axiom is equivalent to levying the (`obvious') condition that all $F$-matrices $F^{abc}_{d}$ where any of $a,b,c$ coincide with $0$, are the identity.
\end{enumerate}
\end{remark}

\subsection{Leg-bending operators and pivotal coefficients}
\label{lbopsec}
For all $a,b,c$, there exist linear maps $K^{ab}_{c}$,$L^{ab}_{c}$,$K^{c}_{ab}$,$L^{c}_{ab}$ as defined in (\ref{lbmdeq}). We call them the  \textit{leg-bending operators}.
\begin{subequations}
\label{lbmdeq}
\begin{align}
\raisebox{-10mm}{\includegraphics[width=0.85\textwidth]{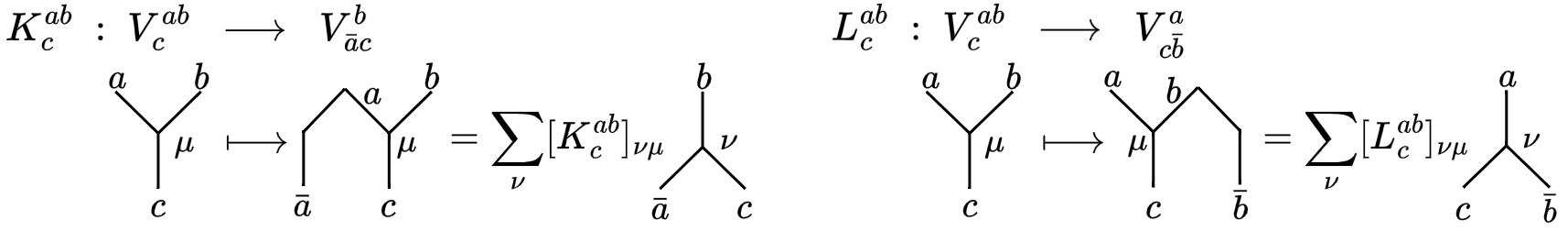}}
\label{tgcfren}
\end{align}
\vspace{-2mm}
\begin{align}
\raisebox{-10mm}{\includegraphics[width=0.85\textwidth]{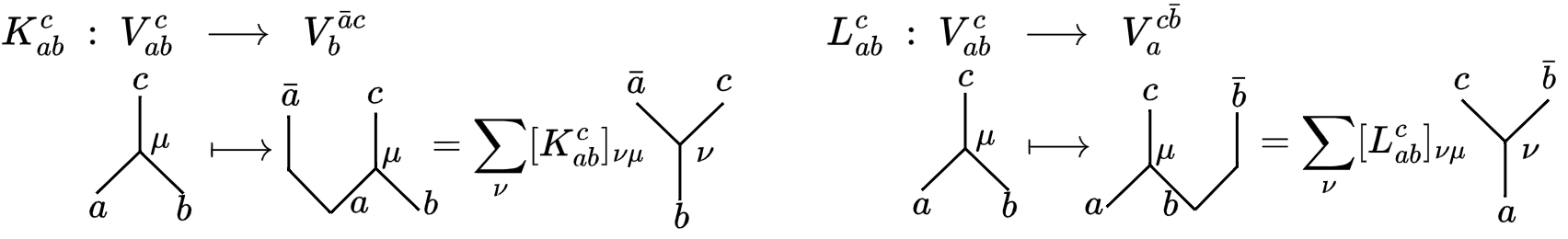}} 
\label{tgcfren2}
\end{align}
\label{tgceqs}
\end{subequations}

\noindent The leg-bending operators are isometries (unitary matrices) and satisfy (\ref{cptcoherence}) \cite[Theorem E.6]{kitaev06}. This commutative diagram can be seen as a compatibility condition for CPT transformations.\footnote{Here, CPT symmetry is the preservation of inner products under an inversion of charge, parity and time. CPT transformations can be realised via two orientations, which are not isotopic in $(2+1)$D; equality is given by (\ref{cptcoherence}). CPT transformations can be performed on any Hom-space via leg-bending, with consistency ensured by (\ref{cptcoherence}).}$^{,}$\footnote{It may be helpful to refer to Section \ref{anytheosec} while reading this note. A 6j fusion system whose leg-bending operators are isomorphisms satisfying \eqref{cptcoherence} is said to be \textit{pivotal}, and is equivalent to a so-called pivotal fusion category. The referenced theorem of \cite{kitaev06} shows that a unitary fusion category is always pivotal (see also \cite[Proposition 8.23]{ENO}), whence a theory of anyons always possesses pivotal structure (i.e. CPT symmetry). It is conjectured that even if unitarity is relaxed, any 6j fusion system is pivotal; or in other words, that any fusion category admits a pivotal structure \cite[Conjecture 2.8]{ENO}. The concrete definition of pivotality used here is shown in \cite[Section 3.6]{bartlett} to be equivalent to the definition more commonly used in the mathematical literature (namely, a fusion category is called pivotal if there exists a natural isomorphism between the identity functor and double dual endofunctor).} Equation (\ref{pivolini}) is called the \textit{pivotal identity}, where equalities (i) and (ii) can respectively be seen by completing an anticlockwise and clockwise loop of (\ref{cptcoherence}).\footnote{\eqref{pivolini} is also implied by the triviality of $2\pi$-twisting three anyons whose total charge is $0$ \cite[Chapter 14.8]{simonbook}.}
\vspace{1mm}

\begin{subequations}
\noindent\centering
\begin{minipage}{0.6\linewidth}
\noindent
\begin{align}
\noindent
\hspace{-10mm}
\raisebox{-17.5mm}{\includegraphics[width=0.55\textwidth]{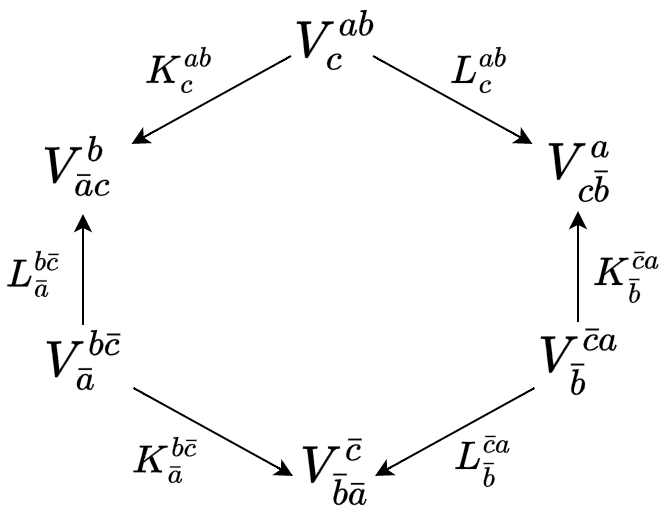}} 
\label{cptcoherence}
\end{align}
\end{minipage}\hfill
\hspace{-13.5mm}
\begin{minipage}{0.6\linewidth}
\noindent
\begin{align}
\noindent
\hspace{-10mm}
\raisebox{-8mm}{\includegraphics[width=0.6\textwidth]{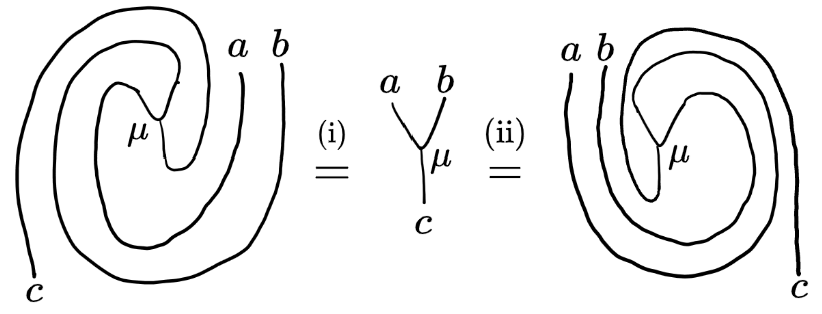}}
\label{pivolini}
\end{align}
\end{minipage}
\end{subequations}

\vspace{1mm}

\noindent The entries of leg-bending matrices are given by $F$-symbols (\ref{bendywendy1})-(\ref{bendywendy2}).\footnote{Derivations in the multiplicity-free case are found in \cite[Section 6.2]{wolf}; they follow analogously with multiplicity.}

\begin{subequations}
\label{bendywendy1}
\noindent\centering
\begin{minipage}{0.48\textwidth}
\begin{align}
[K^{ab}_{c}]_{\nu\mu}=\sqrt{\dfrac{d_a d_b}{d_c}}[F^{\bar{a}ab}_{b}]^{*}_{(c,\mu,\nu),0}
\label{kabby}
\end{align}
\end{minipage}
\hfill
\begin{minipage}{0.48\textwidth}
\begin{align}
[L^{ab}_{c}]_{\nu\mu}=\sqrt{\dfrac{d_a d_b}{d_c}}[F^{ab\bar{b}}_{a}]_{0,(c,\mu,\nu)}
\label{labby}
\end{align}
\end{minipage}
\end{subequations}

\begin{subequations}
\label{bendywendy2}
\noindent\centering
\begin{minipage}{0.48\textwidth}
\begin{align}
K^{c}_{ab}=(K^{ab}_{c})^{*}
\label{kabby2}
\end{align}
\end{minipage}
\hfill
\begin{minipage}{0.48\textwidth}
\begin{align}
L^{c}_{ab}=(L^{ab}_{c})^{*}
\label{labby2}
\end{align}
\end{minipage}
\end{subequations}
\vspace{3mm}

\noindent The quantity $t_{a}\in\mathbb{C}$ from (\ref{dollypow}) is called the \textit{pivotal coefficient} of $a$.
\begin{equation}
\raisebox{-8.5mm}{
\includegraphics[width=0.9\textwidth]{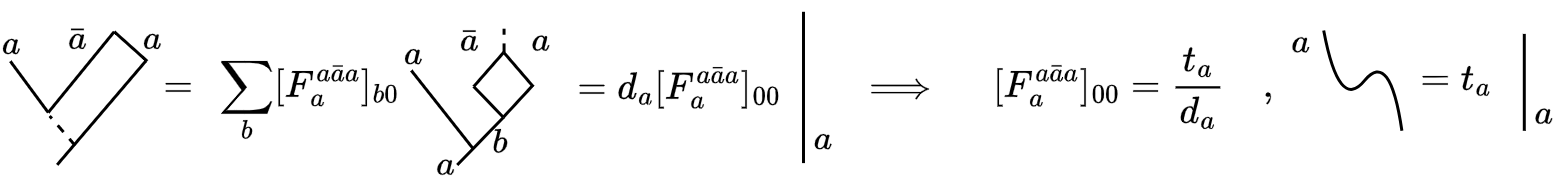}}
\label{dollypow}
\end{equation}
\noindent Then noting that $L^{a\bar{a}}_{0}=d_{a}[F^{a\bar{a}a}_{a}]_{00}^{*}=t_{a}$, we see that
\begin{equation}
\raisebox{-3.5mm}{
\includegraphics[width=0.375\textwidth]{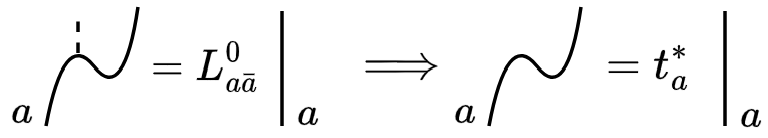}}
\label{gumption}
\end{equation}

\noindent For any $a$, the pivotal coefficients satisfy (\ref{pivcefs}). Equation (\ref{pivcef1}) follows from the unitary of the leg-bending operators, and (\ref{pivcef2}) from commutative diagram (\ref{cptcoherence}) with one of $a$ or $b$ set to $0$.

\begin{subequations}
\label{pivcefs}
\noindent\centering
\begin{minipage}{0.48\textwidth}
\begin{align}
|t_{a}|=1
\label{pivcef1}
\end{align}
\end{minipage}
\hfill
\begin{minipage}{0.48\textwidth}
\begin{align}
t_{\bar{a}}=t_{a}^{*}
\label{pivcef2}
\end{align}
\end{minipage}
\end{subequations}
\vspace{1mm}

\noindent When $a$ is self-dual, its pivotal coefficient is called the \textit{Frobenius-Schur indicator} of $a$ and is written as $\varkappa_{a}$. Note that $\varkappa_{a}=\pm1$ and $\varkappa_{0}=1$.

\begin{remark} Zigzags of the form in (\ref{dollypow}) and (\ref{gumption}) are straight lines (up to a scalar). For this reason, one does not assign the usual normalisation factor of $d_{a}^{-1/2}$ to their trivalent vertices.
\end{remark}

\begin{remark}[\textbf{Quantum dimensions revisited}]\phantom{boo}
    \begin{enumerate}[label=(\roman*)]

        \item By \eqref{dollypow} and \eqref{pivcef1}, $d_{a}=|[F^{a\bar{a}a}_{a}]_{00}|^{-1}$. \label{qdimfeq}
    
        \item Identity (\ref{qdimeq2}) for quantum dimensions is implied by the unitarity of the matrices in (\ref{bendywendy1}).  \label{smeagol}

    \end{enumerate}
    \label{lavat}
\end{remark}

\vspace{-8mm}
\subsection{Topological spin and modularity}
Clockwise $2\pi$-rotating an anyon $a$ accumulates a phase factor $\vartheta_{a}\in\U(1)$, known as the \textit{topological spin} or \textit{twist factor} of $a$. We always have $\vartheta_{0}=1$. 

\begin{figure}[H]
    \centering
    \includegraphics[width=0.55\textwidth]{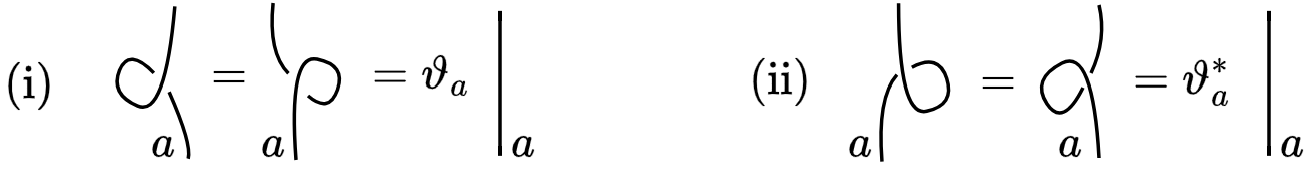}
    \caption{\textit{Kinks} represent (i) clockwise and (ii) anticlockwise $2\pi$-twists of $a$.}
    \label{twistydiags}
\end{figure}
\vspace{-1mm}

\noindent The endomorphism $M^{ab}:=R^{ba}\circ R^{ab}$ describes the (clockwise) \textit{monodromy} of $a$ and $b$, and $M^{ab}_{c}$ is its restriction to $V^{ab}_{c}$. Topological spins always satisfy (\ref{tspineq1})-(\ref{tspineq2}). A result of Vafa tells us that any topological spin is a root of unity \cite{vafa}.\footnote{A proof of Vafa's result is also found in \cite[Theorem E.10]{kitaev06}.} Thus for any $a,b$, $(M^{ab})^{n}=\id$ for some $n\geq1$.

\vspace{2mm}

\begin{subequations}
\label{frukt}
\noindent\centering
\begin{minipage}{0.6\linewidth}
\noindent
\begin{align}
           [M^{ab}_{c}]_{\mu\nu}=\sum_{\lambda}[R^{ba}_{c}]_{\mu\lambda}[R^{ab}_{c}]_{\lambda\nu}=\frac{\vartheta_{c}}{\vartheta_{a}\vartheta_{b}}\delta_{\mu\nu}
           \label{tspineq1}
\end{align}
\end{minipage}\hfill
\begin{minipage}{0.35\linewidth}
\noindent
\begin{align}
\vartheta_{a}=\vartheta_{\bar{a}}
       \label{tspineq2}
\end{align} 
\end{minipage}
\end{subequations}
\vspace{-4mm}
\begin{figure}[H]
    \centering
    \includegraphics[width=0.75\textwidth]{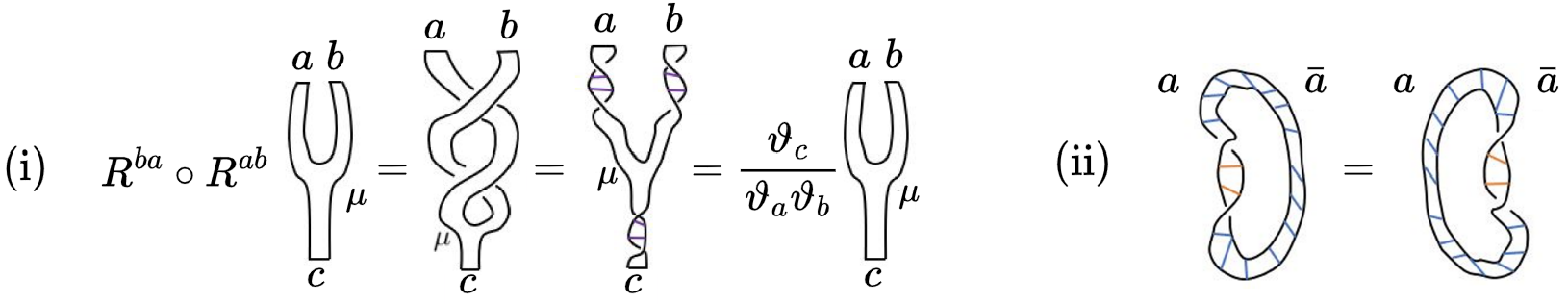}
    \vspace{-2mm}
    \caption{Promoting worldlines to worldribbons, kinks in Figure \ref{twistydiags} can be tautened to $2\pi$-twists. We respectively illustrate identities (\ref{tspineq1})-(\ref{tspineq2}) using worldribbons. (i) The middle equality is seen by pulling taut the ribbon from the tops and bottom. (ii) We must be able to push twists around a closed loop. If we do the same but with kinks and worldlines, we recover the equivalence of kinks as in Figure \ref{twistydiags}.}
    \label{ribthick}
\end{figure}

\vspace{-2mm}
\noindent Identities (\ref{tspinreveqs}) show that the $F$ and $R$-symbols of a theory encode its topological spins. Identity (\ref{rteq1}) can be seen by replacing the crossing in a clockwise kink with its expansion (\ref{strobius})(ii) in the standard basis; then, the result is obtained by applying leg-bending matrices and orthogonality relation (\ref{fyxl})(i) as appropriate. Identity (\ref{rteq2}) follows from \eqref{manywish}.

\begin{subequations}
\label{tspinreveqs}
\noindent\centering
\begin{minipage}{0.48\textwidth}
\begin{align}
\vartheta_{a}=\frac{1}{d_{a}}\sum_{c}d_{c}\tr(R^{aa}_{c})
\label{rteq1}
\end{align}
\end{minipage}
\hfill
\begin{minipage}{0.48\textwidth}
\begin{align}
\vartheta_{a}=t_{\bar{a}}(R^{a\bar{a}}_{0})^{*}=t_{a}(R^{\bar{a}a}_{0})^{*}
\label{rteq2}
\end{align}
\end{minipage}
\end{subequations}
\begin{equation}
\raisebox{-6mm}{
\includegraphics[width=0.475\textwidth]{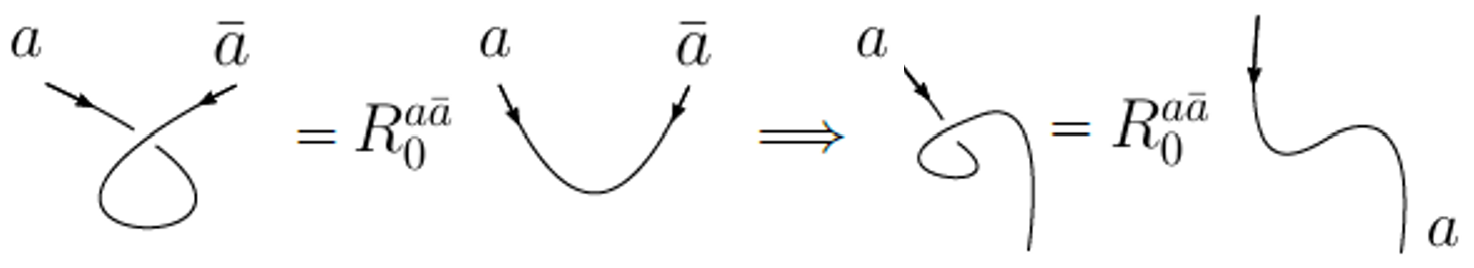}}
\label{manywish}
\end{equation}
\noindent We also see in \eqref{dekinkeq} that the composition of oppositely oriented kinks can be transformed to the identity strand using the $2^{\text{nd}}$ and $3^{\text{rd}}$ Reidemeister moves, along with the pivotal identity (\ref{pivolini}).
\begin{equation}
\raisebox{-6mm}{
\includegraphics[width=0.525\textwidth]{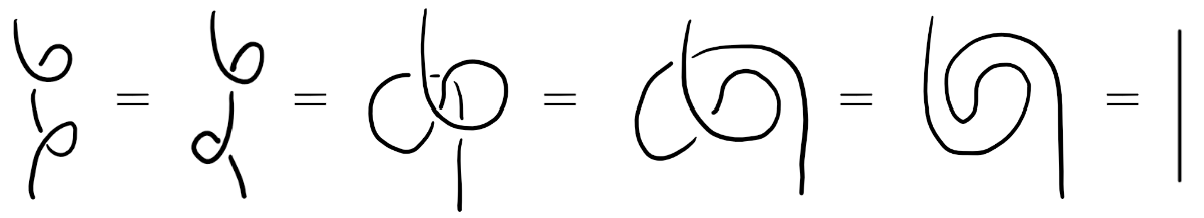}}
\label{dekinkeq}
\end{equation}
\noindent The $S$-\textit{matrix} of a theory of anyons is given by (\ref{normsmat}), where $\mathcal{D}=\sqrt{\sum_{c}d_{c}^{2
}}$ is the \textit{total quantum dimension} of the theory.\footnote{Caveat: conventions for the labelling of components in \eqref{normsmat} vary in the literature. Here, we follow \cite{kitaev06}.} The second equality in \eqref{normsmat} can be verified by expanding the clockwise monodromy of $a$ with $\bar{b}$ in the basis of jumping jacks, and then applying leg-bending matrices and orthogonality relation \eqref{fyxl}(i) as appropriate.
\begin{equation}[S]_{ab}:=\frac{1}{\mathcal{D}}\raisebox{-7mm}{\includegraphics[width=0.15\textwidth]{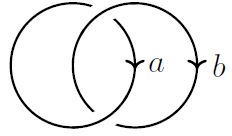}}=\frac{1}{\mathcal{D}}\sum_{c}\frac{\vartheta_{c}}{\vartheta_{a}\vartheta_{b}}N^{a\bar{b}}_{c}d_{c} \ \ , \ \ a,b\in\mathfrak{L}\label{normsmat}\end{equation}
\begin{remark}[\textbf{Braiding nondegeneracy and modularity}]\phantom{boo}
    \begin{enumerate}[label=(\roman*)]
        \item A charge $x$ is called \textit{transparent} if its monodromy with any charge $a$ is trivial, and a theory of anyons said to have a \textit{nondegenerate braiding} if the only transparent charge is $0$.
        
        \item Let $\bm{s}_{x}$ denote a column vector of $S$, i.e. $[\bm{s}_{x}]_{a}=[S]_{ax}$. A charge $x$ is transparent if and only if $\bm{s}_{x}=d_{x}\bm{s}_{0}$ \cite[Lemma E.13]{kitaev06}. It follows that $\nul(S)\geq t$, where $t$ is the number of nontrivial transparent charges. Hence, it is necessary to have $t=0$ in order to have $S$ invertible. In fact, it turns out that this is actually sufficient: a theory is called \textit{modular} if its $S$-matrix is unitary, which is true if and only if it has a nondegenerate braiding \cite[Section E.5]{kitaev06}.
    \end{enumerate} 
\end{remark}

\noindent The $T$-matrix of an anyon theory is defined $[T]_{ab}:=\vartheta_{a}\delta_{ab}$. Together, the $S$ and $T$-matrix of a theory are called its \textit{modular data}. 
See e.g. \cite{simonbook}-\cite{kitaev06} for further details.

\subsection{Gauge transformations}
\label{gaugesec}
Explicit $F$ and $R$-matrices are obtained by fixing an orthonormal basis on triangular spaces, which gives rise to redundancy in our algebraic description. That is, for any space $V^{ab}_{c}$, there is a $\U(N^{ab}_{c})$-freedom in the choice of basis. A change of (orthonormal) basis is called a \textit{gauge transformation}. Let $u^{ab}_{c}$ denote a gauge transformation on $V^{ab}_{c}$, where
\begin{equation}
\raisebox{-5mm}{
\includegraphics[width=0.35\textwidth]{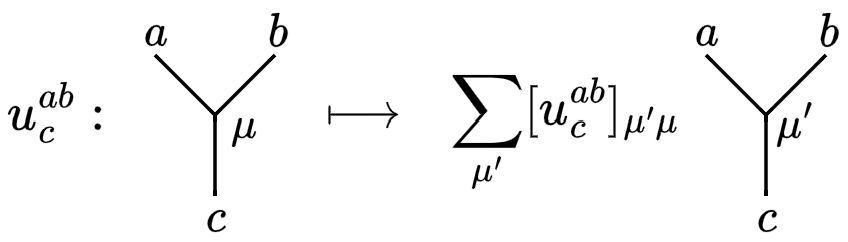}}
\label{gtransvertdiag}
\end{equation}
Gauge transformations of $F$-symbols and $R$-matrices are given by 
\begin{gather*}
[(F^{abc}_{d})']_{(f,\nu'_{1},\nu'_{2}),(e,\mu'_{1},\mu'_{2})}=\sum_{\mu_{1},\mu_{2},\nu_{1},\nu_{2}}[u^{af}_{d}]_{\nu'_{2}\nu_{2}}[u^{bc}_{f}]_{\nu'_{1}\nu_{1}}[F^{abc}_{d}]_{(f,\nu_{1},\nu_{2}),(e,\mu_{1},\mu_{2})}[(u^{ab}_{e})^{\dagger}]_{\mu_{1}\mu'_{1}}[\left(u^{ec}_{d}\right)^{\dagger}]_{\mu_{2}\mu'_{2}} \\
(R^{ab}_{c})'=u^{ba}_{c}R^{ab}_{c}(u^{ab}_{c})^{\dagger}
\end{gather*}

\noindent which in the absence of multiplicity, become

\begin{subequations}
\label{mfgtrans}
\noindent\centering
\begin{minipage}{0.48\textwidth}
\begin{align}
[(F^{abc}_{d})']_{fe}=\dfrac{u^{af}_{d}u^{bc}_{f}}{u^{ab}_{e}u^{ec}_{d}}[F^{abc}_{d}]_{fe}\label{mfgtransF}
\end{align}
\end{minipage}
\hfill
\begin{minipage}{0.48\textwidth}
\begin{align}
(R^{ab}_{c})'=\dfrac{u^{ba}_{c}}{u^{ab}_{c}}R^{ab}_{c}\label{mfgtransR}
\end{align}
\end{minipage}
\end{subequations}
\vspace{1mm}

\noindent Transformations satisfy (\ref{trivgt}) in accord with the triviality of fusion and braiding with the vacuum.
\begin{equation}
u^{a0}_{a}=u^{0a}_{a}=1
\label{trivgt}
\end{equation}

\begin{remark}[\textbf{Gauge-invariance}]
Physically observable quantities are necessarily \textit{gauge-invariant}. Such quantities include e.g. (i) monodromies; (ii) $R$-symbols $R^{aa}_{b}$ for $N^{aa}_{b}=1$; (iii) topological spins; (iv) $F$-symbols $[F^{abc}_{b}]_{bb}$; (v) quantum dimensions; (vi) Frobenius-Schur indicators. 
\end{remark}

\subsection{Anyon theories and fusion categories}
\label{anytheosec}

\noindent Let $(\mathfrak{L},\times)$ denote a set of labels with fusion coefficients as defined above. A \textit{theory of anyons} is defined by a corresponding set of unitary $F$ and $R$-matrices up to gauge equivalence; that is, a gauge class of unitary solutions to the pentagon and hexagon equations. If fusion rules $(\mathfrak{L},\times)$ admit at least one theory of anyons (i.e. a unitary solution to the pentagon and hexagon equations), then the fusion rules $(\mathfrak{L},\times)$ constitute an \textit{anyon model}, and the \textit{rank} of such a theory is $|\mathfrak{L}|$.\\

\noindent A theory of anyons is precisely a \textit{unitary braided fusion category} $\mathcal{C}$, and the underlying fusion rules together with the associated set of $F$ and $R$-symbols is called the \textit{skeletal data} of $\mathcal{C}$. This skeletal data uniquely specifies the category $\mathcal{C}$, and the label set $\mathfrak{L}$ corresponds to the isomorphism classes of simple objects in $\mathcal{C}$. Several examples of prototypical anyon theories are discussed in \cite{simonbook}, and skeletal data for some of these is found in \cite{bonderson-thesis,rsw}. See also \cite{anyonwiki} for a growing repository (thus far compiled mostly by J.K. Slingerland and G. Vercleyen) of multiplicity-free fusion rules, fusion categories and associated data. In Section \ref{tyanyonsec}, we speak more generally of fusion categories: to this end, we briefly summarise below some salient points in terms of the framework introduced above. See \cite{egno} for a contemporary treatise on the relevant mathematical formalism. 

\begin{itemize}

\item A set of fusion rules $(\mathfrak{L},\times)$ together with a solution to the pentagon equation (that is, a consistent set of $F$-matrices) yields a set of data $\{\mathfrak{L},\mathcal{N},\mathcal{F}\}$, where $\mathcal{N}$ are the fusion coefficients and $\mathcal{F}$ are the $F$-symbols. This set of data consitutes a well-defined planar algebra called a \textit{6j fusion system}, or a \textit{skeletal fusion category}. The gauge class of this data uniquely specifies a fusion category. We say that $(\mathfrak{L},\times)$ admits (fusion) categorification. \\

\item The hexagon equations take as input a valid 6j fusion system $\{\mathfrak{L},\mathcal{N},\mathcal{F}\}$. If a solution exists (that is, a consistent set of $R$-matrices), we now have a set of data $\{\mathfrak{L},\mathcal{N},\mathcal{F},\mathcal{R}\}$ where $\mathcal{R}$ are the $R$-symbols. This set of data constitutes a well-defined braided planar algebra called a \textit{braided 6j fusion system}, or a \textit{skeletal braided fusion category}. The gauge class of this data uniquely specifies a braided fusion category.\footnote{If additionally, the associated $S$-matrix is unitary, then the category is called a \textit{modular fusion category}.} We say that $(\mathfrak{L},\times)$ admits (braided fusion) categorification, or that the fusion category specified by $\{\mathfrak{L},\mathcal{N},\mathcal{F}\}$ admits a braiding.\\

\item If all of the $F$-matrices of a 6j fusion system are unitary, we say that it is \textit{unitary}. The set of data $\{\mathfrak{L},\mathcal{N},\mathcal{F}\}$ uniquely specifies a \textit{unitary fusion category} $\mathcal{C}$. If $\mathcal{C}$ admits a braiding, it was shown in \cite{Gal14} that this braiding is necessarily unitary: in other words, the constituent $R$-matrices of every associated solution set to the hexagon equations must also be unitary. The set of data $\{\mathfrak{L},\mathcal{N},\mathcal{F},\mathcal{R}\}$ uniquely specifies a \textit{unitary braided fusion category}.\\

\item One can allow for \textit{noncommutative} fusion rules $(\mathfrak{L},\times)$ by relaxing constraint \eqref{twissta}(i) for the fusion coefficients. Although such fusion rules may admit fusion categorification (i.e. a consistent set of $F$-matrices that solve the pentagon equation), it is not possible for such a category to further admit a braiding (i.e. a consistent set of $R$-matrices that solve the hexagon equations); hence, noncommutative fusion rules cannot describe anyonic systems.\\

\item A given set of fusion rules $(\mathfrak{L},\times)$ (commutative or otherwise)  may admit $k_{1}\geq0$ distinct fusion categorifications; furthermore, a given fusion category may admit $k_{2}\geq0$ distinct braidings. By a result known as \textit{Ocneanu rigidity} (see \cite{ENO,bartlett,GHS}), $k_{1}$ and $k_{2}$ are finite.\footnote{A rank-finiteness theorem \cite[Theorem 4.6]{rfgxbfc} tells us that there are finitely many braided fusion categories of a given rank. Consequently, we know that there are finitely many anyon theories of a given rank. While this theorem subsumes Ocneanu rigidity in the braided case, it remains open as to whether rank-finiteness can be extended to include fusion categories that do not admit a braiding.}

\end{itemize}

\noindent 

\section{\textbf{Tambara-Yamagami Anyons}}
\label{tyanyonsec}
\noindent Let $G$ be a \textit{finite abelian group} of order $d\geq2$. The \textit{Tambara-Yamagami} fusion rules are given by $(\mathfrak{L}^{\TY}_{G},\times)$, where for $g,h\in G$ we have
\begin{equation}
    \mathfrak{L}^{\TY}_{G}=\{q\}\sqcup G \ \ , \ \ q\times q=\sum_{g\in G}g \ \ , \ \ q\times g=g\times q=q \ \ , \ \ g\times h=gh
\label{tyfusrules}
\end{equation}

\noindent We will use $0$ to denote the identity element of $G$, in keeping with the notation for the vacuum element of $\mathfrak{L}$. It was shown by Tambara and Yamagami that $(\mathfrak{L}^{\TY}_{G},\times)$ always admits fusion categorification \cite{TYog}.\footnote{One might ask whether noncommutative fusion rules $(\mathfrak{L}^{\TY}_{G},\times)$ admit fusion categorification for $G$ nonabelian. The result of \cite{TYog} shows that $(\mathfrak{L}^{\TY}_{G},\times)$ admits fusion categorification if and only if $G$ is abelian.} Any such fusion category is called \textit{Tambara-Yamagami}, and we denote it by $\TY(G)$. In particular, $\TY(G)$ is necessarily \textit{unitary}.\footnote{Unitarity follows from \cite[Theorem 5.20]{ghr11}, which asserts that every \textit{nilpotent} fusion category is unitary. Let us recall nilpotency as defined in \cite{ghr11}. Given a fusion category $\C$, its adjoint category $\C_{\text{ad}}$ is the smallest fusion subcategory containing all objects $X\otimes X^{*}$, where $X\in\C$ is a simple object and $X^{*}$ is its dual. The upper central series of $\C$ is given by $\C^{(0)}\supseteq \C^{(1)} \supseteq \C^{(2)} \supseteq \cdots$, with defining relations $\C^{(0)}=\C$ and $\C^{(n)}=(C^{(n-1)})_{\text{ad}}$ for $n\geq1$. We say that $\C$ is nilpotent if its upper central series converges to the trivial fusion category $\text{FdVec}_{\mathbb{C}}$, which has underlying label set $\mathfrak{L}=\{0\}$. Indeed, $\TY(G)$ is nilpotent since it has upper central series $\TY(G)\supset G\supset \text{FdVec}_{\mathbb{C}}$.} The notation $\TY(G)$ can be ambiguous, e.g. $(\mathfrak{L}^{\TY}_{\mathbb{Z}_2},\times)$ admits 2 distinct fusion categorifications, each of which further admit 4 distinct braidings: this results in a total of 8 distinct anyon theories, each of which we refer to as an \textit{Ising theory} (see Section \ref{isingsec}).\footnote{In other words, $(\mathfrak{L}^{\TY}_{\mathbb{Z}_2},\times)$ admits 8 distinct braided fusion categorifications.} $\TY(G)$ shall henceforth refer to a theory of anyons, whence we know that a braiding is always assumed. For the purposes of our exposition, it will usually be unimportant to specify precisely which theory we are referring to once $G$ is specified. Necessary and sufficient conditions on $G$ for the existence of a braiding on $\TY(G)$ are presented immediately below in Section \ref{tyquditsec}.

\subsection{Fusion qudits}
\label{tyquditsec}
Recall that if a unitary fusion category admits a braiding, then this braiding is necessarily unitary \cite{Gal14}: whence the resulting category is a theory of anyons. 

\begin{tcolorbox}
\begin{convention}[\textbf{Assumptions on $G$}]
    Let $G$ be a finite abelian group of order $d$. The category $\TY(G)$ admits a braiding if and only if $G\cong\mathbb{Z}_{2}^{n}$ for all $n\geq1$ \cite[Theorem 1.2 (1)]{siehler}. We shall thus assume $G\cong\mathbb{Z}_{2}^{n}$ (consequently, $d=2^{n}$) unless stated otherwise. Furthermore, it will always be assumed that $\TY(G)$ is braided.
    \label{tyconv}
\end{convention}    
\end{tcolorbox}

\noindent Under Convention \ref{tyconv}, we say that $\TY(G)$ realises a \textit{Tambara-Yamagami theory} of anyons. We call an anyon of charge $q$ satisfying fusion rules of the form (\ref{tyfusrules}), a \textit{Tambara-Yamagami anyon} or `\textit{TY-anyon}' for short. All charges in the theory are self-dual. While $q$ is a nonabelian charge with quantum dimension $d_{q}=\sqrt{d}=2^{n/2}$, all $G$-valued charges are abelian.\\

\noindent Let $V^{q^{2p}}$ denote the fusion space of $2p$ TY-anyons, and note that 
\begin{equation}
    V^{q^{2p+1}}\cong V^{q^{2p}}\cong \bigoplus_{h_{1},\ldots,h_{p}\in G} V^{qq}_{h_1}\otimes\cdots\otimes V^{qq}_{h_p} \quad , \quad p\geq 1
\label{paired-basis}
\end{equation}
\noindent Thus, $\dim(V^{q^{2p+1}})=\dim(V^{q^{2p}})=d^{p}$, and $N\geq2$ TY-anyons encode $\lfloor\frac{N}{2}\rfloor$ qudits. Given a pair of TY-anyons, we can realise a qudit whose logical basis correspond to their $d=2^{n}$ possible fusion outcomes (and may thus be enumerated by all $n$-bit strings). Such a state is written
    \begin{equation}
        \ket{\varphi}=\sum_{g\in G}\gamma_{g}\ket{qq\to g}=\sum_{g\in\mathbb{Z}_{2}^{n}}\gamma_{g}\ket{g}
        \label{tyqudit}
    \end{equation}
\noindent Similarly, a triple of TY-anyons also encodes a single qudit $\ket{\varphi}\in V^{qqq}$. In this context, the notation used in \eqref{tyqudit} implicitly assumes that we have specified one of the two possible fusion bases as our computational basis.

\subsection{Simplest example: Ising anyons}
\label{isingsec}
The simplest example of a Tambara-Yamagami theory is $\TY(\mathbb{Z}_{2})$. Here, the fusion rules \eqref{tyfusrules} take form
\begin{equation} 
q \times q = 0 + 1 \ \ , \ \ q \times 1 = 1 \times q = q  \ \ , \ \  1 \times 1 = 0
\end{equation}
There are $8$ theories of anyons that arise from the model $(\mathfrak{L}^{\TY}_{\mathbb{Z}_{2}},\times)$, all of which are modular. Half of these are called \textit{Ising} theories and have $\varkappa_q=1$, while the other half have $\varkappa_q=-1$ and are called $\SU(2)_{2}$ theories. By a slight abuse of terminology, we will refer to all $8$ theories as Ising, and call $q$ an \textit{Ising anyon} in each of these. In this setting, note that the state $\ket{\varphi}$ from \eqref{tyqudit} is a \textit{qubit}. Some skeletal data for the Ising theories is given in Section \ref{skelesec}, although this will not be needed to prove our main theorems. 

\section{\textbf{Rungs and Branches}}
\label{standsec}
\noindent We introduce notation and terminology for diagrams that appear frequently in the sequel; these are string diagrams in a Tambara-Yamagami theory, and so for convenience of presentation, we apply Convention \ref{tyconvention}. We will consider the diagrams below as linear operators in $\End(\mathbb{C}^{d})$, i.e. as acting on a single qudit: indeed, they are all unitary (Remark \ref{unimark}) and may thus be thought of as quantum gates in this context. Further properties are presented in Sections \ref{furtherpropsec}-\ref{paulisansbrsec}. 

\begin{tcolorbox}
\begin{convention}[\textbf{Anyon worldlines in Tambara-Yamagami theories $\TY(G)$}] Solid black worldlines will represent Tambara-Yamagami anyons $q$, and solid blue worldlines will represent anyons whose charge takes value in $G\cong\mathbb{Z}_{2}^{n}$.
\label{tyconvention}
\end{convention}    
\end{tcolorbox}

\noindent We refer to the diagrams in \eqref{rungeq1} as \textit{rungs} (or $g$-rungs). We will later see that $A_{g}$ and $B_{g}$ are Hermitian conjugates of one another (Remark \ref{unimark}). Applying the leg-bending operators from \eqref{tgceqs}, we see that $g$-rungs satisfy relations \eqref{rungeq2}: since these diagrams differ only by a global phase factor, their action on a qudit will be identical. It is therefore useful to introduce the notation $\mathcal{O}_{g}$ for a $g$-rung, where $\mathcal{O}_{g}\in\{A_{g},B_{g},C_{g},C'_{g}\}$. 
\vspace{-2mm}
\begin{subequations}
\begin{align}
    \raisebox{-5mm}{\includegraphics[width=0.575\textwidth]{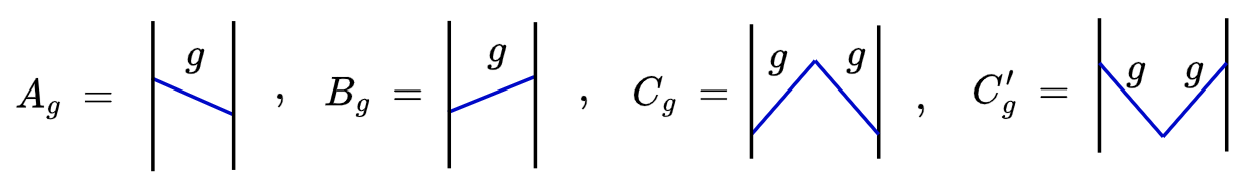}} \label{rungeq1} \\
    L^{qg}_{q}A_{g}=C_{g}=K^{gq}_{q}B_{g} \ \ , \ \ (L^{qg}_{q})^{*}B_{g}=C'_{g}=(K^{gq}_{q})^{*}A_{g} \label{rungeq2}
\end{align}
\label{rungeqs}
\end{subequations}
\noindent The diagrams in \eqref{brancheq1} shall be referred to as \textit{branches} (or $g$-branches), and satisfy relations \eqref{brancheq2}.
\vspace{-4mm}
\begin{subequations}
\begin{align}
    &\raisebox{-5mm}{\includegraphics[width=0.675\textwidth]{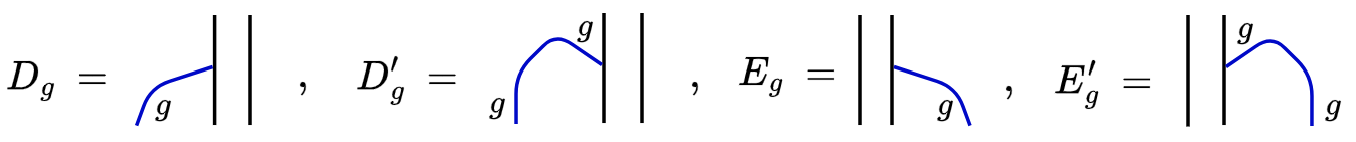}} \label{brancheq1} \\[2mm]
    & \hspace{31mm} D'_{g}=K^{gq}_{q}D_{g} \quad , \quad \ E'_{g}=L^{qg}_{q}E_{g}
    \label{brancheq2}
\end{align}
\label{brancheqs}
\end{subequations}
\noindent Since $V^{gq}_{q}, V^{qg}_{q}$ are 1-dimensional spaces, $K^{gq}_{q},L^{qg}_{q}$ are phases. As linear operators, rungs lie in $\End(V^{qq})$, and branches in $\Hom(V^{qq},V^{gqq})$ and $\Hom(V^{qq},V^{qqg})$; hence, they all act on $d$-dimensional spaces, where $d=2^{n}$. Thus, rung and branch operators act on a qudit encoded in the pair of TY-anyons. Although the particle processes corresponding to the rungs in \eqref{rungeq1} are distinct, \eqref{rungeq2} shows that their action on the qudit is physically indistinguishable, as it differs only by a global phase factor;  by \eqref{brancheq2}, the same is true of $D_{g},D'_{g}$ (and separately, $E_{g},E'_{g}$).\footnote{It is later shown that the action of $D_{g}$ and $E_{g}$ on a qudit is also physically indistinguishable (Proposition \ref{branchswitch}). \label{tacokatt}} 
\vspace{-2mm}
\begin{equation}
   \raisebox{-6mm}{\includegraphics[width=0.825\textwidth]{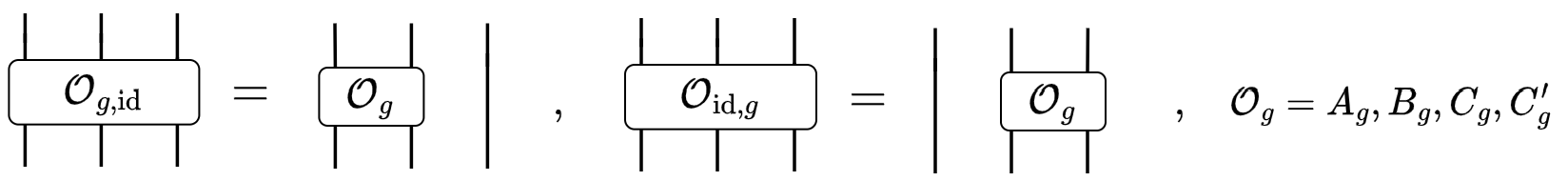}} \label{og_def} 
\end{equation}
We may also consider the action of a $g$-rung on a qudit encoded in the fusion space of three TY-anyons; diagrammatically, such an operator is given by a coupon $\mathcal{O}_{g,\id}$ or $\mathcal{O}_{\id,g}$ as defined in \eqref{og_def}. Its representation depends on the fusion (i.e. computational) basis of the qudit. Let $\Lambda[\mathcal{O}]$ and $\mathrm{P}[\mathcal{O}]$ respectively denote the representation of any $\mathcal{O}\in\End(V^{qqq})$ in the fusion basis where the leftmost and rightmost pair of anyons are the first to be fused. Then,
\begin{subequations}
    \begin{align}
        \Lambda[\mathcal{O}_{g,\id}]=\mathcal{O}_{g} \quad &, \quad \mathrm{P}[\mathcal{O}_{\id,g}]=\mathcal{O}_{g} \label{diagrep} \\
        \mathrm{P}[\mathcal{O}_{g,\id}]=F^{qqq}_{q}\mathcal{O}_{g}G^{qqq}_{q} \quad &, \quad \Lambda[\mathcal{O}_{\id, g}]=G^{qqq}_{q}\mathcal{O}_{g}F^{qqq}_{q} \label{nondiagrep}
    \end{align}
\label{lrreps}    
\end{subequations}
\noindent Note that $\mathcal{O}_{g}$ is represented as a diagonal matrix on $V^{qq}$, since $\End(V^{qq})=\bigoplus_{g\in G}\End(V^{qq}_{g})$. On the other hand, the representations in \eqref{nondiagrep} will be non-diagonal.\footnote{Proposition \ref{branchswitch} shows that the two representations in \eqref{nondiagrep} differ only by a phase factor, whence their action is physically indistinguishable; also, said action is indistinguishable from that of a $g$-branch (see footnote \ref{tacokatt}).} 

\section{\textbf{Teleportation Without Braiding}}
\label{telechap}

\noindent The primary goal of this section is to prove the following theorem.

\begin{thm}
    Consider a Tambara-Yamagami theory of rank $d+1$ where $d=2^{n}$. The fusion state of $N$ Tambara-Yamagami anyons (that is, a $\lfloor\frac{N}{2}\rfloor$-qudit state) can be teleported via the procedure shown in Figure \ref{palominoscruff}. In particular, no braiding is required. 
    \label{brfreetelethm}
\end{thm}

\noindent In Section \ref{tynatsec}, we posit that Tambara-Yamagami theories are a natural setting for anyonic teleportation; consequently, we stay within the setting of $\TY(G)$ throughout the rest of the paper (and so Convention \ref{tyconvention} applies to all string diagrams from Section \ref{telecorrsec} onwards). In brief, we start with an arbitrary anyon theory $\mathcal{C}$ as a candidate setting, and introduce a constraint \ref{t1} together with Heuristic \ref{the-heuristic}. Jointly, these imply that a constraint \ref{t2} should be satisfied, which leads us to the conclusion that we should set $\mathcal{C}=\TY(G)$.\\
In Section \ref{maxentangsec}, we establish a necessary and sufficient condition for maximal entanglement between the two pairs of nonabelian anyons in the yellow and green regions of the system highlighted below. This allows us to verify that Alice and Bob share an e-dit in our proposed setup.
\vspace{-4mm}
\begin{figure}[H]
    \includegraphics[width=0.2\textwidth]{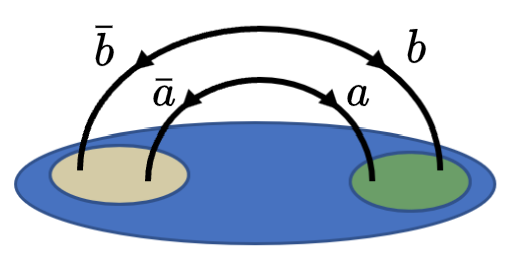}
    \caption*{}
\end{figure}
\vspace{-12mm}
\noindent In Section \ref{telecorrsec}, we explain how Bob, upon receipt of Alice's measurement outcomes, is able to apply corrections to the state of his anyons (without braiding) in order recover her initial fusion state. In Section \ref{1qtelesum}, this is generalised to a scenario where Alice wishes to teleport the state of $N\geq2$ TY-anyons. We summarise the procedure and present the proof of Theorem \ref{brfreetelethm}. The proof of the theorem is diagrammatic and exploits the following ingredients: (i) the \textit{pivotal property} of anyons (namely, that we can straighten out zigzags at the cost of a phase factor); (ii) a simple \textit{graphical move \eqref{keyid1}}; (iii) \textit{`branch decoupling'} (Section \ref{branchsec}); and (iv) \textit{`rung-cancelling'} (Lemma \ref{key-prop}).\\
In Section \ref{telefreesec}, we mention a few obvious variations of the braid-free teleportation process. 

\subsection{Tambara-Yamagami theories as a natural setting for teleportation}
\label{tynatsec}

\noindent Let $\C$ be a theory of anyons with underlying label set $\mathfrak{L}$. Suppose Alice wants to send a qudit 
\[ \ket{\varphi}=\sum_{c}\gamma_{c}\ket{ab\to c}\in\bigoplus_{c}V^{ab}_{c}=V^{ab}\]
to Bob, where $a,b,c\in\mathfrak{L}$. As usual, $d>1$ for a qudit, and here we have $d:=\dim(V^{ab})=\sum_{c}N^{ab}_{c}$; this means that both $a,b\in\mathfrak{L}$ must be \textit{nonabelian} charges. We have also assumed \ref{t1}, the first of our two constraints. In addition to simplifying the analysis that follows, this assumption is bolstered by the observation that most anyon models of physical interest are multiplicity-free. \\ 

\begin{enumerate}[label=\textbf{(T\arabic*)},series=teleconditions]
    \item For simplicity, we assume that $\C$ is \textit{multiplicity-free}, i.e. $N^{ab}_{c}\in\{0,1\}$ for all $a,b,c\in\mathfrak{L}$. \\
    \label{t1}
\end{enumerate}

\begin{figure}[H]
 \begin{minipage}[c]{0.4\textwidth}
    \hspace{5mm}
    \includegraphics[width=\textwidth]{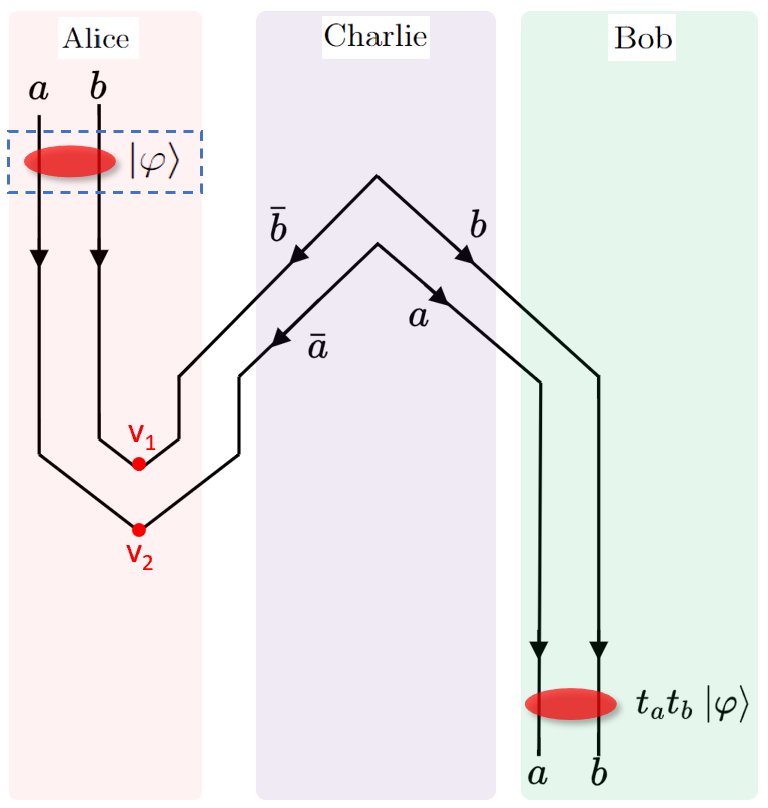}
\end{minipage}\hfill
  \begin{minipage}[c]{0.59\textwidth}    
    \caption{\small Within the dashed blue box, we indicate that $\ket{\varphi}$ is the fusion state of the anyons contained in the spatial region shaded in red at the given timeslice. In the illustrated process, Alice transmits a fusion state $\ket{\varphi}$ to Bob by fusing her anyons with those sent by Charlie (at the highlighted red vertices $v_{1},v_{2}$): Bob receives Alice's state immediately after the fusion at $v_2$. In particular, both fusions must be \textit{annihilations}, and so her transmission may fail with nonzero probability. As usual, the global phase factor $t_{a}t_{b}$ can be ignored; this factor comes from \eqref{dollypow}. The two zigzags can be thought of as forming a `pipeline' along which Alice's state flows to Bob. This pipeline is formed immediately after the fusion at $v_2$.}
     \label{TQT-scaffold}
    \end{minipage}
\end{figure}

\noindent The pivotal structure possessed by anyon theories appears to serve as an obvious foundation for constructing a teleportation protocol: the zigzags seen in Figure \ref{TQT-scaffold} act as a natural conduit of information. Specifically, this relies on the property that a zigzag labelled by $a\in\mathfrak{L}$ can be \textit{`yanked straight'} at the cost of a global phase factor $t_{a}$ (namely, the pivotal coefficient of $a$; see Section \ref{lbopsec}).  

\begin{tcolorbox}
 \begin{heuristic}[\textbf{Information `pipeline'}] Our guiding heuristic in the design of the protocol is to exploit the zigzags seen in Figure \ref{TQT-scaffold}: by analogy, these form a `\textit{pipeline}' along which we envision the `flow' of quantum information in spacetime from Alice to Bob.   
 \label{the-heuristic}
 \end{heuristic}
\end{tcolorbox} 

\vspace{-3mm}
\vspace{-3mm}

\begin{figure}[H]
    \includegraphics[width=0.8\textwidth]{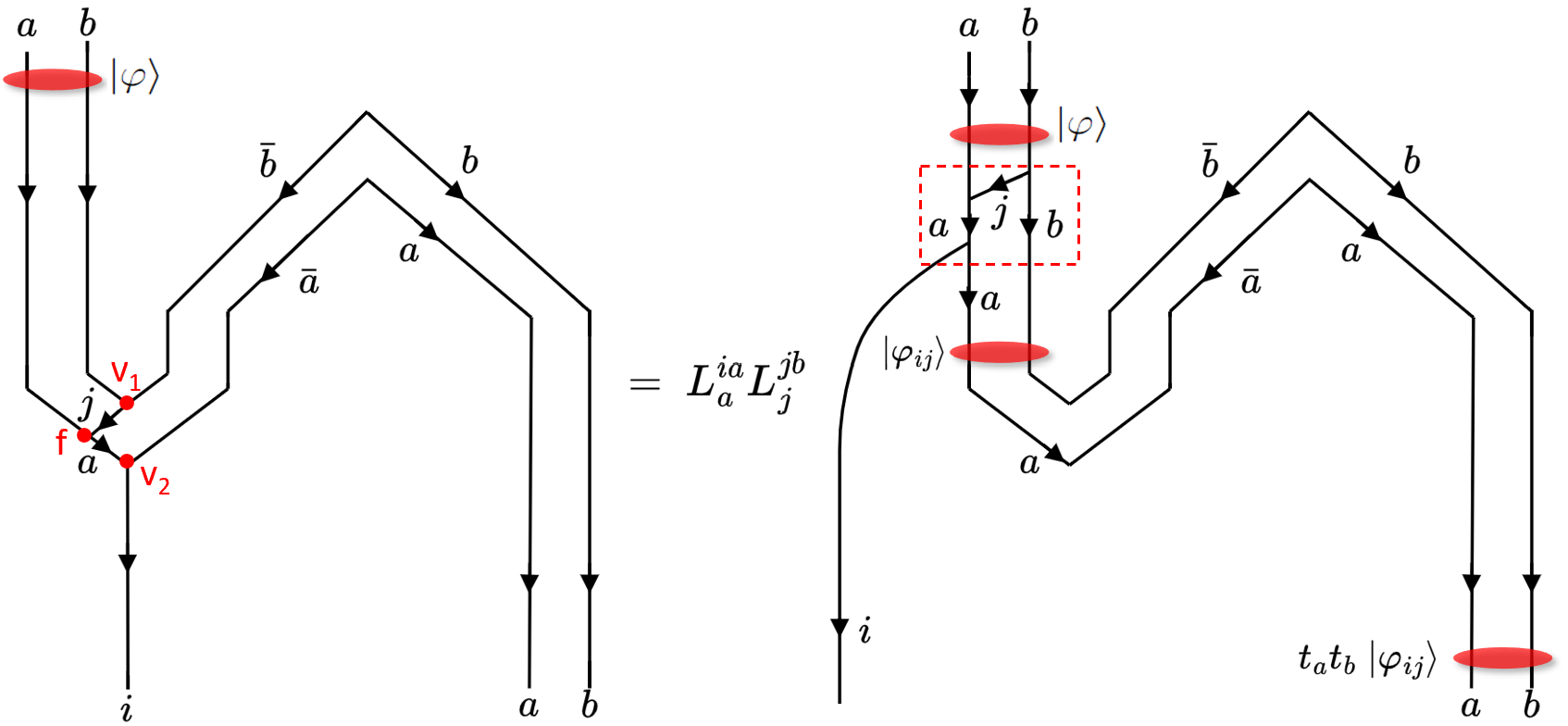}
    \vspace{-3mm}
    \caption{\small Applying \eqref{keyid1} at vertices $v_{1},v_{2}$ yields the right-hand side. The dashed red box encloses a process that scrambles Alice's qudit; we denote its effect by $U_{ij}:\ket{\varphi}\mapsto\ket{\varphi_{ij}}$. }
    \label{keyid1-corrol}
\end{figure}

\noindent Bob's receipt of Alice's state $\ket{\varphi}$ (Figure \ref{TQT-scaffold}) requires annihilations at fusion vertices $v_{1},v_{2}$. Since $a,b$ are nonabelian, we have that $\sum_{c}N^{a\bar{a}}_{c},\sum_{c}N^{b\bar{b}}_{c}>1$ (see Remark \ref{abvnab}). Hence, there is a nonzero chance of nontrivial fusion outcomes at points $v_{1},v_{2}$. Such outcomes act as \textit{obstructions} to a successful instantaneous transmission (following fusion at $v_2$) of Alice's state to Bob, and will be referred to as \textit{exceptions} in the context of the protocol. The protocol must be able to \textit{handle} exceptions $(j,i)\neq(0,0)$, where $j,i$ respectively denote the fusion outcomes at $v_{1},v_{2}$. Although it may seem that exceptions are bound to obstruct the pipeline, applying the identity \eqref{keyid1} as in Figure \ref{keyid1-corrol} indicates otherwise.\footnote{Identity \eqref{keyid1} simply follows directly from the definition of the leg-bending operators in \eqref{tgcfren2}.}
\vspace{-1mm}
\begin{equation}
 \raisebox{-9mm}{\includegraphics[width=0.63\textwidth]{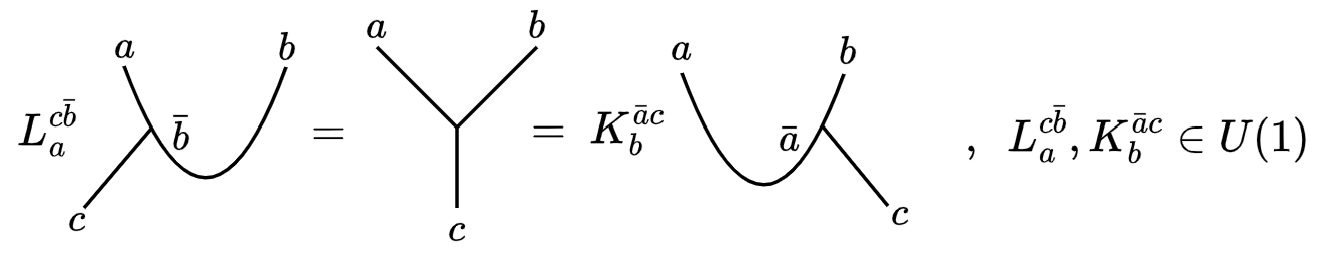}}
\label{keyid1} 
\end{equation}

\vspace{-3mm}
\vspace{-3mm}

\noindent The left-hand side of Figure \ref{keyid1-corrol} is identical to Figure \ref{TQT-scaffold} except for the depiction of possible exceptions occurring at $v_1, v_2$, and Alice's fusion of a nontrivial fusion outcome $j$ with her leftmost anyon $a$ at point $f$. Indeed, in order for fusion $v_2$ (of $a$ with $\bar{a}$) to take place, we must demand that the anyon $j$ is first \textit{absorbed} by either $a$ or $\bar{a}$ (as we wish to avoid braiding); we assume the former without loss of generality (see Section \ref{telefreesec}). The equivalence in Figure \ref{keyid1-corrol} reveals that if Alice performs fusions $v_{1},f,v_{2}$, then Bob's pair of anyons will be in the state $\ket{\varphi_{ij}}$ immediately after the fusion at $v_2$. The pipeline is thus effectively preserved, with the catch that Alice's qudit is \textit{scrambled} by some $U_{ij}\in\End(\mathbb{C}^{d})$ before travelling along the pipeline (noting $U_{00}=\id$). Recalling the 1-qudit operators defined in Section \ref{standsec}, Alice's qudit is (up to a global phase as seen in Figure \ref{keyid1-corrol}) scrambled by a $j$-rung and then an $i$-branch: 
\vspace{-1mm}
\begin{equation} U_{ij}: \ket{\varphi}\xmapsto{B_{j}} \ket{\varphi_{j}} \xmapsto{D_{i}} \ket{\varphi_{ij}}\label{scrambleparse}\end{equation}
As usual, we expect that Bob should somehow be able to \textit{correct} this scrambling upon receipt of dits $i,j$ from Alice. This is explained in Section \ref{telecorrsec}.\\

\noindent Crucially, demanding absorption of $j$ at point $f$ levies an additional constraint \ref{t2} on $(\mathfrak{L},\times)$.\\

\begin{enumerate}[resume*=teleconditions]
    \item Any fusion outcome of $b,\bar{b}$ must be absorbed by $a$ with probability $1$. That is,  it must hold that $N^{aj}_{x}N^{b\bar{b}}_{j}=\delta_{ax}$ for all $j\in\mathfrak{L}$ such that $N^{b\bar{b}}_{j}=1$. \\
    \label{t2}
\end{enumerate}

\noindent Firstly, \ref{t2} implies that all fusion outcomes $j$ of $b,\bar{b}$ must be \textit{abelian}. Secondly, we note from application of \eqref{twissta} that $N^{aj}_{a}=N^{\bar{a}a}_{\bar{j}}=N^{\bar{a}a}_{j}$, whence it \textit{suffices} to additionally set $b:=\bar{a}$. We are thus led to seek an anyon model $(\mathfrak{L},\times)$ where there exists  
\begin{enumerate}[label=(\alph*)] 
\item $G=\{0,\ldots\}\subset\mathfrak{L}$ where $|G|=:d\geq2$ and all charges in $G$ are abelian 
\item $a\in\mathfrak{L}$ such that $\bar{a}\times a=\sum_{g\in G}g$
\end{enumerate}
\noindent Note that $(G,\times)$ is a \textit{finite group} of order $d\geq2$. The obvious family of candidates for $(\mathfrak{L},\times)$ is $(\mathfrak{L}^{\TY}_{G},\times)$, as defined in Section \ref{tyanyonsec}. Recall that $(\mathfrak{L}^{\TY}_{G},\times)$ defines an anyon model if and only if $G\cong\mathbb{Z}_{2}^{n}$, and that the corresponding anyon theories are denoted by $\TY(G)$.

\begin{tcolorbox}
    \textbf{Conclusion.} We henceforth set $\mathcal{C}=\TY(G)$ and $a,b=q$.
\end{tcolorbox}

\subsection{Interlude: maximal entanglement}
\label{maxentangsec} At this point, a sanity check may be in order. For teleportation to be possible, Bennett's laws \eqref{benloix} indicate that Alice's pair of anyons (received from Charlie) should be \textit{maximally entangled} with Bob's pair. Since our teleportation protocol uses TY-anyons $q$, Proposition \ref{maxentangprop} below tells us that this is indeed the case: take $a,b=q$, and recall that all fusion outcomes of $q$ with itself are abelian (i.e. they all have quantum dimension $1$).

\begin{prop} Suppose we pair-create nonabelian anyons $\bar{b},\bar{a},a,b$ as shown in Figure \ref{TQT-scaffold}. Assuming \ref{t1} (i.e. fusion is multiplicity-free), the pair $\bar{b},\bar{a}$ is maximally entangled with pair $a,b$ if and only if the quantum dimension $d_{e}$ is constant over all possible fusion outcomes $e$ (of $a$ with $b$). 
\label{maxentangprop}
\end{prop}

\begin{proof}
Recall that $\sum_{c}N^{ab}_{c}=:d$ and by \eqref{twissta} that $N^{\bar{b}\bar{a}}_{\bar{c}}=N^{ab}_{c}$. Since $a$ and $b$ are nonabelian, we have $d\geq2$. Pair-creations of the form (i) below initialise fusion state (ii) in $V^{\bar{b}\bar{a}ab}_{0}$, which we shall express in a fusion basis of form (iii). 
\vspace{-3mm}
\begin{figure}[H]
\includegraphics[width=0.8\textwidth]{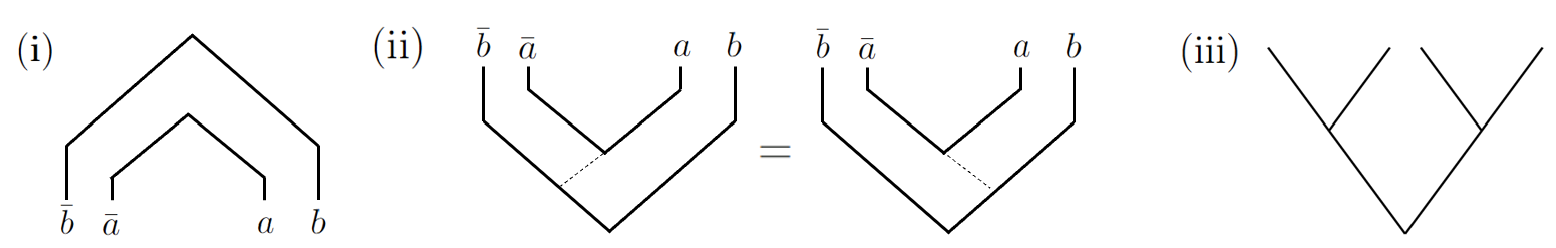}
\end{figure}
\vspace{-3mm}
\noindent Recoupling, we get
\vspace{-2mm}
\begin{figure}[H]
\includegraphics[width=0.86\textwidth]{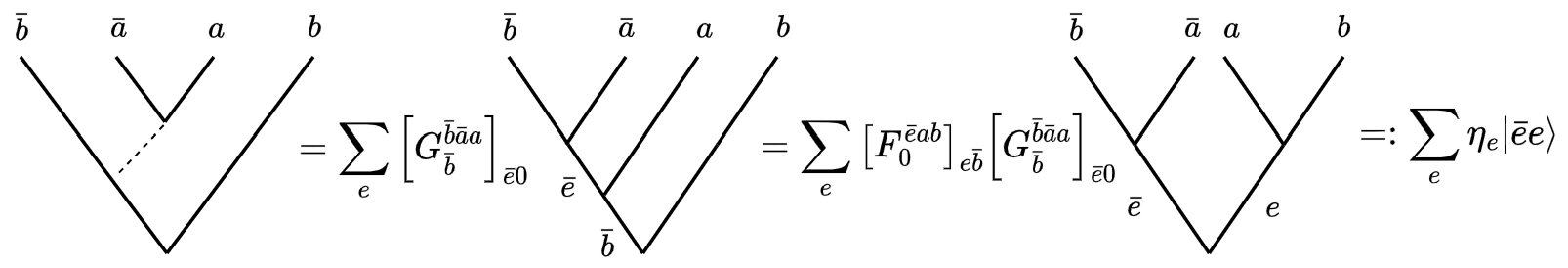}
\end{figure}
\vspace{-2mm}
\noindent where $[F^{\bar{e}ab}_{0}]_{e\bar{b}}\in\U(1)$, and where $\left|[G^{\bar{b}\bar{a}a}_{\bar{b}}]_{\bar{e}0}\right|=\sqrt{\frac{d_e}{d_a d_b}}$ by unitarity of $L_{\bar{e}}^{\bar{b}\bar{a}}$. Thus, $|\eta_{e}|=\sqrt{\frac{d_e}{d_a d_b}}$. The two pairs of anyons are maximally entangled if $|\eta_{e}|=d^{-1/2}$ for all fusion outcomes $e$ (of $a$ with $b$). The quantity $|\eta_{e}|$ is constant for all $e$ if and only if $d_{e}$ is constant for all such $e$; indeed, if this is the case, then we see via \eqref{qdimeq2} that $|\eta_{e}|=d^{-1/2}$.
\end{proof}

\subsection{The correction step}
\label{telecorrsec}
Once Bob has received Alice's dits $i,j\in\mathbb{Z}_{d}$ (signalling her measurement outcomes), he must accordingly apply a transformation $U^{-1}_{ij}\in\End(\mathbb{C}^{d})$ in order to unscramble the fusion state $\ket{\varphi_{ij}}$ of his pair of TY-anyons. By doing so, Bob recovers Alice's initial qudit $\ket{\varphi}$. In this section, we formulate a procedure for applying this transformation without the need for any braiding; our prescription relies on two observations:
\begin{enumerate}[label=(\arabic*)]
    \item \textit{`Branch decoupling'} (Section \ref{branchsec})
    \item \textit{`Rung-cancelling'} (Section \ref{ladsec})
\end{enumerate}
\begin{tcolorbox}
    Recall that we set $\mathcal{C}=\TY(G)$. Hence, Convention \ref{tyconvention} applies to all diagrams in the sequel.
\end{tcolorbox}

\subsubsection{Branch decoupling: "Recouple to decouple!"}
\label{branchsec}
We refer to a worldline connected to the exterior of the `pipeline' as a branch (see Section \ref{standsec}). Our branches will always carry charge valued in $G\cong\mathbb{Z}_{2}^{n}$. We know that the occurrence of such a (nontrivial) branch from the pipeline is partly responsible for the scrambling of Alice's qudit, and thus seek to neutralise its effect upon Bob's receipt.

\vspace{-3mm}
\begin{figure}[H]
\centering
\includegraphics[width=0.8\textwidth]{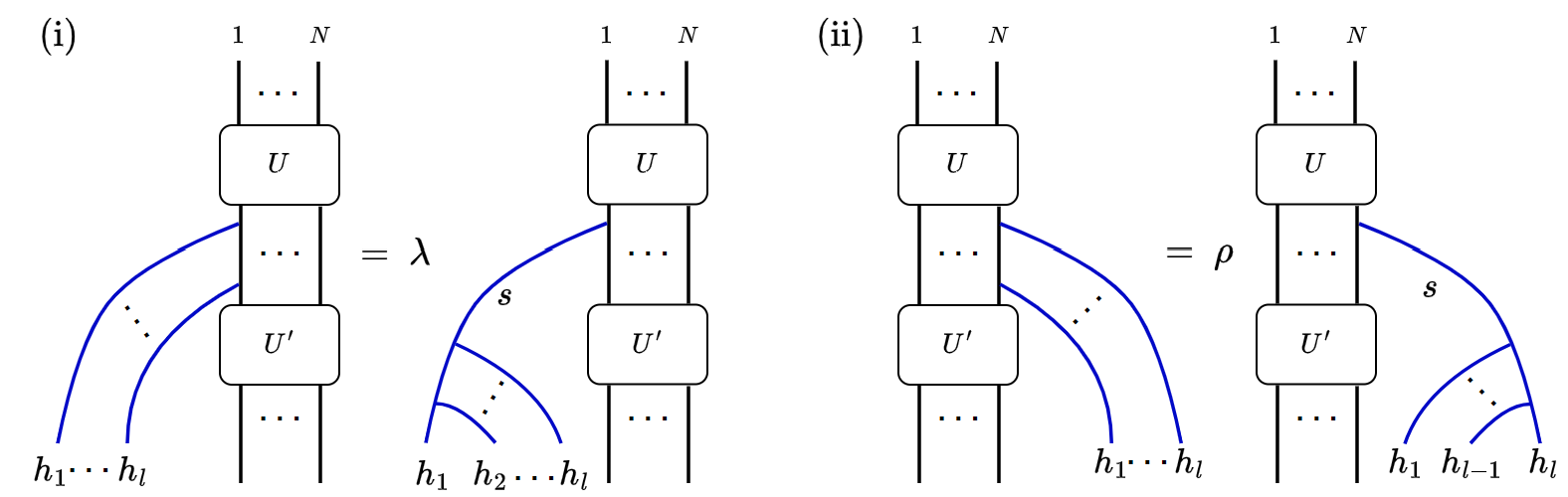}
\vspace{-3mm}
\caption{\small Coupons $U,U'\in\End\left(V^{q\cdots q}\right)$ each represent some particle process on $N$ TY-anyons. Equations (i)-(ii) can respectively be seen as a change of fusion basis on the $1$-dimensional spaces $V^{q}_{h_{1}\cdots h_{l}q}$ and $V^{q}_{qh_{1}\cdots h_{l}}$, whence $\lambda,\rho\in\U(1)$.}
\label{stumped}
\end{figure}
\vspace{-3mm}

\begin{figure}[H]
\centering
\includegraphics[width=0.7\textwidth]{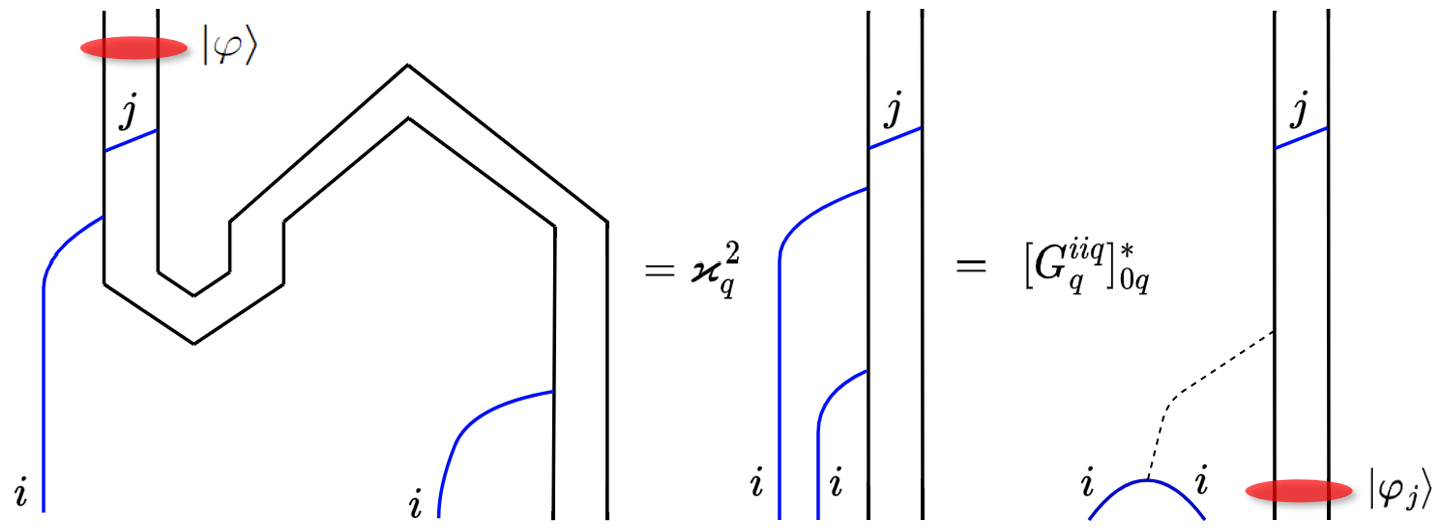}
\vspace{-3mm}
\caption{\small The left-hand side continues from the right-hand side of Figure \ref{keyid1-corrol}, but with the following changes: (a) we now specialise to the setting of TY-anyons; and (b) we add a second $i$-branch. The first equality amounts to straightening both zigzags (incurring a scaling factor of $\varkappa^{2}_{q}=1$), and the second is a decoupling. Hence, after Bob applies an $i$-branch, his pair of TY-anyons is (up to a global phase) in state $\ket{\varphi_j}=B_{j}\ket{\varphi}$.}
\label{decouplefig1}
\end{figure}

\noindent In Figure \ref{stumped}, the phases $\lambda,\rho$ are each products of $1$-dimensional $F$-matrices. The equations demonstrate that (up to a global phase) the occurrence of multiple branches carrying charge $h_{1},\ldots,h_{l}$ is equivalent to that of a single branch carrying charge $s=h_{1}\cdots h_{l}$. Since the charge carried by a branch is abelian, the order in which branches are applied does not matter (i.e. branch operators commute up to a global phase factor). When $s=0$, we say that the $l$ branches are effectively \textit{decoupled} from the system of TY-anyons, since this is physically equivalent to having no branches at all.\footnote{That is, $\Pi_{i=1}^{l}D_{h_i}$ and $\Pi_{i=1}^{l}E_{h_i}$ coincide with the identity operator (up to a global phase factor) when $\Pi_{i=1}^{l}{h_i}=0$.} This observation finds an obvious application in correcting for the effect of an $i$-branch: Bob simply has to generate an $i$-branch (i.e. split an $i$-anyon from his leftmost TY-anyon) upon receipt of Alice's dits $i,j$; see Figure \ref{decouplefig1}. 

\vspace{-3mm}
\vspace{-3mm}

\subsubsection{Rung-cancelling}
\label{ladsec}
After Bob has corrected for the effect of the $i$-branch, there solely remains the task of correcting for the effect of the $j$-rung $B_{j}$ from Figure \ref{decouplefig1}. To this end, Lemma \ref{key-prop} is key: Bob can apply $A_{j}:\ket{\varphi_{j}}\mapsto \ket{\varphi}$. The $1$-qudit teleportation protocol is summarised in Figure \ref{TQT-1dit}.

\begin{lemma}[\textbf{Rung-cancelling}]
For each $h\in G$, we have that $A_{h}B_{h}=B_{h}A_{h}=\id_{V^{qq}}$. That is,
\begin{equation}
\raisebox{-7mm}{\includegraphics[width=0.25\textwidth]{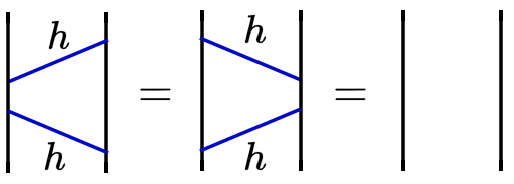}}
\label{key-prop-diag}
\end{equation}
    \label{key-prop}
\end{lemma}
\vspace{-3mm}
\begin{proof}
In the following, we assume that all blue edges are labelled by $h\in G\cong\mathbb{Z}_{2}^{n}$.
\vspace{-3mm}
\begin{subequations}
\begin{align}
    &\raisebox{-5mm}{\includegraphics[width=0.65\textwidth]{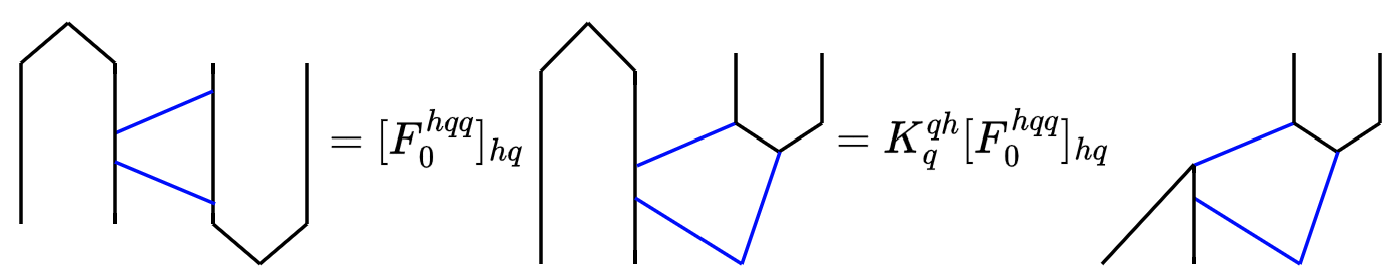}} \label{rungcaneq1} \\
    & \raisebox{-5mm}{\includegraphics[width=0.65\textwidth]{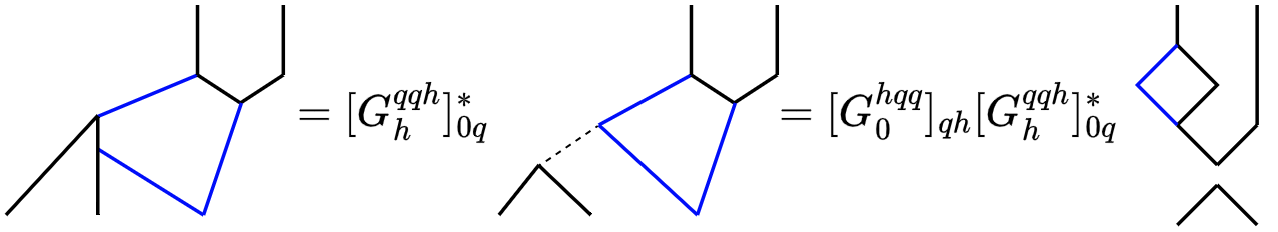}} \label{rungcaneq2}
\end{align}
\label{rung-can-eqset-1}
\end{subequations}
\noindent Plugging \eqref{rungcaneq2} into \eqref{rungcaneq1} and simplifying, we obtain
\begin{equation}
    \raisebox{-8mm}{\includegraphics[width=0.225\textwidth]{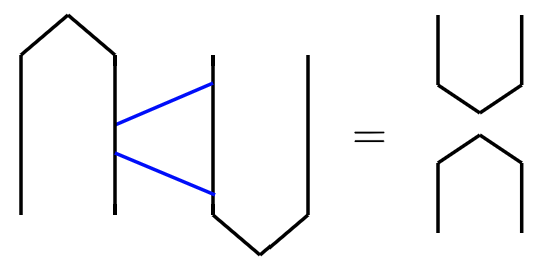}}
\label{rung-can-eqset-2}    
\end{equation}
This allows us to deduce \eqref{rung-can-eqset-3}, where the first and third equalities simply follow from $\varkappa_{q}^{2}=1$, and the second equality from \eqref{rung-can-eqset-2}.
\begin{equation}
    \raisebox{-8mm}{\includegraphics[width=0.5\textwidth]{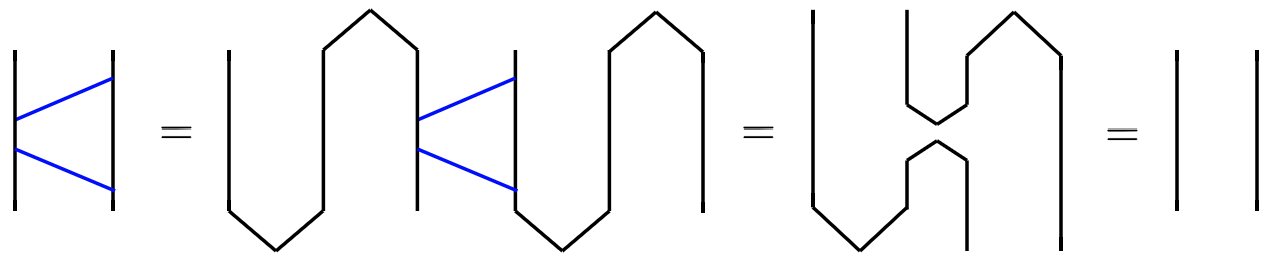}}
\label{rung-can-eqset-3}      
\end{equation}
Finally, we need $[A_{h},B_{h}]=0$. This immediately follows from the fact that $\End(V^{qq})$ is isomorphic to $\bigoplus_{g\in G}\End(V^{qq}_{g})$ (where $N^{qq}_{g}=1$ for all $g\in G$), implying all endomorphisms on $V^{qq}$ commute.
\end{proof}

\subsection{Protocol summary and proof of theorem}
\label{1qtelesum}

\noindent In Section \ref{tynatsec}, we began with Heuristic \ref{the-heuristic} as a foundation for the anyonic teleportation protocol. In order to handle `exceptions', we were led to the setting of Tambara-Yamagami theories, and 
then appealed to some diagrammatic observations; namely, identity \eqref{keyid1}, `branch decoupling' (Section \ref{branchsec}) and `rung-cancelling' (Section \ref{ladsec}). Altogether, the initial pipeline template from Figure \ref{TQT-scaffold} is recast as in Figure \ref{TQT-1dit} below.   

\vspace{-1mm}
\begin{figure}[H]
 \begin{minipage}[c]{0.33\textwidth}
    \hspace{5mm}
    \includegraphics[width=1.1\textwidth]{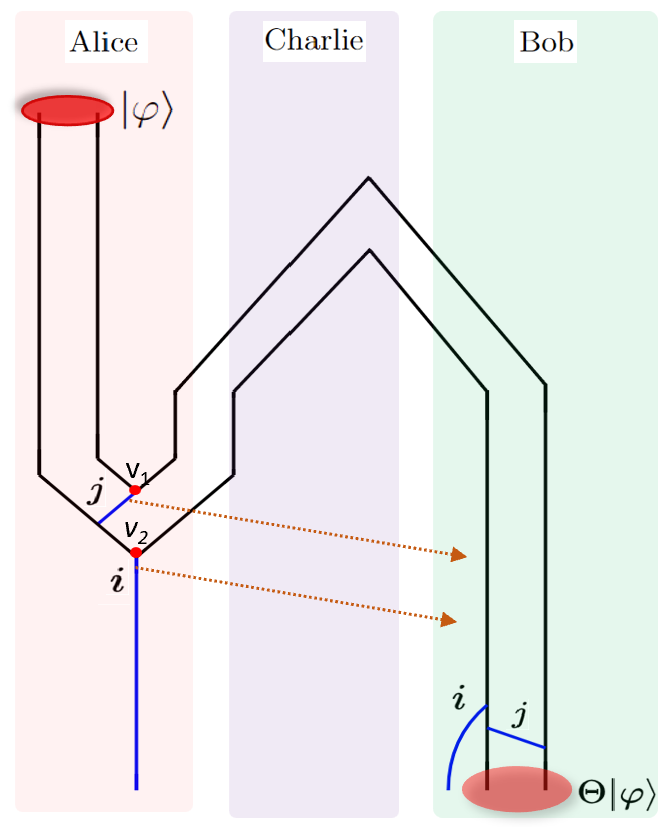}
\end{minipage}\hfill
  \begin{minipage}[c]{0.66\textwidth}    
    \caption{\small Braid-free single qudit ($d=2^{n}$) teleportation protocol using anyon theory $\TY(\mathbb{Z}_{2}^{n})$. Worldlines presented according to Convention \ref{tyconvention}. Dotted orange lines represent classical $1$-dit transmissions from Alice to Bob following fusion measurements at points $v_1,v_2$. Immediately after the measurement at $v_2$, Bob's pair of anyons is in state $L^{iq}_{q}L^{jq}_{q}\ket{\varphi_{ij}}$ (see Figure \ref{keyid1-corrol}). Bob applies the corrections described in Section \ref{telecorrsec} upon receipt of Alice's dits $i,j\in\mathbb{Z}_{d}$. The resulting state of his pair of anyons is $\Theta\ket{\varphi}$, where $\Theta=[F^{iiq}_{q}]_{q0}L^{iq}_{q}L^{jq}_{q}$. Since $\Theta$ is a global phase factor (in fact, it is gauge-dependent when $i,j\neq0$), it can be disregarded. An animated example of this process using Ising anyons ($n=1$) and with Approach \ref{pconly} employed during Bob's correction step, is found at \cite{sjvyt1}.}
     \label{TQT-1dit}
    \end{minipage}
\end{figure}
\vspace{-1mm}

\begin{figure}[H]
\includegraphics[width=0.8\textwidth]{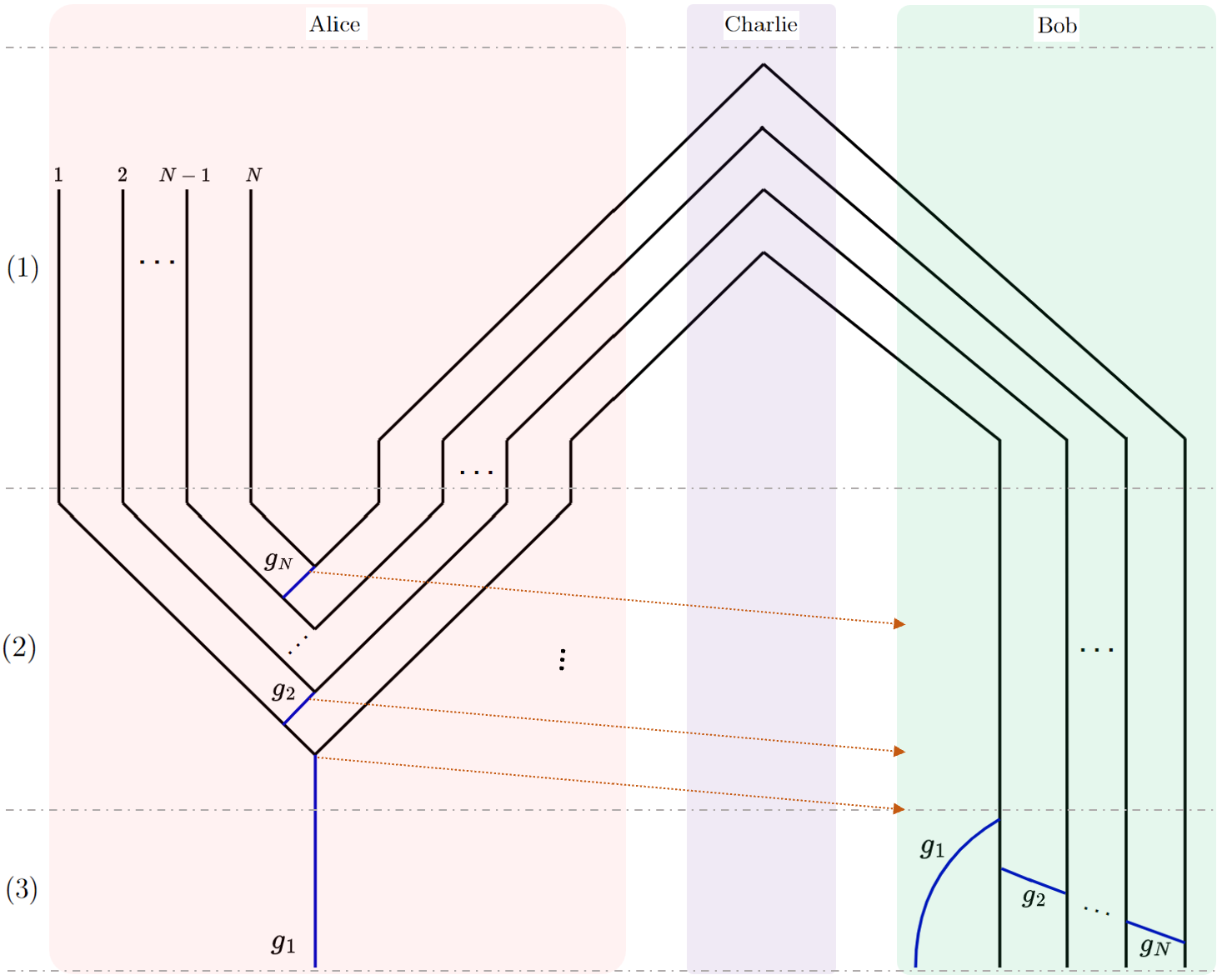}
\vspace{-2mm}
\caption{\small Braid-free $\lfloor\frac{N}{2}\rfloor$-qudit ($d=2^{n}$) teleportation protocol using Tambara-Yamagami anyon theory $\TY(G)$ where $G\cong\mathbb{Z}_{2}^{n}$. Worldlines presented according to Convention \ref{tyconvention}. Dotted orange lines represent classical $1$-dit transmissions from Alice to Bob.}
\label{palominoscruff}
\end{figure}

\noindent The above procedure readily generalises to the case of multiple qudits. Recall from Section \ref{tyquditsec} that $N\geq2$ TY-anyons encode $\lfloor\frac{N}{2}\rfloor$ qudits. Suppose Alice wishes to teleport the state of $N\geq 2$ TY-anyons to Bob; this can be achieved without braiding, as illustrated in Figure \ref{palominoscruff}.\footnote{Conversely, if Alice and Bob share $p$ e-dits, Alice can teleport (at most) the state of $2p+1$ TY-anyons.} Let $f:G\to\mathbb{Z}_{d}$ be a bijection, and let $\ket{\varphi}$ be Alice's state in some fusion basis $\mathcal{B}$. Alice and Bob have agreed on  bijection $f$ and computational basis $\mathcal{B}$ beforehand. We summarise the procedure according to the three phases highlighted between the horizontal dashed lines. 
\vspace{1mm}
\begin{enumerate}[label=(\arabic*)]
\item Charlie pair-creates and distributes $2N$ TY-anyons between Alice and Bob. In doing so, he equally shares the halves of $\lfloor\frac{N}{2}\rfloor$ e-dits between them.\footnote{Recall from Section \ref{maxentangsec} that $1$ e-dit is encoded in two pairs of TY-anyons (consecutively pair-created by Charlie): the left pair goes to Alice and the right pair to Bob; each pair encodes one of the two maximally entangled qudits.} 
\vspace{1mm}
\item Alice fuses the rightmost of her $N$ TY-anyons with the leftmost of those sent by Charlie, producing an anyon of charge $g_{N}\in G$ with probability $1/d$ (see Remark \ref{bartok}). She sends dit $f(g_{N})$ to Bob,\footnote{Since $d=2^n$, $1$ dit can be sent in the form of $n$ bits.} and fuses $g_{N}$ with a TY-anyon to the immediate left. This continues until outcome $g_1$ is obtained and $f(g_1)$ is sent to Bob. Alice is left with ancillary anyon $g_{1}$.
\vspace{1mm}
\item After Bob receives $N$ consecutive $1$-dit transmissions from Alice, he performs a sequence of splittings (according to the received dits) and fusions as shown in Figure \ref{palominoscruff}. This leaves Bob with $N$ TY-anyons in state $\Theta\ket{\varphi}$ (see \eqref{dawgphase}), along with an ancillary anyon of charge $g_1$ to their left.
\end{enumerate}

\noindent That Bob's fusion state is $\ket{\varphi}$ (following the process shown in \mbox{Figure \ref{palominoscruff}}) is seen using the same observations as for the $1$-qudit case above.

\begin{proof}[Proof of Theorem \ref{brfreetelethm}]
Let $D$ denote the diagram on the left-hand side of \eqref{telepfdiag}, and $P$ the string diagram in Figure \ref{palominoscruff} (ignoring annotated regions and dit transmissions). Then \mbox{$P=\varkappa_{q}^{N}\cdot\Pi_{j=1}^{N}L^{g_{j}q}_{q}\cdot D$;} starting with $P$, this equality is seen by (a) applying identity \eqref{keyid1} to outcomes $g_{N},\ldots,g_{1}$, and then (b) straightening the $N$ zigzags labelled by $q$. The first equality of \eqref{telepfdiag} is just a recoupling on a 1-dimensional space (i.e. `branch decoupling' as in Section \ref{branchsec}), and the second equality follows from successive application of the `rung-cancelling' Lemma \ref{key-prop} for $g_{2},\ldots,g_{N}$.
\begin{equation}
\raisebox{-20mm}{\includegraphics[width=0.85\textwidth]{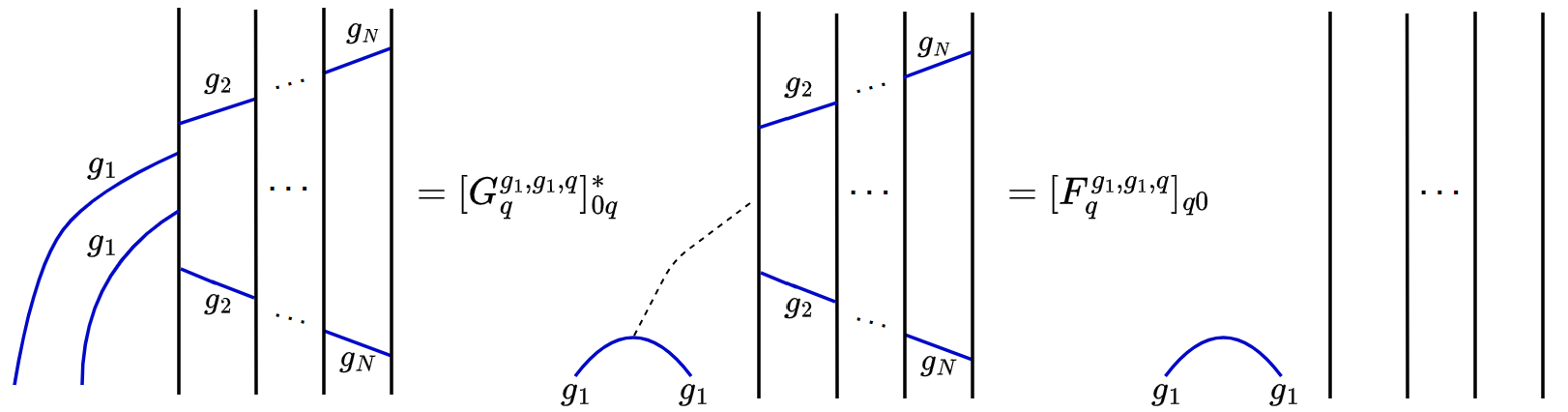}}
\label{telepfdiag}
\end{equation}
\noindent Suppose Alice's $N$ TY-anyons are initially in state $\ket{\varphi}$ (in some fixed fusion basis) at the start of the process shown in Figure \ref{palominoscruff}. By the above, the state of Bob's $N$ TY-anyons (in the same fusion basis) immediately after this process is $\Theta\ket{\varphi}$, where
\begin{equation}
    \Theta=[F^{g_{1},g_{1},q}_{q}]_{q0}\cdot\varkappa_{q}^{N}\cdot\left(\Pi_{j=1}^{N}L^{g_{j}q}_{q}\right)   
    \label{dawgphase}
\end{equation}
\noindent Since $\Theta$ is a global phase, it may be disregarded.
\end{proof}

\begin{remark}[\textbf{Outcome probability}] The probability that Alice measures a given $g\in G$ for an outcome $g_{j}$ is $\frac{1}{d}$. To see this, apply (\ref{fester}) in Figure \ref{palominoscruff}, noting $\left|[G^{qqq}_{q}]_{g0}\right|=d_{q}^{-1}$ by unitarity of $L^{qq}_{0}$.
\vspace{-1mm}
\begin{equation}
    \raisebox{-7mm}{\includegraphics[width=0.37\textwidth]{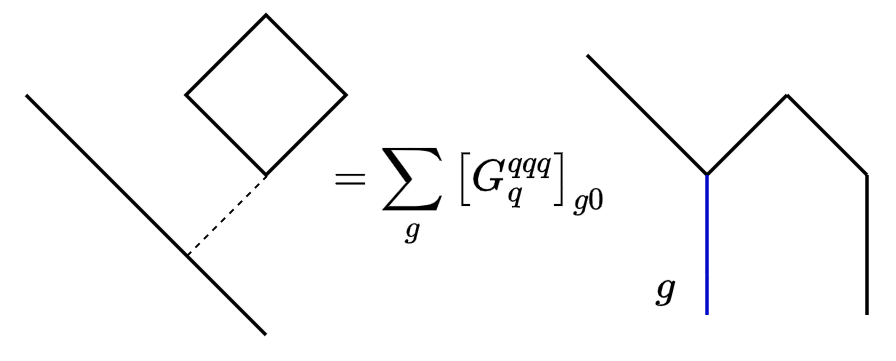}}
    \label{fester}
\end{equation}
\label{bartok}
\end{remark}

\subsection{Freedoms in procedure}
\label{telefreesec}
The process for braid-free teleportation using Tambara-Yamagami anyons is not restricted to the form shown in Figure \ref{palominoscruff}. Below, we summarise some of the possible variations in how the protocol might be realised. As is a common theme of our exposition, the idea is often to show that two processes are equivalent up to a global phase factor.

\begin{enumerate}[label=(\arabic*)]

    \item Alice may instead fuse any one of her measured fusion outcomes $g_{j}\in G$ ($j>1$) with the TY-anyon to the \textit{right}: the process differ only by a global phase factor of $\left[F^{q,g_{j},q}_{g_{j-1}}\right]_{qq}$. \label{whichwayfree}

    \item Recall the relationship amongst rung and branch operators respectively from \eqref{rungeqs} and \eqref{brancheqs}. Thus, during Bob's correction step, any of his rung operators $A_{g_j}$ may be replaced with $B_{g_j}$, $C_{g_{j}}$ or $C'_{g_j}$ for $j>1$; and his $g_{1}$-branch $D_{g_1}$ with $D'_{g_1}$.  

    \item Suppose $N=2$ as in Figure \ref{TQT-1dit}. During Bob's corection step, his $g_{1}$-branch $D_{g_1}$ may also be replaced with $E_{g_1}$ or $E'_{g_{1}}$. This follows from a result proved later on (Proposition \ref{branchswitch}).

    \item In the case $G=\mathbb{Z}_{2}$, there is further freedom in how Bob corrects the bit-flip errors on his qubits. This is explained in Section \ref{locatesec}. \label{x-freedoms}

    \item During Bob's correction step, the order in which the worldlines of charges $g_{j},g_{j+1}$ (where $j\leq 1 \leq N-1$) make contact with that of the TY-anyon separating them does not matter: it is clear that the two processes differ only by a global phase factor. For instance, in Approach \ref{pconly}, the two processes differ by a phase factor of $[F^{g_{j},q,g_{j+1}}_{q}]_{qq}$, and so the order in which Bob performs his pair-creations and fusions does not matter.\label{the_var}

     \item If Bob is instead to the left of Alice, it is easily verified that the mirror image of the process shown in Figure \ref{palominoscruff} achieves teleportation.

    \item Alice or Bob assumes Charlie's role, or they shared their EPR halves in situ before parting.
    
\end{enumerate}

\noindent Some of these freedoms (e.g. \ref{whichwayfree} and \ref{the_var}) may be seen as boons in the sense that the protocols can tolerate certain procedural inaccuracies.

\begin{approach}[\textbf{Only pair-creations during correction}] 
Bob can opt solely to pair-create $g_{j}$-anyons as described by $D'_{g_1}$ and $C_{g_{j}}$ ($j>1$). See also \ref{the_var} in Section \ref{telefreesec}.
\label{pconly}
\end{approach}

\section{\textbf{Superdense Coding Without Braiding}}
\label{sdcchap}
\noindent Recall from Bennett's laws \eqref{benloix} that superdense coding is `dual' to quantum teleportation: roughly, we should be able to obtain one protocol from the other through a rearrangement of resources and operations. Indeed, working within the setting of Tambara-Yamagami anyons from Section \ref{telechap}, we prove Theorem \ref{brfreesdcthm} below. The duality of the protocols is graphically manifest in that Figure \ref{sdcbrfreediag} can be obtained through a rearrangement of Figure \ref{TQT-1dit}. 

\begin{thm}
   Consider a Tambara-Yamagami theory of rank $d+1$ where $d=2^{n}$.  We can realise the $d$-ary superdense coding protocol using Tambara-Yamagami anyons as shown in Figure \ref{sdcbrfreediag}. In particular, no braiding is required.
    \label{brfreesdcthm}
\end{thm}

\subsection{Protocol summary and proof of theorem}
\begin{figure}[H]
\begin{minipage}[c]{0.4\textwidth}
\includegraphics[width=\textwidth]{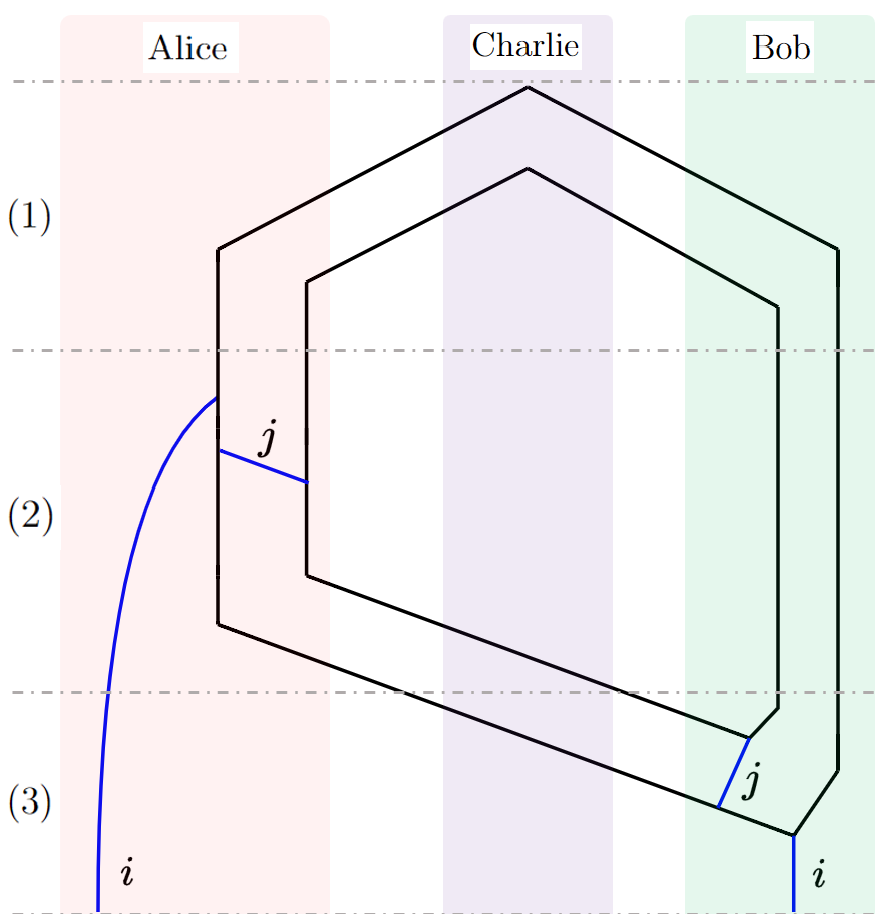}
\end{minipage}
\hfill
\begin{minipage}[c]{0.59\textwidth}
  \caption{\small Braid-free $d$-ary superdense coding protocol using anyon theory $\TY(\mathbb{Z}_{2}^{n})$, where $d=2^{n}$. Worldlines are presented according to Convention \ref{tyconvention}. Alice encodes two dits in the EPR pair using only her half, which she then sends to Bob who is able to recover the dits. Alice and Bob are each left with an ancillary anyon of charge $i$ at the end of the procedure. Observe that this string diagram can be obtained by rearranging the one for teleporting a qudit in Figure \ref{TQT-1dit}. An animated example of this process using Ising anyons ($n=1$) and with Approach \ref{pconly_2} employed during Alice's encoding step, is found at \cite{sjvyt2}.}
\label{sdcbrfreediag}  
\end{minipage}
\end{figure}

Suppose Alice and Bob each possess half of a maximally entangled pair of TY-qudits. Alice wishes to encode a $2$-dit string in her half and then send it to Bob. This can be achieved without braiding, as illustrated in Figure \ref{sdcbrfreediag}; we summarise the procedure according to the three phases highlighted between the horizontal dashed lines. Alice and Bob have agreed on a bijection $f:G\to\mathbb{Z}_{d}$ beforehand.
\vspace{1mm}
\begin{enumerate}[label=(\arabic*)]
\item Charlie has shared the halves of $1$ e-dit between Alice and Bob. 
\vspace{1mm}
\item Alice encodes dits $f(i),f(j)$ in her pair of TY-anyons, which she then transmits to Bob. 
\vspace{1mm}
\item Bob receives Alice's qudit and performs two measurements in order to recover dits $f(i),f(j)$. 
\end{enumerate}

\begin{proof}[\textit{Proof of Theorem \ref{brfreesdcthm}}]
\noindent Applying orthogonality relation \eqref{fyxl}(i) to the final diagram in the sequence of equalities below, we see that the only permissible labelling is $(x,y)=(j,i)$. 
\vspace{-2mm}
\begin{equation*}
\raisebox{-15mm}{
\includegraphics[width=\textwidth]{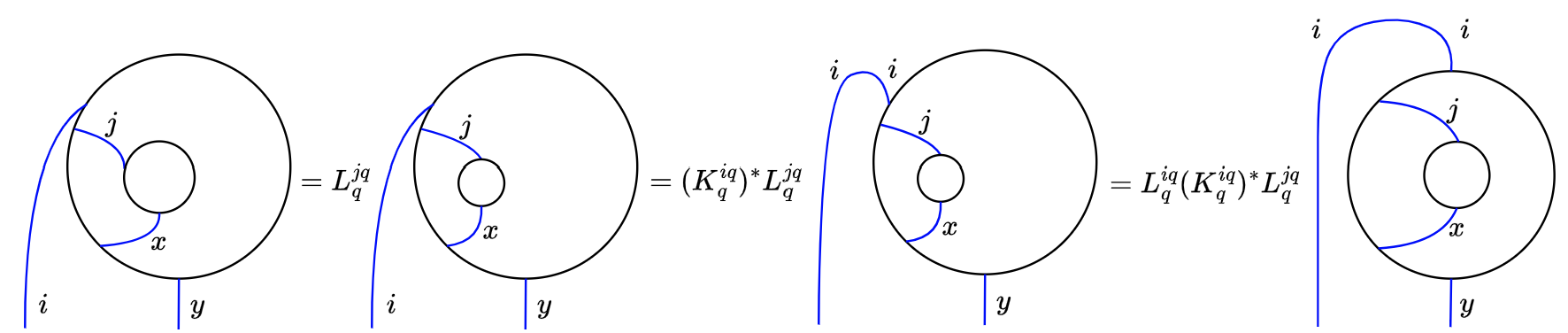}}
    \label{sdcpfeq}
\end{equation*}
\end{proof}

\subsection{Freedoms in procedure}
\label{sdcfreesec}
The process for braid-free superdense coding using Tambara-Yamagami anyons is not restricted to the form shown in Figure \ref{sdcbrfreediag}. Variations in the realisation of this protocol follow in a similar fashion to the teleportation protocol, as presented in Section \ref{telefreesec}. We recapitulate some of these below.

\begin{enumerate}[label=(\arabic*)]

    \item Bob may instead fuse his measured fusion outcome $j\in G$ with the TY-anyon to the right.
    
    \item During Alice's encoding step, she may replace the rung operator $A_{j}$ with $B_{j}$, $C_{j}$ or $C'_{j}$. Also, her $i$-branch $D_{i}$ may be replaced by $D'_{i}$.

    \item During Alice's encoding step, the order in which the worldlines of charges $i$ and $j$ make contact with that of the TY-anyon separating them does not matter: it is clear that the two processes differ only by a global phase factor. For instance, in Approach \ref{pconly_2}, changing the order in which $i$ and $j$ are abosrbed by the TY-anyon between them alters the process by a phase factor of $[F^{iqj}_{q}]_{qq}$, and so the order in which Alice performs her pair-creations and fusions does not matter. \label{the_var_2}
    
    \item In the case where Bob is to the left of Alice, it is easily verified that the mirror image of the process shown in Figure \ref{sdcbrfreediag} achieves superdense coding.

    \item Alice or Bob assume  Charlie's role, or they shared their EPR halves in situ before parting.

\end{enumerate}

\begin{approach}[\textbf{Only pair-creations during encoding}] 
Alice can opt solely to pair-create the $i,j$-anyons as described by $D'_{i}$ and $C_{j}$. See also \ref{the_var_2} in Section \ref{sdcfreesec}.
\label{pconly_2}
\end{approach}

\section{\textbf{Additional Results}}
\label{techsec}
In contrast to their original circuit model formulations, our proofs of the teleportation and superdense coding protocols in the previous sections required no knowledge of the explicit matrix representations of operators; instead, we relied only on the algebraic properties of anyon theories, and leveraged the utility of their graphical calculus. Nonetheless, it is natural to seek matrix representations for the rung and branch operators that underlie our protocols.\\
In Section \ref{furtherpropsec}, we resume the discussion from Section \ref{standsec} in order to learn more about how rungs and branches act on a fusion qudit. These observations permit us a finer understanding of the individual phases of the anyonic teleportation and superdense coding procedures, and are put to further use in the subsections that follow. In Section \ref{scrambsec}, we deduce properties of the effects that Alice's measurements have on a qudit which she is teleporting to Bob. Finally, Sections \ref{paulisansbrsec}-\ref{locatesec} present results within the context of Ising theories $\TY(\mathbb{Z}_{2})$; namely, the realisation of Pauli gates without braiding, the braided analogues of the teleportation and superdense coding protocols, and an analysis of correcting for the phase-flip and bit-flip errors that occur during the teleportation protocol (which highlights extra freedoms in how the correction step is performed).

\subsection{Further properties of rungs and branches}
\label{furtherpropsec}

\noindent We prove the propositions listed below, which offer some insights into the components comprising the protocols from the previous sections. They will also be useful for making the observations in the subsections that follow. 
\begin{enumerate}[label=(\roman*)]
    \item \textit{Proposition \ref{rungexprop}} yields diagonal matrix representations of rung operators $\mathcal{O}_{g}\in\End(V^{qq})$ in terms of the $F$-symbols of $\TY(G)$.
    \item \textit{Proposition \ref{rungpropop}} establishes some basic properties of rung operators. Additionally, we deduce that both rungs and branches act as unitary operators (Remark \ref{unimark}).
    \item \textit{Proposition \ref{branchswitch}} shows that for some fixed $g$, the matrix representations of a $g$-branch, $\mathrm{P}[\mathcal{O}_{g,\id}]$ and $\Lambda[\mathcal{O}_{\id,g}]$ differ at most by a scaling phase. In Lemma \ref{quasireg}, we show that all of these have a matrix representation of the form $\hat{\sigma}(g)$ in \eqref{phafirst}. 
\end{enumerate}
\noindent First, we introduce a few basic lemmas that will be of use. 

\begin{lemma} Both an $i$-branch and a $j$-rung are self-inverse up to a phase factor.
\begin{proof} By the observation in Section \ref{branchsec}, $D_{i}$ is self-inverse up to a global phase factor (and similarly for $E_{i}$): Bob's $i$-branch together with the scrambling $i$-branch result in a net evolution of $D_{i}^{2}=[F^{iiq}_{q}]_{q0}\cdot\id$ (or $E_{i}^{2}=[F^{qii}_{q}]^{*}_{0q}\cdot\id$ if Alice is to Bob's right).\footnote{Moreover, these global phase factors are gauge-dependent for $i\neq0$, meaning they can be disregarded altogether.}  Then by \eqref{brancheq2}, we have that any branch operator is self-inverse up to a phase factor. Finally, by combining the rung-cancelling Lemma \ref{key-prop} with \eqref{rungeq2}, we deduce that any rung operator is self-inverse up to a phase factor.
\end{proof}
\label{selly-mlem}
\end{lemma}

\begin{lemma}
    Let $g\neq0$. The action of (a) a $g$-branch, or (b) $\mathrm{P}[\mathcal{O}_{g,\id}]$ or $\Lambda[\mathcal{O}_{\id,g}]$ on a fusion qudit $\ket{\varphi}=\sum_{h\in G}\gamma_{h}\ket{h}\in\mathbb{C}^{d}$, is given by an operator of the form $\hat{\sigma}(g)$, where\\
\begin{subequations}
\begin{minipage}[c]{0.55\textwidth}
\vspace{0pt}
\begin{align}
\hat{\sigma}(g)=\sum_{h\in G}\alpha^{(g)}_{h}\ket{gh}\bra{h} \ , \ \alpha^{(g)}_{h}\in\U(1) \label{phafirst} 
\end{align}
\end{minipage}
\hfill
\begin{minipage}[c]{0.4\textwidth}
\vspace{-4mm}
\begin{align}
\alpha_{gh}^{(g)}\alpha_{h}^{(g)} \text{ is the same for all $h$} \label{phasecond}
\end{align}
\end{minipage}
\end{subequations}
\begin{proof}
    That the operators in (a)-(b) are of the form $\hat{\sigma}(g)$ in \eqref{phafirst} follows directly from conservation of charge (see Section \ref{chargeconsec}). That the entries of this matrix must satisfy \eqref{phasecond} follows from the observation that operators (a)-(b) are self-inverse up to a global phase factor (Lemma \ref{selly-mlem}).   
\end{proof}
\label{quasireg}
\end{lemma}

\noindent If $\alpha_{h}^{(g)}=1$ for all $g,h$, then $g\mapsto \hat{\sigma}(g)$ coincides with the regular representation of $G$, and each $\hat{\sigma}(g)$ is a self-inverse permutation matrix: this happens for $G=\mathbb{Z}_{2}$ given an appropriate choice of gauge.

\begin{lemma} $\left|[F^{qqq}_{q}]_{ij}\right|=d_{q}^{-1}$ for all $i,j\in G$.
    \begin{proof}
Solely within the scope of this proof, we will relabel Tambara-Yamagami charge $q$ by $x$ so as to avoid a clash of variables. Take pentagon equation \eqref{mfpenteq} for $(a,b,c,d,e)=(x,x,\overbar{j},x,x)$, $(p,q)=(j,0)$, $(r,s)=(x,i)$. That is,
\[
[F^{xxx}_{x}]_{ij}[F^{j\overbar{j} x}_{x}]_{x0}=
[F^{x\overbar{j}x}_{i}]_{xx}[F^{xxx}_{x}]_{i0}[F^{xx\overbar{j}}_{0}]_{xj}
\]
\noindent We note that $[F^{j\overbar{j} x}_{x}]_{x0}$, $[F^{x\overbar{j}x}_{i}]_{xx}$, $[F^{xx\overbar{j}}_{0}]_{xj}\in\U(1)$ since they act on $1$-dimensional spaces. Thus, $f_{ij}=\omega\cdot f_{i0}$ where $\omega\in\U(1)$. By unitarity of $K^{xx}_{i}$, $|f_{i0}|=d_{x}^{-1}$. The result follows.     
    \end{proof}
\label{gtoqrn}
\end{lemma} 

\begin{prop}[\textbf{Expression for rung operators}]
    Let  $f_{gh}=[F^{qqq}_{q}]_{gh}$ and $f_{gh}=|f_{gh}|\cdot\omega_{gh}$ where $\omega_{gh}\in\U(1)$. Then for $h\in G$,
    \begin{center}
        (i) \ \ $A_{h}=(K^{qh}_{q})^{*}\sum_{g\in G}K^{qq}_{g}\omega_{gh}\ket{g}\bra{g}$ \quad , \quad (ii)  \ \ $B_{h}=(L^{hq}_{q})^{*}\sum_{g\in G}L^{qq}_{g}\omega_{gh}\ket{g}\bra{g}$
    \end{center}
\begin{proof}
We give the proof for (i); that of (ii) follows similarly. Expressing (i) diagrammatically,
\vspace{-1mm}
\begin{equation}
\raisebox{-5mm}{
\includegraphics[width=0.625\textwidth]{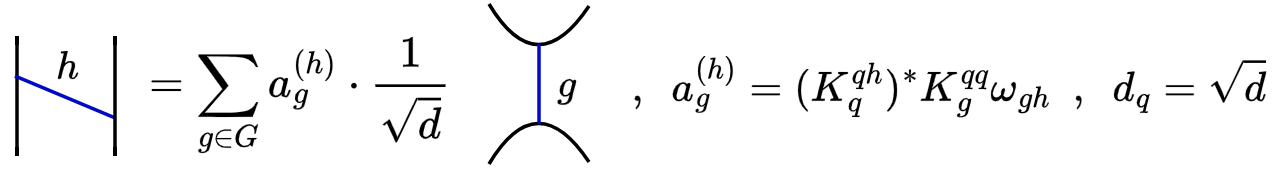}}
\label{rungpfdiag1}
\end{equation}
where for clarity, we have explicitly written the nontrivial normalisation factors of all trivalent vertices. Equation \eqref{rungpfdiag1} simply expresses $A_{h}$ in terms of the standard basis for $\End(V^{qq})$.  We now verify the expression for $[A_{h}]_{gg}=:a^{(h)}_{g}$. Continuing from the diagrammatic equation in \eqref{rungpfdiag1}, 
\vspace{-1mm}
\begin{equation*}
\includegraphics[width=0.825\textwidth]{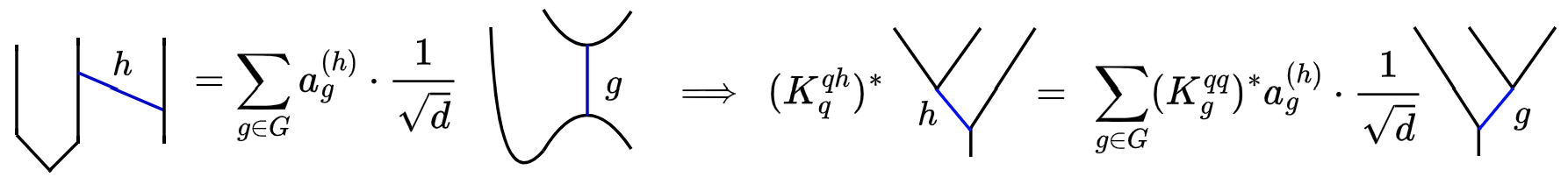}
\end{equation*}
\noindent Both sides of the final equation are related by a transformation under $F^{qqq}_{q}$. Comparing coefficients, 
$a_{g}^{(h)}=(K^{qh}_{q})^{*}K^{qq}_{g}f_{gh}\sqrt{d}$. By Lemma \ref{gtoqrn}, $f_{gh}=\omega_{gh}/\sqrt{d}$. 
\end{proof}
\label{rungexprop}
\end{prop}

\begin{remark}[\textbf{Unitarity}] We see from \eqref{phafirst} that $\hat{\sigma}(g)$ is unitary. Hence, as expected, $g$-branches and $g$-rungs act as unitary operators in $\U(2^n)$. Alternatively, one can verify unitarity by using Proposition \ref{rungexprop}. Combining unitarity with rung-cancelling Lemma \ref{key-prop}, we see that $B_{g}=A^{\dagger}_{g}$. 
\label{unimark}
\end{remark}

\begin{prop}[\textbf{Rung operators: properties and orthogonal basis}]
Let $\mathcal{S}_{0}$ be the set of $d \times d$ real matrices that are diagonal, self-inverse, and traceless. These matrices have their diagonals populated by elements of $\{\pm1\}$, with an equal number of $+1$s and $-1$s. Let $\mathcal{S}$ be the set of quantum gates given by $\mathcal{S}_{0}$ modulo equivalence under a $-1$-scaling. Then,
\begin{enumerate}[label=(\roman*)]
\item Up to a global phase factor, a rung operator $\mathcal{O}_{g}$ is gauge-invariant. \label{gtransrung}
\item Up to a global phase factor, the action of a $g$-rung $\mathcal{O}_{g}$ (for $g\neq0$) on a fusion qudit encoded in a pair of Tambara-Yamagami anyons, is given by an element of $\mathcal{S}$. \label{p1corr7}
\item The correspondence described in \ref{p1corr7} is an injection $\{\mathcal{O}_{g}\}_{g\in G\setminus\{0\}} \hookrightarrow \mathcal{S}$. \label{injitem}
\item The set $\{\mathcal{O}_{i,\id}\circ\mathcal{O}_{\id, j}\}_{i,j\in G}$ of $1$-qudit gates define an orthogonal basis for $\End(V^{qqq}_{q})$ with respect to the trace inner product.\footnote{The same is true of the set $\{\mathcal{O}_{\id, j}\circ\mathcal{O}_{i,\id}\}_{i,j\in G}$. In fact, note that the two operators commute up to a phase factor $[F^{iqj}_{q}]_{qq}$ which is gauge-variant (and thus unphysical) for $i\neq j$. \label{6j-foot}} \label{orthogoscramble}
\end{enumerate}
\begin{proof}\phantom{boo}
\begin{enumerate}[label=(\roman*)]

    \item Using Proposition \ref{rungexprop} and \eqref{mfgtransF}, it can be checked that under a gauge transformation, $A_{g}\mapsto(u^{qg}_{q}/u^{gq}_{q})A_{g}$ and $B'_{g}\mapsto(u^{gq}_{q}/u^{qg}_{q})B_{g}$. The result follows.

    \item By \eqref{rungeq2}, it suffices to show that $A_{g}$ acts as a gate in $\mathcal{S}$ for $g\neq0$. By Proposition \ref{rungexprop}, $A_{g},B_{g}$ are unitary and diagonal; then rung-cancelling Lemma \ref{key-prop} implies that $B_{g}=A^{\dagger}_{g}=A_{g}^{*}$. By \eqref{rungeq2}, $B_{g}=zA_{g}$ where $z:=(K^{gq}_{q})^{*}L^{qg}_{q}$, whence $z^{1/2}A_{g}$ is real (and gauge-invariant up to a global sign). Finally, $\tr(A_{g})=0$ for $g\neq0$ follows from noting that $(F^{qqq}_{q})^{-1}A_{g}F^{qqq}_{q}$ is traceless as it is of the form $\hat{\sigma}(g)$ in \eqref{phafirst}.
    
    \item We must show that no two $\mathcal{O}_{g},\mathcal{O}_{h}$ are related by a scaling factor when $g\neq h$. Proposition \ref{rungexprop} tells us that if this were true, then the $g^{th}$ and $h^{th}$ column of $F^{qqq}_{q}$ would be proportional, contradicting the unitarity of $F^{qqq}_{q}$. Alternatively, the result follows from \ref{sullipeg}.

    \item Without loss of generality, let us fix the fusion basis where the rightmost pair of anyons are the first to be fused, and let $F:=F^{qqq}_{q}$. Then $\mathcal{O}_{i,\id}\circ\mathcal{O}_{\id,j}$ is represented by 
    \begin{equation}
        \mathrm{P}(\mathcal{O}_{i,\id}\circ\mathcal{O}_{\id,j})=\mathrm{P}(\mathcal{O}_{i,\id})\cdot\mathrm{P}(\mathcal{O}_{\id,j})=F\mathcal{O}_{i}F^{\dagger}\cdot\mathcal{O}_{j} =:U_{ij}
    \end{equation}
    Then for distinct pairs $(g_{1},h_{1}),(g_2,h_2)\in G\times G$,
    \[\tr(U^{\dagger}_{g_{1},h_{1}}\cdot U_{g_{2},h_{2}})
    =\tr(\mathcal{O}^{\dagger}_{h_1}F\mathcal{O}^{\dagger}_{g_1}F^{\dagger}\cdot F\mathcal{O}_{g_2}F^{\dagger}\mathcal{O}_{h_2})
    =\tr(\mathcal{O}_{h_2}\mathcal{O}^{\dagger}_{h_1}\cdot F\mathcal{O}^{\dagger}_{g_1}\mathcal{O}_{g_2}F^{\dagger})
    \]
    Suppose $g_{1}\neq g_{2}$. Note that $F\mathcal{O}^{\dagger}_{g_1}\mathcal{O}_{g_2}F^{\dagger}$ is of the form $\hat{\sigma}(g_{1}g_{2})$ in \eqref{phafirst}, and that $\mathcal{O}_{h_2}\mathcal{O}^{\dagger}_{h_1}$ is diagonal, whence the trace vanishes. If $g_{1}=g_{2}$, then we must have $h_{1}\neq h_{2}$, and the inner product amounts to evaluating $\tr(\mathcal{O}_{h_2}\mathcal{O}^{\dagger}_{h_1})$. This quantity vanishes, since the argument of the trace may be conjugated by $F$ to obtain an operator of the form $\hat{\sigma}(h_2 h_1)$ in \eqref{phafirst}.
    \label{sullipeg}
    
\end{enumerate}
\vspace{-3mm}
\end{proof}
\label{rungpropop}
\end{prop}

\noindent In the case where $q$ is an Ising anyon, the gates from Proposition \ref{rungpropop}\ref{orthogoscramble} act on a qubit and define an orthogonal basis for $\End(\mathbb{C}^{2})$; in Corollary \ref{brfrepaul}\ref{paulibasis}, it is shown that this basis is given by the Pauli gates $\{\id,X,Y,Z\}$, whose definition is recalled in \eqref{pmatz}. 

\begin{prop}[\textbf{Relating branches to rungs}] Rung operators $\mathrm{P}[\mathcal{O}_{g,\id}],\Lambda[\mathcal{O}_{\id,g}]$ and $g$-branches $D_{g},D'_{g},E_{g},E'_{g}$ all realise the same quantum gate on a fusion qudit $\ket{\varphi}\in\End(\mathbb{C}^{d})$.
\label{branchswitch}   
\vspace{-1mm}
\begin{proof}
By \eqref{rungeq2}-\eqref{brancheq2}, it suffices to show the following:
\begin{enumerate}[label=(\roman*)]
\item the action of $D_{g}$ on a qudit is given (up to a global phase) by $\mathrm{P}[B_{g,\id}]$ and $\mathrm{\Lambda}[B_{\id,g}]$.
\item the action of $E_{g}$ on a qudit is given (up to a global phase) by $\mathrm{P}[A_{g,\id}]$ and $\mathrm{\Lambda}[A_{\id,g}]$.
\end{enumerate}
\vspace{-3mm}
\begin{subequations}
    \begin{minipage}[c]{0.5\textwidth}
         \begin{align}
            \raisebox{-10mm}{
            \includegraphics[width=0.7\textwidth]{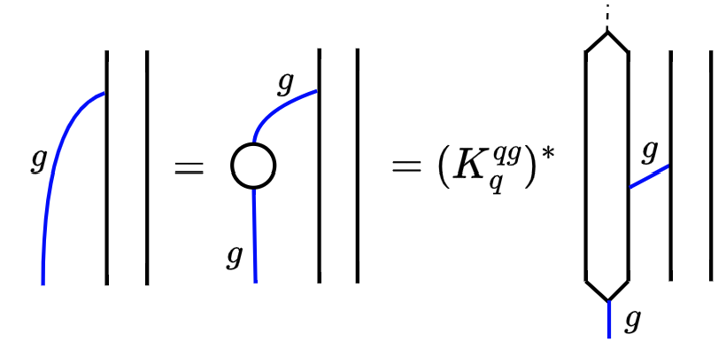}
            }
        \label{flimflam1}    
        \end{align}
    \end{minipage}
    \hfill
    \begin{minipage}[c]{0.45\textwidth}
         \begin{align}
            \raisebox{-12mm}{
            \includegraphics[width=0.575\textwidth]{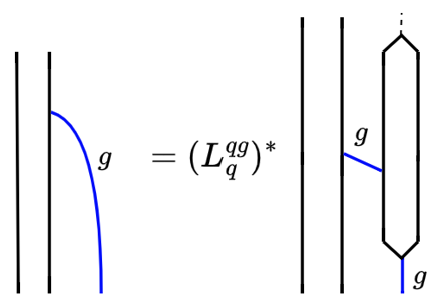}
            }
        \label{flimflam2}    
        \end{align}
    \end{minipage}
\end{subequations}

\noindent Equation \eqref{flimflam2} follows analogously to the manipulation in \eqref{flimflam1}. We give the proof for (i) below by considering \eqref{flimflam1}; the proof of (ii) follows in a similar fashion by considering \eqref{flimflam2}. \\

\noindent The processes on the left and right sides of \eqref{flimflam1} are given by the same element of $\Hom(V^{qq},V^{gqq})$ up to global phase factor $(K^{qg}_{q})^{*}$. This process is annotated in \eqref{annoteq-1}, where we have set the initial state of the pair of TY-anyons to be $\ket{h}$ for any $h\in G$.\footnote{By conservation of charge, the state of the two pairs of TY-anyons is equal (up to a phase factor) to $\ket{g}\otimes\ket{gh}$ immediately after the $g$-rung is applied on the right-hand side of the equation in \eqref{annoteq-1}. \label{consistorama}} By the equivalence of the processes, 
\begin{equation}
    D_{g}\ket{h}=\left(K^{qg}_{q}\right)^{*}\cdot\ket{\psi_h} \ \ , \ \ \ket{\psi_h}=\mathscr{F}_{i}^{-1}B_{g}\mathscr{F}_{i}\ket{h} \ , \ i=1,2
\label{nojuice}    
\end{equation}
where $\mathscr{F}_{i}$ denotes a change of fusion basis on $V^{qqqq}_{h}$ shown in \eqref{annoteq-1}: we know from the pentagon equation \eqref{mfpenteq} that there are two equivalent ways of applying such a change of basis transformation; $\mathscr{F}_{1},\mathscr{F}_{2}$ respectively denote the left path (length $3$) and right path (length $2$) around the pentagon.
\begin{equation}
\raisebox{-15mm}{
\includegraphics[width=0.7\textwidth]{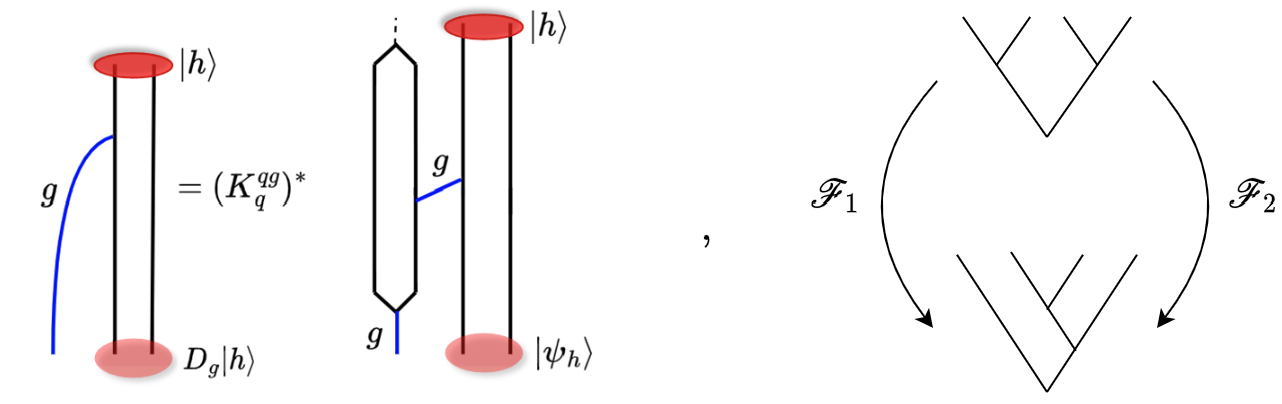}}
    \label{annoteq-1}
\end{equation}
\noindent Taking $i=1$,
\begin{align*}
    \ket{\psi_h}=[F^{gqq}_{h}]_{gh,g}\left(\sum_{x\in G}[G^{qqq}_{q}]_{gx}[G^{qxq}_{h}]_{qq}[B_{g}]_{xx}[F^{qxq}_{h}]_{qq}[F^{qqq}_{q}]_{x0}\right)[G^{0qq}_{h}]_{qh}\ket{h} =[F^{gqq}_{h}]_{gh,g}\cdot\Lambda[B_{\id,g}]\ket{h}
\end{align*}
\noindent where the phase $[F^{gqq}_{h}]_{gh,g}$ is gauge-dependent for $g\neq0$, and is thus set to $1$ as it is unphysical. Taking $i=2$,
\begin{align*}
    \ket{\psi_h}=[G^{q,q,gh}_{h}]_{g,q}\left(\sum_{x}[F^{qqq}_{q}]_{gh,x}[B_{g}]_{xx}[G^{qqq}_{q}]_{xh}]\right)[F^{qqh}_{h}]_{q0}\ket{h}=\left([G^{q,q,gh}_{h}]_{g,q}\cdot [F^{qqh}_{h}]_{q0} \right)\cdot\mathrm{P}[B_{g,\id}]\ket{h}
\end{align*}
\noindent where the phase $[G^{q,q,gh}_{h}]_{g,q}\cdot [F^{qqh}_{h}]_{q0}$ is gauge-dependent for $g\neq0$, and is thus set to $1$ as it is unphysical. Hence, the final state $\ket{\psi_{h}}$ of the pair of TY-anyons is given by $\Lambda[B_{\id,g}]\ket{h}$, or equivalently by $\mathrm{P}[B_{g,\id}]\ket{h}$.\footnote{Using Lemma \ref{quasireg}, we see that both expressions for $\ket{\psi_h}$ are consistent with the observation in footnote \ref{consistorama}.} Plugging the expressions for $\ket{\psi_h}$ into \eqref{nojuice}, statement (i) follows. 
\end{proof}
\end{prop}

\noindent From the above, the action of a branch operators in $\TY(G)$ can be obtained by conjugating the rung operators $\{\mathcal{O}_{g}\}_{g\in G}$ by $F:=F^{qqq}_{q}$. That is, $F$ simultaneously diagonalises the branch operators, and the matrix representation of a $g$-rung is the diagonal matrix of eigenvalues of a $g$-branch. It is then obvious from Lemmas \ref{selly-mlem}-\ref{quasireg} that nontrivial $g$-rungs will be given (up to a global phase) by a traceless matrix of signs, as was shown in Proposition \ref{rungpropop}\ref{p1corr7}.

\subsection{Scrambling operators}
\label{scrambsec}
During the anyon teleportation procedure (Figure \ref{TQT-1dit}), recall that if Alice first measures $j$ and then $i$, the state of Bob's qudit immediately afterwards is $\ket{\varphi_{ij}}=U_{ij}\ket{\varphi}$ (where $\ket{\varphi}$ is the initial qudit that Alice wishes to teleport). The $d^{2}$ possible outcomes (indexed by $(i,j)\in G\times G$) for $\ket{\varphi_{ij}}$ occur with equal probability (Remark \ref{bartok}). We refer to the operators $U_{ij}$ as the \textit{scrambling operators}, and they satisfy the properties listed in Corollary \ref{scrambprop}. When $i=j=0$, no scrambling occurs, meaning that Bob does not have to apply correction gates to recover $\ket{\varphi}$.

\begin{cor}[\textbf{Properties of scrambling operators}]\phantom{boo}
\begin{enumerate}[label=(\roman*)]
 \item The scrambling operators $\{U_{ij}\}_{i,j \in G}$ define an orthogonal basis of $\End(\mathbb{C}^{d})$. \label{genscramb1}
    \item $U_{ij}$ is equivalent to applying a diagonal (and traceless for $j\neq0$) matrix of signs, and then applying a matrix of the form $\mathscr{D}\mathscr{P}$ (where $\mathscr{D}$ is a diagonal unitary, and $\mathscr{P}$ is a self-inverse permutation matrix). $U_{00}$ is the identity matrix, and $(\mathscr{D}\mathscr{P})^{2}=\lambda\cdot\id$ where $|\lambda|=1$. 
\end{enumerate}
\begin{proof}\phantom{boo}
\begin{enumerate}[label=(\roman*)]
    \item Recall from \eqref{scrambleparse} that $U_{ij}=D_{i}\circ B_{j}$. By Proposition \ref{branchswitch}, $D_{i}$ is given (up to a global phase) by $\mathrm{P}[B_{i,\id}]$. Recall from \eqref{diagrep} that $B_{j}=\mathrm{P}[B_{\id,j}]$. Hence, up to a global phase,  $U_{ij}$ is given by $\mathrm{P}[\mathcal{O}_{i,\id}\circ \mathcal{O}_{\id,j}]$. Analogously, it is easy to check that if Bob is instead to the left of Alice, the scrambling operator is given by $\Lambda[\mathcal{O}_{\id,i}\circ \mathcal{O}_{j,\id}]$. By Proposition \ref{rungpropop}\ref{orthogoscramble}, the $d^{2}$ scrambling operators form an orthogonal basis.
    \item The $j$-measurement amounts to the application of a $j$-rung. From Proposition \ref{rungpropop}\ref{p1corr7}, we know that a nontrivial $j$-rung acts as a traceless diagonal matrix of signs on the fusion qudit. The $i$-measurement amounts to the application of an $i$-branch, whose action on the fusion qudit is given by $\hat{\sigma}(i)$ as defined in Lemma \ref{quasireg}. Recall from this lemma that $\hat{\sigma}(i)$ is of the form $\mathscr{D}\mathscr{P}$, and is self-inverse up to a phase factor.
\end{enumerate}
\end{proof}
\label{scrambprop}
\end{cor}
\noindent When $G=\mathbb{Z}_{2}$ (i.e. when Alice is teleporting the state of a pair of Ising anyons), note that $U_{01}$ is equivalent to the Pauli-Z gate. In Corollary \ref{brfrepaul}\ref{paul-x}, we show (for $G=\mathbb{Z}_{2}$) that the matrix $\mathscr{D}$ is of the form $\mathrm{diag}(\omega,\omega^{*})$, and that $\omega$ is gauge-variant (meaning $\mathscr{D}$ has no physical significance). It follows that $U_{10}$ and $U_{11}$ are respectively equivalent to the Pauli-X and Pauli-Y gates.
   
\subsection{Ising anyons: Pauli gates without braiding}
\label{paulisansbrsec}
Let us restrict to an Ising theory ($G=\mathbb{Z}_{2}$) as defined in Section \ref{isingsec}. We can encode a qubit in the fusion state of either a pair or triple of Ising anyons. Following the results of Sections \ref{furtherpropsec}-\ref{scrambsec}, it is easy to check that the Pauli gates $\{\id,X,Y,Z\}$ (whose matrix representations are given in \eqref{pmatz} below) can be realised through the application of rungs and branches. It suffices to show that the Pauli-Z and X gates may be realised in this way, as the Pauli-Y gate may be realised by successive application of the former two gates. 

\begin{equation}
Z=\begin{pmatrix} 1 & 0 \\ 0 & -1 \end{pmatrix}
\ \ , \ \ 
X=\begin{pmatrix}
    0 & 1 \\
    1 & 0
\end{pmatrix} \ \ , \ \ 
Y = iXZ=-iZX
\label{pmatz}
\end{equation}

\vspace{2mm}
\begin{cor}[\textbf{Braid-free Pauli gates using Ising anyons}] \textup{\small (We remind the reader that the definitions and diagrammatic form of the below operators are found in Section \ref{standsec}).}
\begin{enumerate}[label=(\roman*)]
    \item The Pauli-Z gate is realised by a $1$-rung $\mathcal{O}_{1}$. \label{z-rung}
    \item The Pauli-X gate is realised by a $1$-branch, as well as $\mathrm{P}[\mathcal{O}_{1,\id}]$ and $\Lambda[\mathcal{O}_{\id,1}]$. \label{paul-x} 
    \item The Pauli gates are realised by the orthogonal basis $\{\mathcal{O}_{i,\id}\circ\mathcal{O}_{\id,j}\}_{i,j\in\mathbb{Z}_{2}}$ of four $1$-qudit gates acting on $\End(V^{qqq}_{q})$.    \label{paulibasis}
\end{enumerate}    
\begin{proof} The proof below \textit{does not rely on the skeletal data} found in Section \ref{skelesec}: an alternative approach is to use this data together with Proposition \ref{rungexprop}.
\begin{enumerate}[label=(\roman*)]
    \item Applying Proposition \ref{rungpropop}\ref{p1corr7}, we see that $\mathcal{O}_{1}$ will always be equal to $\alpha Z$ where $\alpha$ is a global phase factor. Moreover, this holds true in any gauge. \label{atb}
    \item By Proposition \ref{branchswitch}, it will suffice to show that $\Lambda[\mathcal{O}_{\id,1}]$ realises the $X$-gate. By \eqref{nondiagrep}, $\Lambda[\mathcal{O}_{\id,1}]=F^{\dagger}\mathcal{O}_{1}F$ where $F:=F^{qqq}_{q}$. Then by \ref{atb}, the action of $\Lambda[\mathcal{O}_{\id,1}$] on a qubit is given by $F^{\dagger}ZF=:\mathcal{U}$. Applying Lemma \ref{quasireg} and $\mathcal{U}^{2}=\id$, we get $\mathcal{U}=\omega\ket{0}\bra{1}+\omega^{*}\ket{1}\bra{0}$ where $|\omega|=1$. It only remains to show that the phase $\omega$ is gauge-variant, and may be thus be set to $1$ (as it is unphysical). Expressed in terms of $F$-symbols, $\omega=f_{00}^{*}f_{01}-f_{10}^{*}f_{11}$. Indeed, using \eqref{mfgtransF}, we see that $\omega$ gauge transforms as $\omega\mapsto\omega'=u^{qq}_{0}/(u^{qq}_{1}u^{1q}_{q})$.
    \item Selecting a fusion basis $\Lambda$ (left-to-right fusion) or $\mathrm{P}$ (right-to-left fusion) on $V^{qqq}_{q}$, the result follows from inspection of \eqref{lrreps} and parts \ref{z-rung}-\ref{paul-x}.
\end{enumerate}
\end{proof}
\label{brfrepaul}
\end{cor}

\begin{remark}\phantom{boo}
\begin{enumerate}[label=(\roman*)]
    \item \textit{Textbook teleportation.} Following Corollaries \ref{scrambprop}-\ref{brfrepaul} and the observations in Section \ref{maxentangsec}, it is easily checked for $n=1$ (i.e. when $q$ is an Ising anyon) that the braid-free teleportation procedure in Figure \ref{TQT-1dit} matches the usual qubit teleportation circuit encountered in quantum computing textbooks (e.g. see \cite[Figure 1.13]{ncqc}).
    \vspace{1mm}
    \item \textit{Anticommutativity.} By the observation in footnote \ref{6j-foot}, we  expect all of the quantum gates realised by $\mathcal{O}_{i,\id}$ to commute with those realised by $\mathcal{O}_{\id,j}$, with the possible exception of the instances where $i=j\neq0$ (in which case the gates would only fail to commute by a global phase factor). This is consistent with the results of Corollary \ref{brfrepaul}, where $X$ and $Z$ anticommute: this correctly implies that the gauge-invariant symbol $F^{1q1}_{q}$ is equal to $-1$ in each of the Ising theories $\TY(\mathbb{Z}_{2})$ (see \eqref{ntrivfsymeq}). 
    \vspace{1mm}
    \item \textit{Pauli gates by braiding Ising anyons.} It is easily seen from the skeletal data of Ising theories (Section \ref{skelesec}) that the Pauli gates may be realised through monodromies of Ising anyons.
\end{enumerate}
    
\end{remark}

\subsection{Ising anyons: Teleportation and superdense coding -- \textit{with braiding}}
\label{brcorsec}
Restricting to the Ising theories $\TY(\mathbb{Z}_{2})$ (that is, setting $n=1$), we now present the the braided analogues of Theorems \ref{brfreetelethm} and \ref{brfreesdcthm} as straightforward corollaries of the respective theorems. To this end, it will be useful to first present some of the well-known skeletal data for the Ising theories (Section \ref{skelesec}). In Section \ref{teleisquibrsec}, we show for the $N$-anyon teleportation procedure described in Theorem \ref{brfreetelethm}, that Bob can instead perform his correction step by braiding his Ising anyons (Corollary \ref{brtelecor}). This recovers the main result of \cite{xuzhou}, but with a distinct variation in our teleportation procedure: all braiding operations are deferred to Bob's correction step, which means that no braiding is required at any other stage of the procedure. Similarly, in Section \ref{sdcbrisec}, we show for the superdense coding procedure described in Theorem \ref{brfreesdcthm}, that Alice can instead perform her $2$-bit encoding step by braiding by her Ising anyons (Corollary \ref{brsdccor}).

\subsubsection{Some skeletal data}
\label{skelesec}
 We use the standard column vector representation $(\gamma_0,\gamma_1)^{T}$ for an Ising qubit $\gamma_0\ket{0}+\gamma_1\ket{1}$. Fixing a choice of gauge, the nontrivial $F$-symbols of any Ising theory are presented in \eqref{ntrivfsymeq}, where the Frobenius-Schur indicator $\varkappa_{q}=\pm1$ is a gauge-invariant property of Ising anyon $q$ determined by the given theory. The only nontrivial gauge-invariant $F$-symbols are $[F^{qqq}_{q}]_{00}$ and $[F^{1q1}_{q}]_{qq}$. Note that $F^{qqq}_{q}$ coincides (up to a possible sign) with the Hadamard matrix.
\begin{equation}
F^{qqq}_{q}=\dfrac{\varkappa_{q}}{\sqrt{2}}\begin{pmatrix}
1 & \phantom{-}1 \\
1 & -1
\end{pmatrix}
\ \ , \ \ 
F^{1q1}_{q}=F^{q1q}_{1}=-1
\label{ntrivfsymeq}
\end{equation}
\noindent In any Ising theory, the clockwise monodromy $M^{qq}$ of a pair of Ising anyons is gauge-invariant, and coincides (up to a phase of factor $\vartheta_{q}^{-2}$) with the Pauli-Z matrix. 
\begin{equation}
M^{qq}=\vartheta_{q}^{-2}
\begin{pmatrix}
\vartheta_0 & 0 \\
0 & \vartheta_1
\end{pmatrix}
=\vartheta_{q}^{-2}Z 
\label{zmono}
\end{equation}
Given a triple of Ising anyons, note that we can realise the Pauli gates by winding individual anyons in a loop around their immediate neighbour.

\subsubsection{Teleporting Ising qubits with braiding} 
\label{teleisquibrsec}
\begin{figure}[H]
\includegraphics[width=0.7\textwidth]{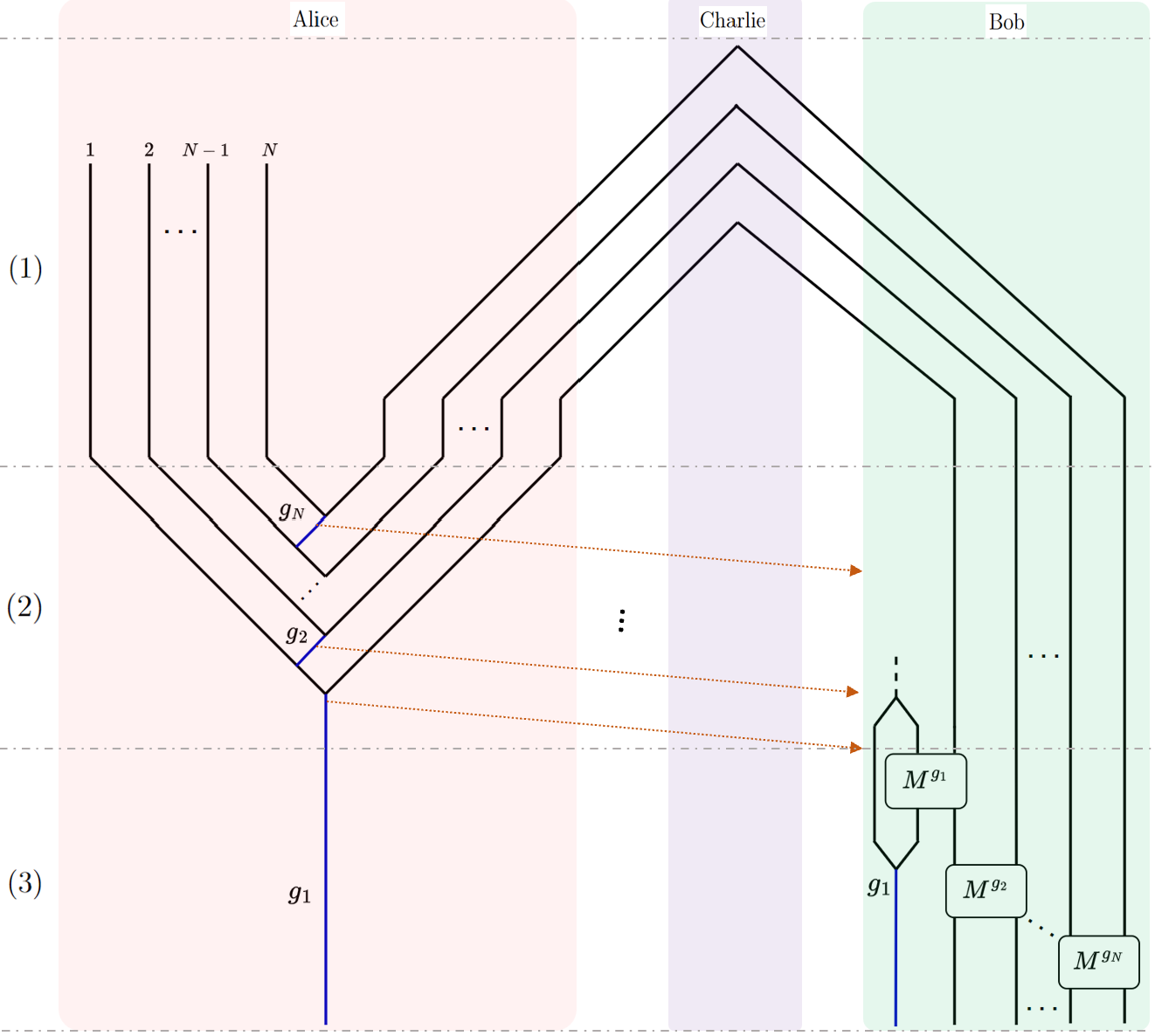}
\vspace{-2mm}
\caption{\small $\lfloor\frac{N}{2}\rfloor$-qubit teleportation protocol using Ising anyons, with worldlines presented according to Convention \ref{tyconvention}. Dotted orange lines are classical $1$-bit transmissions from Alice to Bob. A coupon $M^{g_{i}}$ represents operator $(M^{qq})^{g_i}$ where $g_{i}\in\{0,1\}$; i.e. it is the identity operator if $g_{i}=0$, and is the (clockwise) monodromy operator if $g_{i}=1$.}
\label{palominoscruff-2}
\end{figure}

\begin{cor}
    Consider an Ising theory $\TY(\mathbb{Z}_{2})$. The fusion state of $N$ Ising anyons (that is, a $\lfloor\frac{N}{2}\rfloor$-qubit state) can be teleported via the procedure shown in Figure \ref{palominoscruff-2}. Braiding is not required anywhere other than the correction step, where any necessary corrections are made via monodromies.
    \label{brtelecor}
\end{cor}
\vspace{-2mm}

\begin{proof}(Corollary \ref{brtelecor}). Recall the braid-free teleportation procedure shown in Figure \ref{palominoscruff}: we want to show that the corrections applied by Bob in phase (3) of the protocol coincide with those in Figure \ref{palominoscruff-2}. Firstly, we replace the $g_{1}$-branch with a $g_{1}$-rung using identity \eqref{flimflam1}. That is,
\vspace{-3mm}
\begin{figure}[H]
\includegraphics[width=0.25\textwidth]{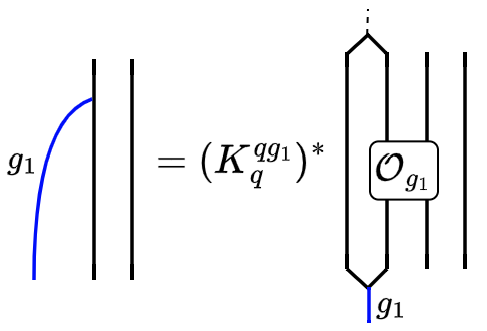}
\end{figure}
\vspace{-4mm}
\noindent It only remains to show that for each $i=1,\ldots,N$, $\mathcal{O}_{g_i}$ is equal to $(M^{qq})^{g_{i}}$ up to a phase factor. Since $q$ is an Ising anyon, we have $\mathcal{O}_{g_i}=(\mathcal{O}_{1})^{g_i}$ where $g_{i}\in\{0,1\}$. The result then follows from equation \eqref{zmono} and Corollary \ref{brfrepaul}\ref{z-rung}.
\end{proof}

\begin{remark}[\textbf{Freedoms in procedure}]
    In addition to the applicable freedoms summarised in Section \ref{telefreesec}, we note that \ref{firstorc}-\ref{lastorc} below apply to the teleportation procedure shown in Figure \ref{palominoscruff-2}. As stated in point \ref{x-freedoms} of Section \ref{telefreesec}, there is freedom in how Bob corrects the bit-flip errors on his qubits, which is explained in Section \ref{locatesec}: if Bob applies corrections by braiding his Ising anyons as above, then Remark \ref{brcoremk} (which is simply a reminder that rungs and branches can be replaced with monodromies in $\TY(\mathbb{Z}_{2})$) applies to the process depicted in \eqref{bcoup}.
    \begin{enumerate}[label=(\roman*)]
        \item Bob may freely choose the orientation (clockwise or anticlockwise) of each monodromy that he applies, since $M^{qq}$ and its inverse differ only by a global phase factor. \label{firstorc}
        
        \item The order in which Bob applies monodromies does not matter, since reordering only incurs a global phase factor of $+1$ or $-1$ (i.e. the coupons may slide past one another).
        
        \item Bob has the freedom to correct for the effect of any given measurement $g_i=1$ using either a rung operator $\mathcal{O}_{1}$ or a monodromy $(M^{qq})^{\pm1}$. \label{lastorc}
    
    \end{enumerate}
    \label{brteleremk}
\end{remark}

\subsubsection{Superdense coding by braiding Ising anyons} 
\label{sdcbrisec}

\begin{figure}[H]
\includegraphics[width=0.4\textwidth]{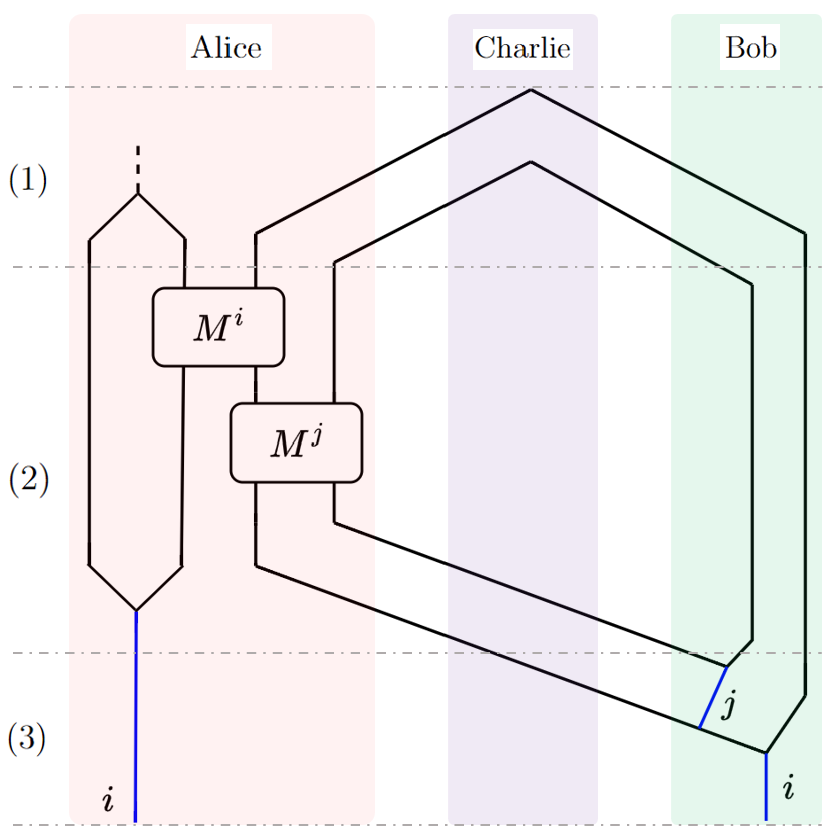}
\vspace{-2mm}
\caption{\small Binary superdense coding protocol using Ising anyons. Worldlines are presented according to Convention \ref{tyconvention}. The coupons are defined in the same way as Figure \ref{palominoscruff-2}.}
\label{sdcbrdiag}
\end{figure}

\begin{cor}
   Consider an Ising theory $\TY(\mathbb{Z}_{2})$.  We can realise the binary superdense coding protocol using Ising anyons, as shown in Figure \ref{sdcbrdiag}. Alice encodes her chosen bits $i,j\in\mathbb{Z}_{2}$ by performing monodromies on her anyons. Braiding is not required anywhere else.
    \label{brsdccor}
\end{cor}

\begin{proof}
    (Corollary \ref{brsdccor}). Recall the braid-free superdense coding procedure shown in Figure \ref{sdcbrfreediag}: we want to show that the encodings performed by Alice in phase (2) of the protocol coincide with those in Figure \ref{sdcbrdiag}. The remainder of the proof is identical to that of Corollary \ref{brtelecor}. 
\end{proof}

\noindent In addition to the applicable freedoms summarised in Section \ref{sdcfreesec}, the procedure in Figure \ref{sdcbrdiag} enjoys monodromy freedoms analogous to those described in Remark \ref{brteleremk}.\\

\subsection{Ising anyons: Locating bit-flip errors}
\label{locatesec}
    In Proposition \ref{branchswitch}, we saw for $p=1$ that the effect of a $g$-branch on the state of $2p$ TY-anyons does not depend on whether it is applied to their left or their right. Let $G=\mathbb{Z}_{2}$ so that the TY-anyons are Ising anyons. For convenience, we fix the gauge from Section \ref{skelesec} so that the matrix representations of a $1$-rung and a $1$-branch are respectively given by the Pauli-$Z$ and Pauli-$X$ matrices. 
    \vspace{-1mm}

    \begin{question}\textit{When $p\geq1$, can we transform a $1$-branch from one side of the $2p$ Ising anyons all the way to the other side?}\footnote{Recall from \eqref{paired-basis} that $V^{q^{2p+1}}\cong V^{q^{2p}}$ for $p\geq1$. That is, given $2p+1$ Ising anyons, we can discard one of the outermost anyons without affecting the encoded $p$-qubit fusion state. We thus take $N=2p$ in this section.}
    \label{brantranq}
    \end{question}

    \noindent \textit{Move 1:} \raisebox{-8mm}{
\includegraphics[width=0.2\textwidth]{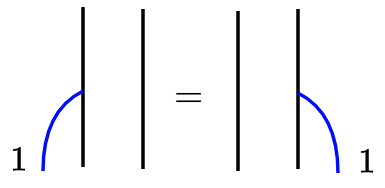}} \ \ , \ \ \textit{Move 2a:}  \raisebox{-10mm}{
\includegraphics[width=0.15\textwidth]{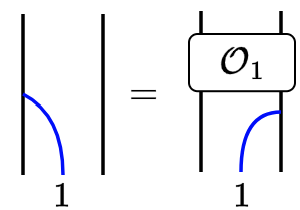}} \ \ , \ \ \textit{Move 2b:}  \raisebox{-10mm}{
\includegraphics[width=0.15\textwidth]{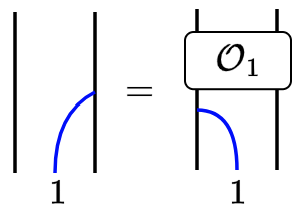}}

    \noindent We can apply the desired transformation through application of Moves 1 and Moves 2a (or 2b). While Move 1 is due to Proposition \ref{branchswitch}, Move 2a follows from equation \eqref{spannereq}.\footnote{Independently of the choice of gauge, the final equality in \eqref{spannereq} is well-defined as a physical equivalence, as the left-hand side is scaled by a global phase factor of $u^{q1}_{q}/u^{1q}_{q}$ under a gauge transformation. By Proposition \ref{rungpropop}\ref{gtransrung}, the right-hand side also scales by a global phase factor under a gauge transformation. Hence, both sides of the equality differ at most by a global phase factor.} Move 2b is implied by Move 2a together with $\mathcal{O}_{1}^{2}=\id$.
    \vspace{-3mm}

\begin{equation}
\raisebox{-10mm}{
\includegraphics[width=0.65\textwidth]{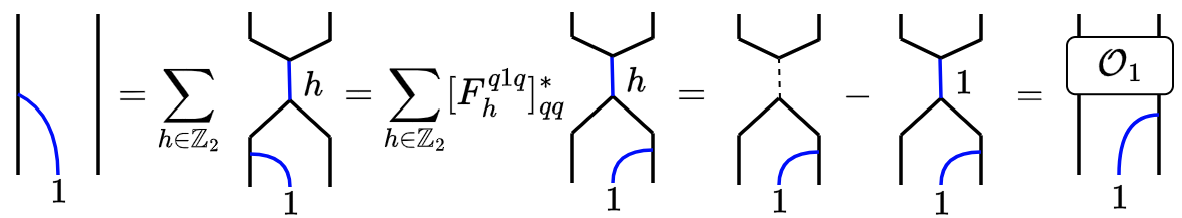}}
    \label{spannereq}
\end{equation}
\vspace{-3mm}

\begin{figure}[H]
\includegraphics[width=0.35\textwidth]{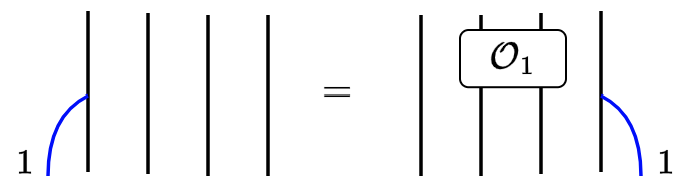}
\vspace{-2mm}
\caption{\small \textit{Example:} we transform a $1$-rung from the left to the right of a $2$-qubit system by applying Moves 1, 2a, and then 1 again.}
\label{2qu1ru}
\end{figure}

\noindent This can be applied as a freedom in how Bob performs his corrections in the $p$-qubit teleportation protocol for Ising anyons (see Figure \ref{palominoscruff}, setting $N=2p$ and $G=\mathbb{Z}_{2}$). Let us relabel Alice's measurement outcomes as $g_{2k-1}=:x_{k}$ and $g_{2k}=:z_{k}$, where $1\leq k \leq p$. We also introduce the variables $x'_{k},z'_{k}$ for $1 \leq k \leq p$ and $x'_{p+1}$, according to which Bob applies his corrections. Let us denote Bob's correction process by a coupon as in \eqref{bcoup} that depends on these variables.
\begin{equation}
\raisebox{-25mm}{
\includegraphics[width=0.85\textwidth]{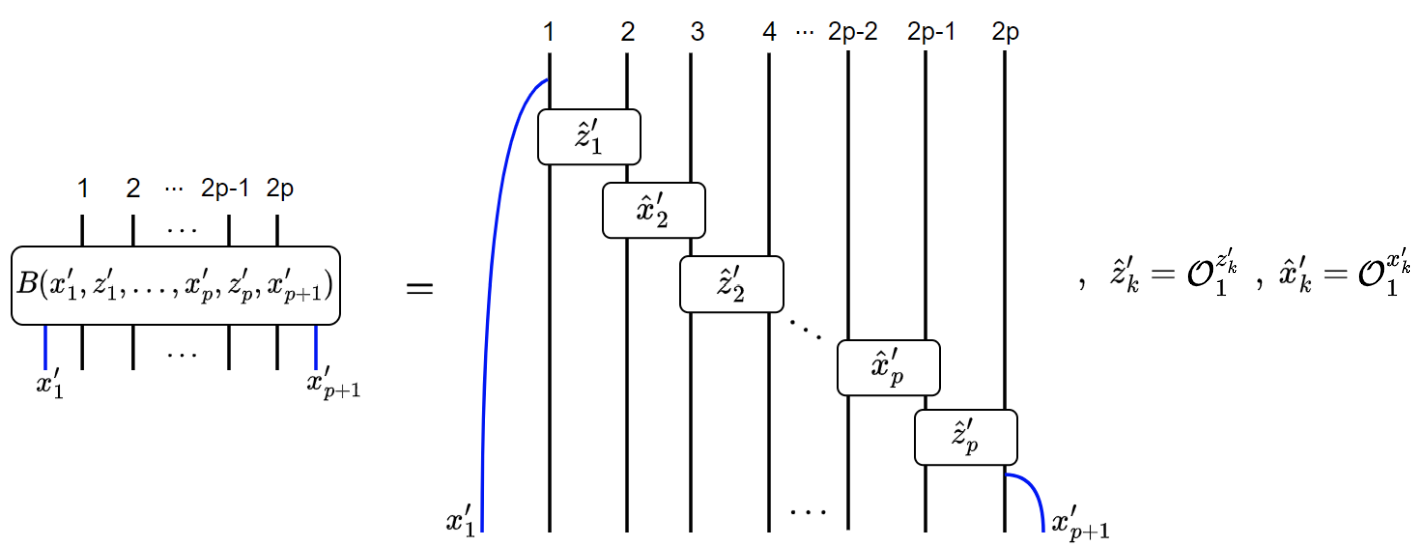}}
    \label{bcoup}
\end{equation}

\begin{remark}[\textbf{Braided case}]
Suppose Bob wishes to apply corrections by braiding his Ising anyons (as in Figure \ref{palominoscruff-2}, Section \ref{teleisquibrsec}). In this case, the process shown in \eqref{bcoup} is adapted as follows. As shown in identities \eqref{flimflam1}-\eqref{flimflam2}, the $x'_{1}$ and $x'_{p+1}$-branches can respectively be replaced by rung operators $\mathcal{O}_{1}^{x'_{1}}$ and $\mathcal{O}_{1}^{x'_{p+1}}$, each of which are attached to an ancillary pair of Ising anyons. To finish, we recall that $\mathcal{O}_{1}$ can be replaced by a monodromy operator $(M^{qq})^{\pm1}$.\\
    \label{brcoremk}
\end{remark}

\noindent We refer to the bit string $e$ of Alice's measurement outcomes as the \textit{error string},
\begin{equation}  
    e:=(x_1,z_1,\ldots,x_p,z_p)\in\mathbb{Z}_{2}^{2p}
\end{equation} 
and the argument $c$ of Bob's coupon as a \textit{correction string},
\begin{equation}
    c:=(x'_1,z'_1,\ldots,x'_p,z'_p,x'_{p+1})\in\mathbb{Z}_{2}^{2p+1}
\end{equation}
\noindent By Theorem \ref{brfreetelethm}, we know for any given $e$ that \eqref{canonical-correction} constitutes a valid choice of correction string: we denote this `canonical' choice by $c_0$.
\begin{equation} x'_{k}=x_{k} \ \ , \ \ z'_{k}=z_{k} \ \ , \ \ x'_{p+1}=0 \ \ , \ \ 1\leq k \leq p
\label{canonical-correction}
\end{equation}
\noindent From the discussion above, it is evident that an error string does not always uniquely determine a correction string. For instance, suppose we have some $e$ such that $x_{1}=1$. Then we can use Moves 1 and 2a (together with the freedom to slide adjacent coupons past one another)\footnote{Swapping the order of two adjacent coupons results in a global phase factor of $\pm1$, which does not matter.} to transform the valid pair $(e,c_0)$ into another valid pair $(e,c_1)$. That is, 
\begin{equation}
c_{0}=(1,z_{1},x_{2},z_{2},\ldots,x_{p},z_{p},0) \longmapsto c_{1}=(0,z_{1},\overbar{x}_{2},z_{2},\ldots,\overbar{x}_{p},z_{p},1)
\label{deformer}
\end{equation}
where we use $\overbar{a}$ to denote the additive inverse of $a\in\mathbb{Z}_{2}$ (i.e. the `bit-flip'). 
\begin{remark}
Setting $c_{0}=(1,0,\ldots,0)$ in \eqref{deformer} provides a formal answer to Question \ref{brantranq}.
\end{remark}

\noindent The next natural question is as follows.

\begin{question}
    Given $e\in\mathbb{Z}_{2}^{2p}$, what is the subset $\mathcal{S}_{e}\subset\mathbb{Z}_{2}^{2p+1}$ of valid correction strings?
    \label{finq}
\end{question}
\noindent In addition, one might seek to determine the subset of $\mathcal{S}_{e}$ containing strings of minimal Hamming weight: this would allow Bob to apply corrections using the minimum number of operations.\\

\begin{figure}[H]
\includegraphics[width=0.6\textwidth]{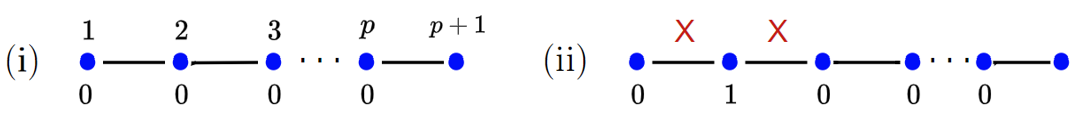}
\vspace{-3mm}
\caption{\small Visual aid for bit-flip errors on a $p$-qubit state. Let $\ket{\varnothing}$ denote some initial $p$-qubit state with no bit-flip errors. Vertices (blue dots) are enumerated from $1$ to $p+1$. For $k<p+1$, the value of Alice's $x_k$ measurement outcome is shown beneath the $k^{th}$ vertex. Adjacent vertices are connected by edges, where the $k^{th}$ edge represents the $k^{th}$ qubit. We usually omit vertex enumeration as in diagram (ii). When the $k^{th}$ qubit is affected by a bit-flip error, we place a red cross above the $k^{th}$ edge. If there are no red crosses, the system is in state $\ket{\varnothing}$ as in diagram (i). A vertex operator $v_{k}$ applies a bit-flip to edges connected to the $k^{th}$ vertex. For example, diagram (ii) shows the state $v_{2}\ket{\varnothing}$.}
\label{oxfig1}
\end{figure}

\begin{figure}[H]
\includegraphics[width=0.6\textwidth]{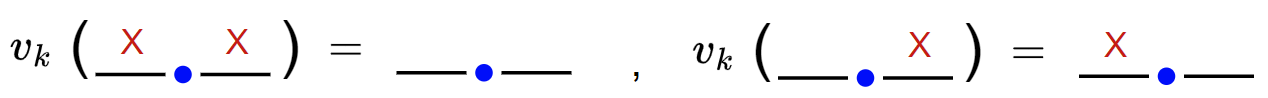}
\vspace{-3mm}
\caption{\small Self-inverse operator $v_{k}$ acting at $k^{th}$ vertex, $1<k<p+1$. \textit{Left:} $v_{k}$ can pair-create and annihilate $X$-errors. \textit{Right:} $v_{k}$ can propagate $X$-errors to the left or right.}
\label{oxfig2}
\end{figure}

\noindent Question \ref{finq} motivates us to better understand the relationship between error string $e$, and the state of Bob's $2p$ Ising anyons immediately before his correction step. We fix the $2p$-anyon fusion basis specified in \eqref{paired-basis}, so that the enumeration of the $p$ qubits goes from left to right. A measurement outcome $z_{k}=1$ results in a $Z$-error on the $k^{th}$ qubit. The only way for Bob to correct such an error is to apply a Pauli-Z gate to his $k^{th}$ pair of Ising anyons, i.e. it is always required that $z'_{k}=z_{k}$ in a valid correction string. Hence, the main task is to unpack how $X$-errors (that is, bit-flip errors) occur and how they are corrected. Let $X_{k}$ denote the $p$-qubit operator that applies a Pauli-X to the $k^{th}$ qubit. For $1<k\leq p$, a measurement outcome $x_{k}=1$ results in an $X$-error on both the $(k-1)^{th}$ and $k^{th}$ qubit. A measurement outcome $x_{1}=1$ results in an $X$-error on solely the first qubit. It will be useful to introduce the \textit{vertex operators} $\{v_{k}\}_{k=1}^{p+1}$, defined as follows. The name of these operators is made clear by the context provided in Figures \ref{oxfig1}-\ref{oxfig2}.
\begin{equation}
    v_{k}=
    \begin{cases}
        X_{k-1}X_{k} \ \ , \ \ 1 < k < p+1 \\ 
        X_{1} \ \ , \ \ k=1 \\
        X_{p} \ \ , \ \ k=p+1
    \end{cases}
\end{equation}

\vspace{-4mm}
\vspace{-2mm}
\noindent Upon Alice's measurements, the $p$-qubit state accumulates $X$-errors according to the operator
\begin{equation}
    A^{(x)}_{e}=v_{1}^{x_{1}}\cdots v_{k-1}^{x_{k-1}}v_{k}^{x_{k}}
\end{equation}
We now obtain a straightforward picture of the relationship between an error string $e$ and the consequent locations of $X$-errors. For $p>1$, note that the effect of $A^{(x)}_{e}$ is as follows.
\begin{enumerate}[label=(\arabic*)]
    \item $X$-errors are pair-created. A pair-creation is possibly followed by leftward propagation of the leftmost error of said pair. \label{gen-1}

    \item Suppose $x_{1}=1$. Then, the total number of $X$-errors must be odd. This can happen in two ways: (i) if $x_{2}=1$, the leftmost $X$-error will have been propagated all the way `out' of the system during the final iteration of \ref{gen-1}; (ii) if $x_{2}=0$, then there is an $X$-error on the first qubit which has not come from a pair-creation, but rather from the action of $v_{1}$ alone.  
\end{enumerate}

\begin{ex}
    We tabulate the location of bit-flip errors on a $4$-qubit system for all possible measurement outcomes $(x_{1},x_{2},x_{3},x_{4})\in\mathbb{Z}_{2}^{4}$.\\
    \begin{centering}
    \includegraphics[width=0.95\textwidth]{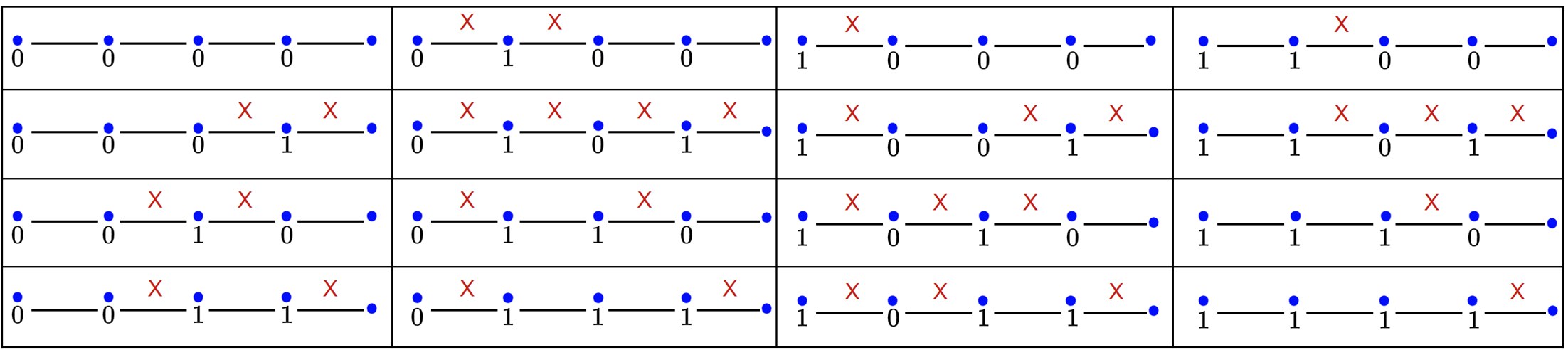}
    \end{centering}
\end{ex}

\noindent Upon Bob's corrections, the $p$-qubit state is acted on by an operator
\begin{equation}
    B_{c}^{(x)}=v_{k}^{x'_{k}}v_{k-1}^{x'_{k-1}}\cdots v_{1}^{x'_{1}}
    \label{bcorop}
\end{equation}

\noindent We know that answering Question \ref{finq} is the same as determining all $c$ such that $B_{c}^{(x)}A_{e}^{(x)}=\id$ for a given $e$. The canonical prescription $c=c_{0}$ from \eqref{canonical-correction} precisely\textit{ reverses the process} by which $X$-errors were accumulated; algebraically, that $B_{c_0}^{(x)}=(A_{e}^{(x)})^{-1}$ is easily seen by recalling that vertex operators are self-inverse.

\begin{remark}[\textbf{Efficient removal of $X$-errors}]
While canonical prescription \eqref{canonical-correction} is a catch-all, it is often possible to apply corrections in a more economical way (i.e. we can minimise the Hamming weight of $c$). Of course, the improvement in thrift becomes more pronounced for $p$ large. To demonstrate this point, we describe a few basic scenarios.
\begin{enumerate}[label=(\roman*)]
    \item As an extreme example, suppose Alice measured $x_{k}=1$ for all $1\leq k \leq p$. The net result is a single bit-flip error on the $p^{th}$ qubit. The canonical correction method generates an $X$-error at edge $1$, propagatse it to edge $p-1$, and then pair-annihilates the errors. On the other hand, we can apply the transformation \eqref{deformer} to $c_0$, giving $B_{c_1}^{(x)}=v_{k+1}$. This latter method instantly propagates the lone $X$-error `out' of the right boundary of the system. 
    \item More generally, given an isolated $X$-error at edge $r$ where $p-r$ is small, it is preferable to propagate this `out' of the system from the right boundary.
    \item Suppose we have $X$-errors at edges $r$ and $s$ (with none in-between and $r<s$) where $p-(s-r)$ is small, and where the canonical prescription would pair-annihilate them. It is preferable to remove them by respectively propagating the errors at sites $r$ and $s$ left and right. 
\end{enumerate}
\end{remark}

\noindent To summarise, a solution to Question \ref{finq} is sketched as follows. Although we must set $z'_{k}=z_{k}$ in any valid $c$ (i.e. the way in which we remove $Z$-errors is fixed), there is freedom in how we remove $X$-errors for $p\geq1$: given error string $e$, we can use the visual aid in Figure \ref{oxfig1} to determine valid correction operators $B_{c}^{(x)}$ for the removal of $X$-errors.


\begin{thebibliography}{99}

\bibitem{halperin} Halperin, B.I. (1984). "Statistics of Quasiparticles and the Hierarchy of Fractional Quantized Hall States", \textit{Physical Review Letters}, 52(18).

\bibitem{arovas} Arovas, D., Schrieffer, J.R. \& Wilczek, F. (1984). "Fractional Statistics and the Quantum Hall Effect", \textit{Physical Review Letters}, 53(7).

\bibitem{nakamura} Nakamura, J., Liang, S., Gardner, G.C. \& Manfra, M.J. (2020). "Direct Observation of Anyonic Braiding Statistics". \textit{Nature Physics}, 16.

\bibitem{bartolomei} Bartolomei, H. \textit{et al.} (2020). "Fractional Statistics in Anyon Collisions". \textit{Science}, 368(6487). 

\bibitem{willett} Willett, R.L. \textit{et al.} "Interference Measurements of Non-Abelian $e/4$ and Abelian $e/2$ Quasiparticle Braiding". \textit{Physical Review X}, 13(1). 

\bibitem{dSmzm} Sarma, S., Freedman, M. \& Nayak, C. (2015). "Majorana Zero Modes and Topological Quantum Computation". \textit{npj Quantum Inf}, 1.

\bibitem{kitaev01} Kitaev, A.Y. (2001). "Unpaired Majorana Fermions in Quantum Wires". \textit{Physics-Uspekhi}, 44(10S).

\bibitem{bondersonMO} Bonderson, P., Freedman, M. \& Nayak, C. (2008). "Measurement-Only Topological Quantum Computation". \textit{Physical Review Letters}, 101(1).

\bibitem{simonbook} Simon, S.H. (2023). "Topological Quantum". \textit{Oxford University Press}. 

\bibitem{wangvideo} Wang, Z. "Reconstructing CFTs from TQFTs". \textit{YouTube}, uploaded by Mathematical Picture Language, 7 October 2020. \url{https://www.youtube.com/watch?v=75M9rtsnmXA}

\bibitem{Bonesteel} Bonesteel, N.E., Hormozi, L., Zikos, G. \& Simon, S.H. (2007). "Quantum Computing with Non-Abelian Quasiparticles". \textit{International Journal of Modern Physics B}, 21(8-9).

\bibitem{sjvyt1} Valera, S.J. "Quantum Teleportation using Majorana Zero Modes (Ising Anyons) -- Without Braiding". \textit{YouTube}, uploaded by Sachin J. Valera, 19 January 2024. \url{https://www.youtube.com/watch?v=9rksGBJBezc}

\bibitem{sjvyt2} Valera, S.J. "Superdense Coding using Majorana Zero Modes (Ising Anyons) -- Without Braiding". \textit{YouTube}, uploaded by Sachin J. Valera, 29 April 2024. \url{https://www.youtube.com/watch?v=ljIF56U-ogI}

\bibitem{interf1} Overbosch, B.J. \& Bais, F.A. (2001). "Inequivalent Classes of Interference Experiments with Non-Abelian Anyons". \textit{Physical Review A}, 64.

\bibitem{interf2} Bonderson, P., Kitaev, A. \& Shtengel, K. (2006). "Detecting Non-Abelian Statistics in the v=5/2 Fractional Quantum Hall State". \textit{Physical Review Letters}, 96.

\bibitem{interf3} Bonderson, P., Shtengel, K. \& Slingerland, J.K. (2008). "Interferometry of Non-Abelian Anyons". \textit{Annals of Physics}, 323(11).

\bibitem{bonderson-thesis} Bonderson, P.H. (2007). "Non-Abelian Anyons and Interferometry". PhD thesis, \textit{Caltech}.

\bibitem{sv24} Sati, H. \& Valera, S.J. "Topological Quantum Computing". (Encylopaedia entry; in preparation).

\bibitem{huang} Huang \textit{et al.} (2021). "Emulating Quantum Teleportation of a MZM Qubit". \textit{Phys. Rev. Lett.}, 126(9). 

\bibitem{xuzhou} Xu, C.Q. \& Zhou, D.L. (2022). "Quantum Teleportation using Ising Anyons". \textit{Physical Review A}, 106(1).


\bibitem{abracoke} Abramsky, S. \& Coecke, B. (2004). "A Categorical Semantics of Quantum Protocols". \textit{Proceedings of the 19th Annual IEEE Symposium on Logic in Computer Science}.

\bibitem{kindergarten} Coecke, B. (2006). "Kindergarten Quantum Mechanics: Lecture Notes". \textit{AIP Conference Proceedings}, 810(1).

\bibitem{vicaryheunen} Heunen, C. \& Vicary, J. (2019). "Categories for Quantum Theory: an Introduction". \textit{Oxford University Press}.

\bibitem{wangbook} Wang, Z. (2010). "Topological Quantum Computation". \textit{American Mathematical Society}.

\bibitem{kitaev06} Kitaev, A. (2006). "Anyons in an Exactly Solved Model and Beyond". \textit{Annals of Physics}, 321(1).

\bibitem{Gal14} Galindo, C. (2014). "On Braided and Ribbon Unitary Fusion Categories". \textit{Canad. Math. Bull.}, 57(3).

\bibitem{val21} Valera, S.J. (2021). "Fusion Structure from Exchange Symmetry in $(2 + 1)$-Dimensions". \textit{Annals of Physics}, 429. 

\bibitem{egno} Etingof, P., Gelaki, S., Nikshych, D. \& Ostrik, V. (2016). "Tensor Categories", \textit{American Mathematical Society}.

\bibitem{ENO} Etingof, P., Nikshych, D. \& Ostrik, V. (2005). "On Fusion Categories". \textit{Annals of Mathematics}.

\bibitem{bartlett} Bartlett, B. (2016). "Fusion Categories via String Diagrams". \textit{Commun. Contemp. Math.}, 18(05).

\bibitem{GHS} Gainutdinov, A.M., Haferkamp, J. \& Schweigert, C. (2023). "Davydov-Yetter Cohomology, Comonads and Ocneanu Rigidity".\textit{ Advances in Mathematics}, 414.

\bibitem{vafa} Vafa, C. (1988). "Towards Classification of Conformal Theories". \textit{Physics Letters B}, 206(3).

\bibitem{wolf} Wolf, R. (2020). "Microscopic Models for Fusion Categories". PhD thesis, \textit{Leibniz Universit\"{a}t Hannover}. 

\bibitem{rfgxbfc} Jones, C., Morrison, S., Nikshych, D. \& Rowell, E.C. (2021). "Rank-Finiteness for G-Crossed Braided Fusion Categories". \textit{Transformation Groups}, 26(3).

\bibitem{rsw} Rowell, E.C., Stong, R. \& Wang, Z. (2009)."On Classification of Modular Tensor Categories". \textit{Communications in Mathematical Physics}, 292(2).

\bibitem{anyonwiki} "AnyonWiki". URL: \url{https://anyonwiki.github.io/} 

\bibitem{TYog} Tambara, D. \& Yamagami, S. (1998). "Tensor Categories with Fusion Rules of Self-Duality for Finite Abelian Groups". \textit{Journal of Algebra}, 209(2).

\bibitem{ghr11} Galindo, C., Hong, S.M. \& Rowell, E.C. (2013). "Generalized and Quasi-Localizations of Braid Group Representations". \textit{International Mathematics Research Notices}, 2013(3).

\bibitem{siehler} Siehler, J.A. (2000). "Braided Near-Group Categories". \textit{Preprint}. \url{https://arxiv.org/abs/math/0011037}

\bibitem{ncqc} Nielsen, M.A. \& Chuang, I.L. (2010). "Quantum Computation and Quantum Information". \textit{Cambridge University Press}.

\end{thebibliography}
\end{document}